\documentclass{aa}
\usepackage{graphicx, color}
\usepackage{xcolor}
\usepackage{natbib}
\usepackage{hyperref}
\usepackage{threeparttable}

\usepackage{txfonts}
\DeclareRobustCommand{\VAN}[3]{#2}
\let\VANthebibliography\thebibliography
\def\thebibliography{\DeclareRobustCommand{\VAN}[3]{##3}\VANthebibliography}

\makeatletter
\renewcommand*\aa@pageof{, page \thepage{} of \pageref*{LastPage}}
\makeatother

\begin{document} 

\title{CONCERTO at APEX -  On-sky performance in continuum}
\subtitle{}

\author{W. Hu\inst{\ref{lam}, \ref{uwc}} 
\and A. Beelen\inst{\ref{lam}}
\and G. Lagache\inst{\ref{lam}}
\and A. Fasano\inst{\ref{lam}, \ref{iac}, \ref{ull}}
\and A. Lundgren\inst{\ref{lam}}
\and P. Ade\inst{\ref{aig}}
\and M. Aravena\inst{\ref{Santiago}}
\and E. Barria\inst{\ref{neel}}
\and A. Benoit\inst{\ref{neel}}
\and M. B$\rm\acute{e}$thermin\inst{\ref{lam}, \ref{stg}}
\and J. Bounmy\inst{\ref{lpsc}}
\and O. Bourrion\inst{\ref{lpsc}}
\and G. Bres\inst{\ref{neel}}
\and C. De Breuck\inst{\ref{eso_germany}}
\and M. Calvo\inst{\ref{neel}}
\and A. Catalano\inst{\ref{lpsc}}
\and F.-X. D$\rm\acute{e}$sert \inst{\ref{ipag}}
\and C. Dubois\inst{\ref{lam}}
\and C.A Dur$\rm\acute{a}$n\inst{\ref{IRAME}}
\and T. Fenouillet\inst{\ref{lam}}
\and J. Garcia\inst{\ref{lam}}
\and G. Garde\inst{\ref{neel}}
\and J. Goupy\inst{\ref{neel}}
\and C. Hoarau\inst{\ref{lpsc}}
\and J.-C. Lambert\inst{\ref{lam}}
\and E. Lellouch\inst{\ref{lesia}}
\and F. Levy-Bertrand\inst{\ref{neel}}
\and J. Macias-Perez\inst{\ref{lpsc}}
\and J. Marpaud\inst{\ref{lpsc}}
\and A. Monfardini\inst{\ref{neel}}
\and G. Pisano\inst{\ref{aig}}
\and N. Ponthieu\inst{\ref{ipag}}
\and L. Prieur\inst{\ref{lam}}
\and D. Quinatoa\inst{\ref{chili_Daysi}}
\and S. Roni\inst{\ref{lpsc}}
\and S. Roudier\inst{\ref{lpsc}}
\and D. Tourres\inst{\ref{lpsc}}
\and C. Tucker\inst{\ref{aig}}
\and M. Van Cuyck\inst{\ref{lam}} 
}

\institute{Aix Marseille Univ, CNRS, CNES, LAM, Marseille, France \label{lam}\\
              \email{wenkai.hu@lam.fr}
        \and 
            Department of Physics and Astronomy, University of the 
            Western Cape, Robert Sobukhwe Road, Bellville, 7535, South Africa\label{uwc} 
        \and 
            Instituto de Astrofísica de Canarias, E-38205 La Laguna, Tenerife, Spain\label{iac}
        \and
            Departamento de Astrofísica, Universidad de La Laguna (ULL), E-38206 La Laguna, Tenerife, Spain\label{ull}
        \and 
            Astronomy Instrumentation Group, University of Cardiff, The Parade, CF24 3AA, United Kingdom\label{aig}
        \and
            Instituto de Estudios Astrof\'{\i}sicos, Facultad de Ingenier\'{\i}a y Ciencias, Universidad Diego Portales, Av. Ej\'ercito 441, Santiago, Chile\label{Santiago}   
        \and
            Univ. Grenoble Alpes, CNRS, Grenoble INP, Institut Néel, 38000 Grenoble, France\label{neel}
        \and
             Université de Strasbourg, CNRS, Observatoire astronomique de Strasbourg, UMR 7550, 67000 Strasbourg, France\label{stg}
        \and
            Univ. Grenoble Alpes, CNRS, Grenoble INP, LPSC-IN2P3, 53, avenue des Martyrs, 38000 Grenoble, France\label{lpsc}   
        \and
            European Southern Observatory, Karl Schwarzschild Straße 2, 85748 Garching, Germany\label{eso_germany}
        \and
            Univ. Grenoble Alpes, CNRS, IPAG, 38400 Saint Martin d’Héres, France\label{ipag}
        \and    
            Instituto de Radioastronom\'ia Milim\'etrica (IRAM), Granada, Spain\label{IRAME}
        \and
            LESIA, Observatoire de Paris, PSL Research University, CNRS, Sorbonne Universit$\rm\acute{e}$, UPMC Univ. Paris 06, Univ. Paris Diderot, Sorbonne Paris Cit$\rm\acute{e}$, 5 place Jules Janssen, 92195 Meudon, France\label{lesia}  
        \and
            Instituto de Física y Astronomía, Universidad de Valparaíıso, Avda. Gran Bretaña 1111, Valparaíso, Chile\label{chili_Daysi} 
            }
\date{Received January 8, 2024; accepted XXX, XXXX}

 
\abstract
   {CONCERTO (CarbON CII line in post-rEionisation and ReionisaTiOn epoch) is a low-resolution mapping spectrometer based on lumped element kinetic inductance detectors (LEKIDs) technology, operating at 130--310\,GHz. It was installed on the 12-meter APEX telescope in Chile from April 2021 to May 2023. CONCERTO's main goals were the observation of [CII]-emission line fluctuations at high redshift and of the SZ signal from galaxy clusters.}
   {We present the data-processing algorithms and the performance of CONCERTO in continuum by analysing the data from the commissioning and scientific observations.}
   {We develop a standard data-processing pipeline to proceed from the raw data to continuum maps.
   Using a large data set of calibrators (Uranus, Mars and quasars) acquired in 2021 and 2022 at the APEX telescope across a wide range of atmospheric conditions, we measure the CONCERTO continuum performance and test its stability against observing conditions. Further, using observations on the COSMOS field and observations targeting a distant sub-millimetre galaxy in the UDS field, we assess the robustness of the CONCERTO performance on faint sources and compare our measurements with expectations.}
   {The beam pattern is characterized by an effective full width at half maximum (FWHM) of $31.9\pm0.6\,\arcsec$ and $34.4\pm1.0\,\arcsec$ for high frequency (HF) and low frequency (LF) bands, respectively. The main beam is slightly elongated with a mean eccentricity of 0.46. 
   Two error beams of $\sim65\arcsec$ and $\sim 130\arcsec$ are characterized, enabling the estimate of a main beam efficiency of $\sim 0.52$. The field of view is accurately reconstructed and presents coherent distortions between the HF and LF arrays. LEKID parameters were robustly determined for 80\% of the read tones. Cross-talks between LEKIDs are the first cause of flagging, followed by an excess of eccentricity for $\sim$10\% of the LEKIDs, all located in a given region of the field of view. On the 44 scans of Uranus selected for the absolute photometric calibration, 72.5$\%$ and 78.2$\%$ of the LEKIDs are selected as valid detectors with a probability $>$70$\%$. By comparing Uranus measurements with a model, we obtain calibration factors of 19.5$\pm$0.6\,[Hz/Jy] and 25.6$\pm$0.9\,[Hz/Jy] for HF and LF, respectively. The point-source continuum measurement uncertainties are 3.0\% and 3.4\% for HF and LF bands, ignoring the uncertainty in the model (which is $<2\%$). This demonstrates the accuracy of the methods we deployed to process the data. Finally, the RMS of CONCERTO maps is verified to evolve as proportional to the inverse square root of integration time. The measured NEFDs for HF and LF are 115$\pm$2\,mJy/beam$\cdot$s$^{1/2}$ and 95$\pm$1\,mJy/beam$\cdot$s$^{1/2}$, respectively, obtained using CONCERTO data on the COSMOS field for a mean precipitable water vapour and elevation of 0.81\,mm and 55.7\,$\deg$, respectively.}
   {CONCERTO has unique capabilities in fast dual-band spectral mapping at $\sim$30 arcsecond resolution and with a $\sim$18.5 arcminutes instantaneous field-of-view. CONCERTO's performance in continuum is perfectly in line with expectations.}

\titlerunning{CONCERTO: on-sky performance in Continuum}
\authorrunning{The CONCERTO collaboration}

\maketitle

  \section{Introduction}

The history of cosmic star formation is essential to understanding galaxy formation and evolution. Over the last two decades, multi-wavelength imaging and spectroscopic galaxy surveys have made spectacular progress in mapping the cosmic history of star formation 
\citep[e.g.][]{2005ApJ...619L..47S,2009ApJ...690..610S,2011A&A...528A..35M,2012A&A...539A..31C,2013MNRAS.432...23G,2013ApJ...768..196S,2013A&A...553A.132M,2015ApJ...803...34B,2020A&A...643A...2B,2023MNRAS.518.6142A,2023arXiv230301658F}.

However, limited by the capabilities of the current generation of telescopes, we know little about the spatial distribution of star formation in large-scale structures beyond z $\geq$ 2 and many uncertainties remain about the nature of the first star-forming galaxies and their impact on the reionisation process \citep{2023ApJS..265....5H,2023MNRAS.518.6011D}. In particular, the difficulty of detecting high-redshift faint and dusty star-forming galaxies using traditional galaxy surveys underscores the urgency for advances in observational techniques.

Line intensity mapping (LIM) is one of these techniques, and a large intensity-mapping [CII] survey targeting high redshift could complement traditional galaxy surveys nicely. Carbon is the fourth most abundant element in the universe. The [CII] emission at 158\,$\mu$m is the dominant cooling line of the interstellar medium \citep{1985ApJ...291..722T} and one of the brightest lines originating from star-forming galaxies in the far-infrared to millimeter range.
Additionally, the [CII] line at 158\,$\mu$m is almost always optically thin and remains unaffected by extinction in most astrophysical environments \citep{2010ApJ...724..957S}. Consequently, this line can be used as a signpost for star formation and diagnostic of the physical conditions of the gas.  
Both simulated \citep{2015ApJ...813...36V,2018A&A...609A.130L,2022arXiv220102636P} and observational \citep{2014A&A...568A..62D,2015ApJ...800....1H,2015Natur.522..455C,2020A&A...643A...3S} studies have proved an empirical nearly-linear correlation between [CII] luminosity and SFR, and the relationship does not evolve with redshift. This makes [CII] emission a valuable tracer of the star formation history. 

The intensity-mapping technique \citep{PhysRevLett.100.091303}, using the bulk emission fluctuations due to galaxy clustering over the surveyed region instead of resolving individual galaxies, can effectively map large-scale structures without costing too much time. 
LIM of emission lines, including [CII], CO, H$\alpha$, [O$_{\rm{II}}$] and [O$_{\rm{III}}$], has the potential to trace the overall cosmic star formation rate density (SFRD) throughout the post-epoch of reionization (EoR) and during the EoR \citep{2004A&A...416..447B,2011ApJ...741...70L,2011ApJ...728L..46G,2012ApJ...745...49G,2014MNRAS.443.3506B, Mashian_2015,2015MNRAS.450.3829Y,2016ApJ...833..153S,2017ApJ...835..273G,2017arXiv170909066K,2019MNRAS.490.1928Y,2019MNRAS.485.3486D,2019MNRAS.488.3014P,2020ApJ...892...51C,2022A&ARv..30....5B,2023A&A...670A..16G}. This method can provide estimates of the total contribution of star-forming galaxies to the energy budget of reionisation.

Several intensity mapping experiments aim to map the redshifted [CII] emission in the sub-millimetre and millimetre band at $z \geq$ 4: the tomographic ionised-carbon intensity mapping experiment (TIME, \citet{2014SPIE.9153E..1WC,2017AAS...22912501C,2022AAS...24031403V}), the Cerro Chajnantor Atacama telescope-prime (CCAT-prime; \citet{2022A&A...659A..12K,2023ApJS..264....7C}), the deep spectroscopic high-redshift mapper/the THz integral field unit with universal nanotechnology (DESHIMA/TIFUUN; \citet{2019JATIS...5c5004E,2022JLTP..209..278T} and the carbon [CII] line in post-reionisation and reionisation epoch project (CONCERTO; \citet{2020A&A...642A..60C}). 

Different from CCAT-prime (based on LEKIDs and Fabry-Perot interferometers), TIME (based on gratings and transition edge bolometers) and TIFFUN (based on integrated superconducting spectrometers), CONCERTO is a millimetre-wave low spectral resolution spectrometer (R=$\nu/\delta\nu \leq 300$), based on lumped element kinetic inductance detectors (LEKIDs, \citealp{article}) coupled with a Martin-Puplett interferometer (MPI, \citealp{MARTIN1970105}). CONCERTO is equipped with 2 focal planes housing 4\,304 LEKIDs. The same detector technology has been utilised and verified on NIKA \citep{2011ApJS..194...24M}, NIKA2 \citep{2018A&A...609A.115A,2020A&A...637A..71P} and KISS \citep{2020JLTP..199..529F}. These detectors are in a $^3$He-$^4$He dilution cryostat, which maintains a base temperature of $\sim$60-70\,mK with 0.1\,mK stability. The presence of a large number of detectors allows it to reach an instantaneous field-of-view of about 20\,arcminutes. Spectra are obtained by a fast room-temperature Martin-Puplett interferometer located in front of the cryostat \citep{2020A&A...642A..60C, 2022SPIE12190E..0QF, 2022JInst..17P0047B}. 
     
The CONCERTO experiment ultimately aims to constrain the three-dimensional fluctuations of the [CII] line emission at $z>5.1$ The experiment also observes the intensity fluctuations of CO emissions from galaxies within the redshift range of 0.3 < z < 2, giving insights into molecular gas's spatial distribution and abundance across a broad range of cosmic time \citep{2023A&A...670A..16G, 2023A&A...676A..62V}. Additionally, CONCERTO is a unique multi-wavelength tool to analyse Sunyaev-Zel’dovitch (SZ) signals \citep{1980ARA&A..18..537S,1980MNRAS.190..413S} with 2D+spectral mapping, and extract mass, temperature and proper motion information about the cluster physics \citep{2020A&A...642A..60C}.

CONCERTO has been installed on the Atacama Pathfinder EXperiment (APEX) 12-meter telescope \citep{2006A&A...454L..13G} and began commissioning observations in April 2021 \citep{2022JLTP..209..751M,2022EPJWC.25700010C,2022SPIE12190E..0QF}. Following a successful commissioning phase that concluded in June 2021, CONCERTO was offered to the scientific community for observations, with a final observation run in December 2022.\\

While initially designed as a spectrometer, CONCERTO can operate in continuum mode (MPI OFF) or spectroscopic mode (MPI ON). 
This paper reviews the CONCERTO performance assessment in the continuum. The presentation of spectroscopic data will be reserved for a forthcoming paper.
Data from commissioning and scientific observations are analysed. We evaluate the stability of performance parameters over time and different atmospheric conditions using a large data set acquired in 2021 and 2022, including observations of calibrators and faint sources spanning the full range of atmospheric conditions at the APEX telescope. We develop a standard data-processing pipeline to generate continuum maps from raw data and assess the robustness of our performance results by comparing our measurements with predictions and previous results from the literature.

The paper is organised as follows: Section~\ref{sec:instrument} summarises the CONCERTO instrumental set-up. Section~\ref{sec:observation} describes the observation modes that have been used for the calibration and the performance assessment. Section~\ref{sec:dataprocessing} describes the data processing. The beam pattern is characterized in Sect.~\ref{sec:beammap}, along with the main beam full-width at half maximum and the beam efficiency. The photometry measurement and stability assessment using Uranus are shown in Sect.~\ref{sec:photometry}. Section~\ref{sec:calibration} presents the absolute photometric calibration and the calibration accuracy and stability. The noise characteristics and the sensitivity are discussed in Sect.~\ref{sec:sensitivity}. Finally, Sect.~\ref{sec:conclusion} summarizes the main measured performance characteristics.

\section{Instrument}
\label{sec:instrument}

A comprehensive description of the CONCERTO instrument can be found in \citet{2020A&A...642A..60C}. In this section, we summarise the aspects related to data reduction and calibration.

The core components in receiving incoming power and converting it to CONCERTO read-out are LEKIDs. CONCERTO has two focal planes (HF and LF) which are single polarisation LEKID arrays containing 2152 LEKIDs each, and covering the frequency ranges 195-310\,GHz for HF and 130-270\,GHz for LF. Each LEKID array has 6 sub-arrays corresponding to 6 feed lines, called "KA" to "KF" for LF and "KG" to "KL" for HF. The superconducting resonators have resonance frequencies that shift linearly with the incoming optical power ($\Delta f$) \citep{2010ApPhL..96z3511S}. Each LEKID is associated with an excitation tone at a specific frequency, and the detection of photons is achieved by measuring the induced LEKID frequency shift $\Delta f$. 

During a typical astronomical observation, the optical load experiences fluctuations with time due to factors such as background sky emission and shifts in elevation angle. Consequently, the resonance frequency of the LEKIDs undergoes temporal changes. For CONCERTO, mitigation techniques including \textsc{sweep}, \textsc{shift}, and \textsc{tuning} have been developed to enhance operational efficiency. 
Comprehensive details can be found in \citet{2022JInst..17P8037B}. To provide a brief overview, the process involves the initial determination of LEKID resonance frequencies through \textsc{sweep}. We have two initial parameter files (i.e. two different \textsc{sweep}) for CONCERTO data: one for 2021 and one for 2022. Subsequently, global adjustments for each electronic box are made to adapt to the LEKIDs' resonance frequency through \textsc{shift}. Lastly, individual LEKIDs' resonance frequency is fine-tuned using \textsc{tuning} \citep{2021A&A...656A.116F}.

Spectra are obtained by a fast room-temperature MPI performing continuous path interferograms at an acquisition frequency ($f_{\rm{acq}}$) of 3.815\,kHz, which is selected to avoid atmospheric drifts during a single interferogram recording. It works by splitting incoming radiation into two beams using a polarizing grid and then recombining them after one beam has passed through a variable-delay path. This creates an interference pattern that can be used (through Fourier transforming) to extract information about the spectral content of the radiation.  

CONCERTO exhibits versatility by operating in two distinct modes: photometric (MPI OFF) and spectroscopic (MPI ON). When in MPI OFF mode, it captures pure continuum data at a rate of approximately 30\,Hz. Conversely, the MPI ON mode generates continuum data from the interferograms and samples it at a frequency of 1.86\,Hz. 

\section{Observations}
\label{sec:observation}

CONCERTO was installed on the APEX telescope, and commissioning observations began at the end of April 2021, followed by scientific observations from July 2021 to December 2022. The primary project for CONCERTO is the intensity mapping experiment on the COSMOS (The Cosmic Evolution Survey, \cite{2007ApJS..172....1S}) field, with additional observations of galaxy clusters (e.g., RXJ1347 and Abell2744) and star-forming regions taken in open time. Most observations use on-the-fly (OTF) scans centred on the targets, with focus and pointing scans performed before and a pointing scan performed after. The observation modes implemented at the APEX telescope include focus, pointing, skydip, OTF, and beam mapping.

The focus observation mode is designed to adjust the telescope focus before pointing by targeting a bright point source. During each focus scan, five successive one-minute raster scans are performed at 5 axial offsets of the third mirror (M3) along the optical axis. The flux and full width at half maximum (FWHM) along each focus's minor and major axes are estimated by elliptical Gaussian fits on the 5 maps. The best axial focus is determined as the maximum of the flux or the minimum of the FWHM using parabolic fits of the 5 measurements. Typically, three focus scans are launched along the X, Y, and Z axis, respectively, and the estimated adjustment of the telescope focus is applied after each focus is finished.

After properly setting the telescope focus, pointing corrections are estimated using a dedicated pointing scan. The telescope performs a back-and-forth scan in azimuth and elevation centred on the observed source. Gaussian profiles are fitted from the timelines of the reference LEKID, and the estimated position of the reference LEKID is used to determine the current pointing offsets of the system in azimuth and elevation. This correction is then applied to subsequent scans.

In the analysis presented here, three types of observations are used:
\begin{itemize}
\item Skydip, which measures the atmospheric transmission at different elevations by scanning across a wide range of elevations at a fixed azimuth. CONCERTO uses skydips to quantify its performance in a large range of atmospheric conditions. 
\item Beam map is a raster scan aiming to map a bright, compact source (such as Uranus, Neptune, or Mars), with each LEKID observing the source during the observation. These scans are crucial for calibrating the focal planes and measuring the beam pattern of the instrument. Combining many LEKIDs with a high $f_{\rm{acq}}$ makes beam-map observations with MPI ON impossible.
\item OTFs, where the telescope moves continuously along a row, column, or spiral pattern with a predetermined velocity. On-the-fly rasters are used to perform mapping observations of relatively large fields. In this paper, observations of Uranus calibration scans, the COSMOS field and a small field around the AS2UDS0001.0 source (from the ALMA–SCUBA2 Ultra Deep Survey, \cite{2019MNRAS.487.4648S}) use the OTF scanning.
\end{itemize}

The CONCERTO observations for AS2UDS and COSMOS were performed in the spectroscopic mode (MPI ON), while the planet calibration scans were carried out in photometric mode (MPI OFF). For the analysis of AS2UDS and COSMOS presented in this paper, the photometric data are thus extracted from the spectroscopic observations. 
For Uranus and Mars pointing scans used for absolute photometric calibration, we utilized only those obtained with a precipitable water vapour (pwv) value lower than 1.5\,mm.

\section{Data processing}
\label{sec:dataprocessing}

A comprehensive data-processing pipeline has been developed for CONCERTO's photometric science. This pipeline allows for the processing of both photometric and spectroscopic data to produce photometric maps or spectral cubes. The pipeline for continuum observations consists of raw-data reading and I/Q calibration, bad LEKIDs/scan masking, flat-field normalisation, opacity correction, correlated noise subtraction and map projection. \\

{\tt LEKIDs electronic calibration:} The raw recorded time-ordered data (TOI) consists of the I (in-phase) and Q (quadrature) components of the excitation signals. For each detector, the I and Q TOI are organized in a succession of blocks, with 2 modulations of the input tone at the beginning of each block.
The raw data is converted to the LEKID frequency shift $\Delta f$ in Hz using the 3-point modulation algorithm \citep{2021A&A...656A.116F,2022JInst..17P8037B}. This method calibrates each block of data independently and produces one continuum point per block. For pure photometric observations, where the MPI mirror is blocked in a fixed position, we can further recover the data within the block to increase the effective sampling of the continuum TOI. In the photometric calibration pipeline (described in Sect.~\ref{sec:calibration}), the $\Delta f$ TOI is further converted into astronomical units in Jy/beam.\\

\begin{figure}
   \centering
   \includegraphics[width=8.1cm]{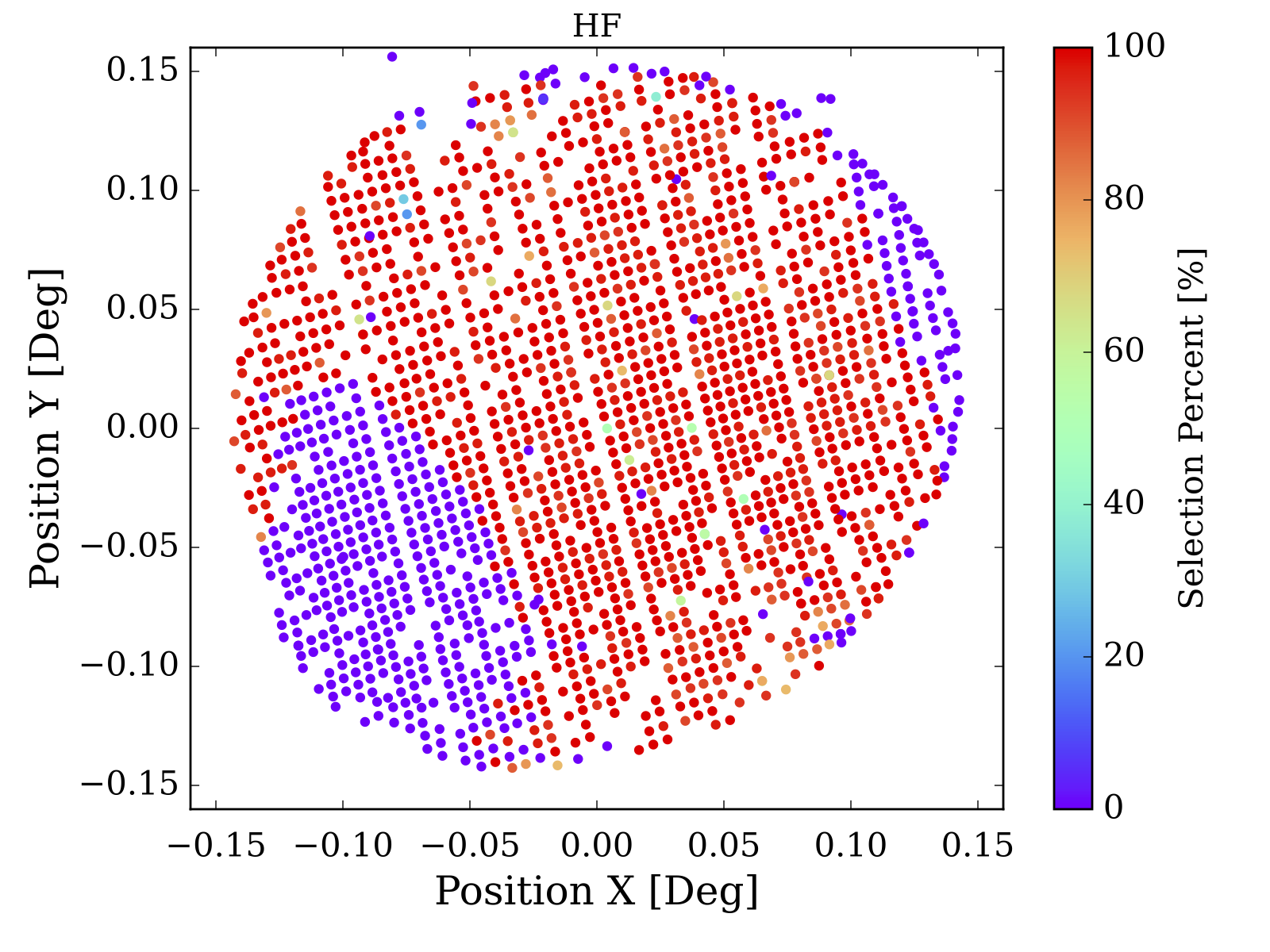}
   \includegraphics[width=8.1cm]{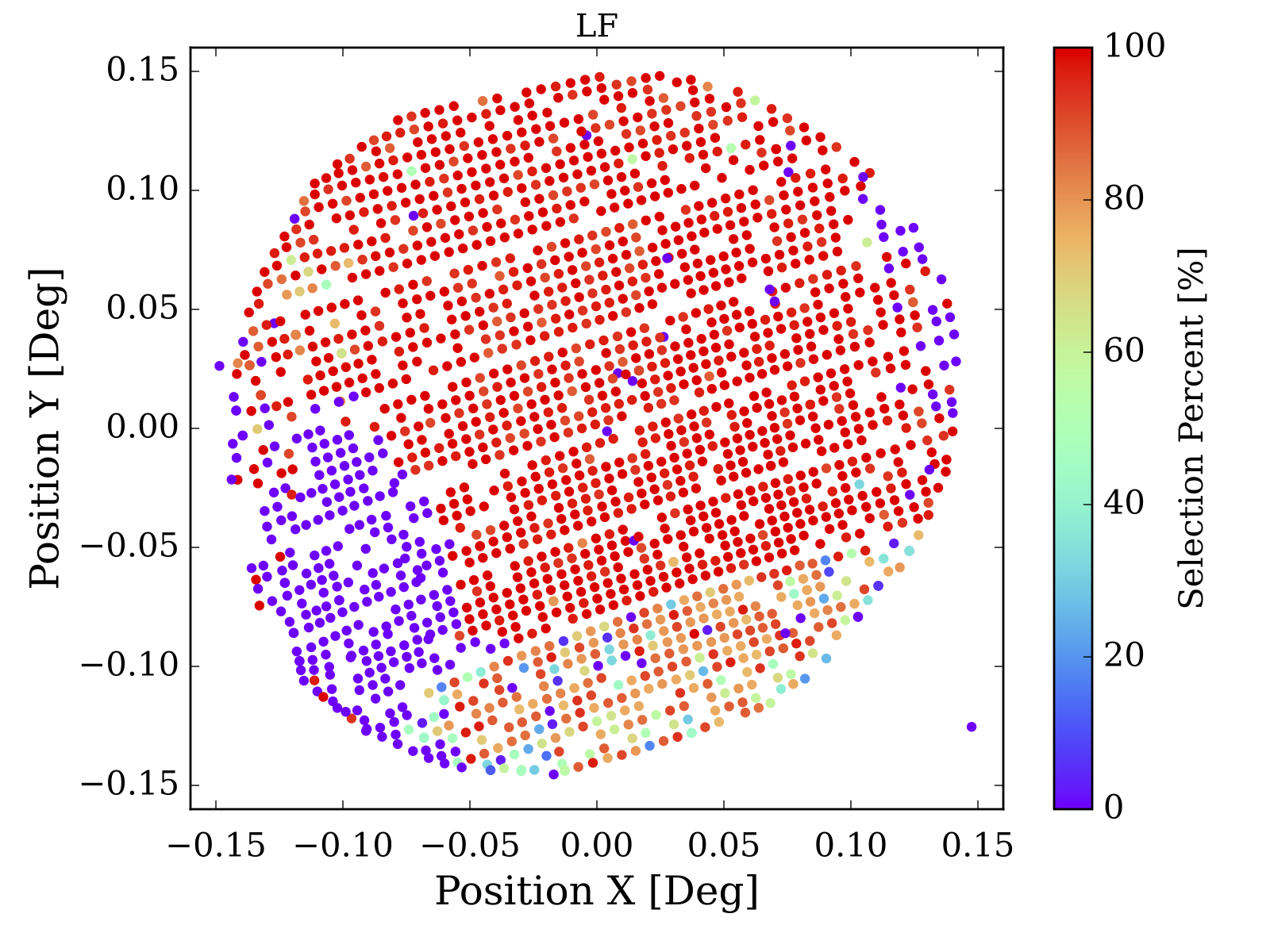}
   \caption{KIDs positions for arrays HF (upper panel) and LF (lower panel) labelled with the fraction of time a LEKID was selected in the data processing of 44 Uranus photometric pointing scans. For LF, the sequence of the 6 feed lines is from KA at the bottom to KF at the top; for HF, it goes from KG on the left to KL on the right.}
   \label{Uranus_mask}
\end{figure}

{\tt LEKIDs selection:} To ensure the final map is high quality, bad LEKIDs must be masked during data processing. Multiple masking steps are applied to select good LEKIDs for producing maps for scientific purposes. First, LEKIDs are flagged if their derived standard deviation of calibration factors (the conversion factor from frequency shift to incoming photon intensity) is more than 5 times the median of the standard deviation of calibration factors among the LEKIDs within the same electronic box. Next, data with poor cryostat regulation or wrong cryostat position (deviation from the median position exceeds 5 times the median absolute deviation) are masked. Then, when the resonance frequencies of LEKIDS are separated by less than 3$f_{\rm{acq}}$ they are also masked. 
Each LEKID is further selected based on its recorded position, beam FWHM, amplitude, and eccentricity information. The default minimum and maximum thresholds for position shift, beam FWHM, amplitude, and eccentricity are (-30\,arcmins, 20\,arcsecs, 0.2, and 0) and (30\,arcmins, 60\, arcsecs, 10, and 0.7), respectively. Finally, LEKIDs are masked if their standard deviation of the continuum exceeds 6 times the median of the standard deviation of the continuum among the LEKIDs within the same electronic box. On strong sources, when an iterative scheme is used to recover their fluxes, an additional LEKID selection is achieved at each iteration, based on the standard deviation of continuum residual from the previous iteration thresholded source model.

Figure.~\ref{Uranus_mask} shows the LEKID focal plane labelled with the percentage of time a LEKID was selected in the data processing of Uranus photometric pointing scans (see Sect.~\ref{sec:photometry}). The number of scans in which each LEKID has been retained after LEKID selection is counted. The six feed lines for each of the two arrays are slightly separated, and spaces between LEKID columns (HF) and LEKID rows (LF) connected to different feed lines are slightly larger than those between LEKID rows of the same feed line. LEKIDs at the lower-left corners and upper-right edges are less often selected than the others and are masked because of large eccentricities ($\geq$ 0.7). In addition, all LEKIDs in KA (at the bottom of the LF array) are less stable than the other feed-line sub-arrays due to frequency separations to the excitation tone frequency being slightly too small ($\leq$ 3$f_{\rm{acq}}$).  
For the HF and LF arrays, we find that 72.5$\%$ and 78.2$\%$ of LEKIDs are selected as valid in at least 70$\%$ of the scans. On the other hand, 46.2$\%$ and 44.9$\%$ of LEKIDs in HF and LF, respectively, are masked as valid in all scans, while 26.3$\%$ and 16.7$\%$ are masked as invalid in all scans.\\


{\tt Flat-field (or gain) correction:} After the LEKIDs selection, the recorded data of each LEKID is normalized by the flat-field.
The flat-field correction aims to equalise the variations in response observed among LEKIDs. The flat-field value for each LEKID is retrieved from the beam-map processing (see Sect.\,\ref{sec:beammap}). \\

{\tt Opacity correction:}
The atmospheric opacity is the primary limitation for (sub-)mm experiments conducted on the ground. Merely a fraction of the source signal reaches the detectors of the CONCERTO instrument. The intensity is corrected for opacity following,
\begin{equation}
I_{\rm{c}} = I\, e^{\tau_\mathrm{eff}\ x},
\label{opacity_correction}
\end{equation}

\noindent where $x$ is the air mass (which depends on the elevation as $x$ = [sin(el)]$^{-1}$), $I$ and $I_{\rm{c}}$ are the measured and corrected (top-of-the-atmosphere) continuum intensities and $\tau_\mathrm{eff}$ is the time-ordered list of derived effective zenith opacity defined as:
\begin{equation}
e^{-\tau_\mathrm{eff}} = \frac{\int e^{-\tau_\nu} F(\nu) \eta(\nu) d\nu}{\int  F(\nu) \eta(\nu) d\nu}
\end{equation}
where $F(\nu)$ is the relative spectral response (i.e., the instrument transmission as a function of frequency) and $\eta(\nu)$ the aperture efficiency (Sect.\,\ref{sec:cali_uranus}). 
The values of $\tau_{\nu}$ are derived using the measured pwv at APEX and the "atmospheric transmission at microwaves" (ATM) model from \cite{982447}. At an elevation of 60 degrees and for a pwv value of 0.8\,mm, the correction for opacity is 9 and 10\,\% for HF and LF, respectively.\\

{\tt Common-mode removal:} Once the opacity is corrected, we estimate and eliminate a common-mode noise. The time-ordered raw data of each LEKID is prone to correlated noise from the atmospheric component (common to all LEKIDs) and electronic noise (common to LEKIDs connected to the same readout electronics). Removing the common mode introduced by the correlated noise is crucial to retrieving the astronomical signals. To achieve this, we have devised various specialized techniques for common-mode removal such as the principal component analysis (PCA, see \citet{2016RSPTA.37450202J} for a review) and the median filtering.\\
To handle observations of bright sources such as planets and quasars, we employ a noise decorrelation method based on median filtering, where the correlated noise is estimated as the median value of all considered LEKIDs per timestamp. On the other hand, when processing data from faint sources like the AS2UDS and COSMOS fields, we use a principal component analysis (PCA) technique to estimate and remove the correlated signals. Compared with median filtering, PCA can reduce the noise to a lower level by removing more principal components. However, using PCA for noise removal can filter more astronomical signals. To account for this possibility, we perform simulations and recover the true signal by a factor derived from these simulations (see below).\\ 

{\tt Continuum maps:} Lastly, the clean and calibrated TOI data are projected into astronomical coordinates and used to generate continuum maps. To reduce the impact of less sensitive LEKIDs, the data from each LEKID is weighted by the inverse of its noise level. The LEKIDs' relative position on the focal planes is derived from beam map analysis (Sect\,\ref{sec:beammap}). The resulting maps are then thresholded at 4.7$\sigma$ to construct a source model which is removed from the raw TOI in an iterative process to recover the source flux from the common-mode removal step. We typically ran 10 iterations, while the process converged quickly after 3 iterations. \\

{\tt Simulations:} We ran simulations to estimate and correct the effects of data reduction and instrument noise on the flux measurements. Artificial sources with various flux densities (ranging from dozens of Jy for planets to several mJy for faint sources) are injected into the raw observational data and then processed using the processing pipeline. The flux densities of these artificial sources after the processing are compared with their original values.
In the case of planet-pointing scans that employed median filtering decorrelation, the differences between the processed and original fluxes are less than one per cent. For faint sources in the AS2UDS and COSMOS fields, the number of removed principal components is carefully selected to ensure that the flux of artificial sources remains unchanged after the noise decorrelation process. As for planet-pointing scans, to quantify the effectiveness of the noise reduction techniques with PCA, artificial point sources with low (0.5Hz) and high (50Hz) intensities were introduced into the AS2UDS dataset. The ratio of output to input flux density was 0.97$\pm$0.04 for the low-intensity sources and 1.01$\pm$0.01 for the high-intensity ones. Similar results are obtained for COSMOS for high-intensity sources (note that low-intensity sources cannot be recovered in COSMOS because of the signal-to-noise ratio is too low). These outcomes indicate that with PCA, we effectively eliminate the correlated noise without affecting our point-source signals. As a further check, we repeat the process in AS2UDS using median filtering, which results in a higher mean and variance in the output/input flux ratio for low-brightness sources. Thus, PCA demonstrated superior performance in handling faint signals.
 

\section{Focal Surface Reconstruction}
\label{sec:beammap}
During the CONCERTO operations, we conducted regular fully sampled observations of strong point-like sources, so-called beam maps. These observations are used to derive several parameters on individual detectors. We employed large on-the-fly maps in horizontal coordinates, covering an area of 22\,arcmin\ by 22\,arcmin, with a scanning speed of 120\,arcsec/s and a vertical step of 6\arcsec. 
Each scan lasted about 40 minutes. These were performed in continuum mode (i.e. with the moving MPI mirror fixed) on Mars, Uranus, or Neptune, depending on their visibility and apparent sizes. Throughout the CONCERTO observing runs, the apparent sizes of those planets were below 4\,arcsec, except for Mars, which showed an increase of apparent size from $\approx$5 arcsec up to 17\,arcsec from September to December 2022.  The data processing followed the procedure described in Sect.~\ref{sec:dataprocessing}, with the necessary following modifications to account for the specificities of beam map observations: i) due to memory constraints, the processing had to be carried out per array; ii) the LEKIDs were not selected on their properties, but instead on their cross-correlation coefficient, which had to be greater than 0.5, selecting between 1\,700 and 1\,950 (used) LEKIDs per array, ensuring that all processed LEKIDs observed the sky emission; iii) no flat-field nor opacity correction was applied; 
iv) a second-order polynomial, applied per sub-scan, was used. Finally, the sampling of these observations allowed the data to be projected onto horizontal coordinates with a pixel size of 7.5\arcsec for each individual LEKID map, ensuring a fine sampling of individual beams.

To analyze each beam map, we conducted 2D elliptical Gaussian fits on the maps of each detector. This process allowed us to determine the relative positions, beam sizes, and gains of each detector. After the initial analysis, certain LEKIDs were flagged if their amplitude, position or beam size values deviated significantly from the overall distribution, indicating them as solid outliers. Furthermore, we visually examined each LEKID map and identified instances of cross-talk signatures, specifically the occurrence of a "double beam". This phenomenon typically occurs between adjacent LEKID resonance frequencies. 
Once identified, these LEKIDs were flagged for further analysis.
The visual inspection process also enabled the exclusion of entire scans for various reasons, such as those made very early during the initial commissioning phase, scans that were too prematurely terminated, scans conducted with the moving mirror engaged, or scans conducted when Mars had an apparent size exceeding 8\arcsec. Of the 53 beam maps observed, 13 scans from 2021 and 15 scans from 2022 were further used in the analysis.

\begin{table*}
\caption{CONCERTO focal plane statistics. The number of used KIDs is given as a maximum interval between the different beam map observations. The second half of the table lists the number of KIDs falling into the different, non-exclusive, flag categories. }
\label{tab:kidsstat}
\centering
\begin{tabular}{l c c c c}
\hline
             & \multicolumn{2}{c}{2021} & \multicolumn{2}{c}{2022} \\
             & LF  & HF  & LF  & HF \\
\hline 
Designed     & \multicolumn{4}{c}{2152} \\
Beammaps & \multicolumn{2}{c}{13} & \multicolumn{2}{c}{15} \\
Read Tones   & 1919 & 1881 & 1998 & 1960 \\
Used KIDs    & 1768 -- 1885 & 1706 -- 1870 & 1746 -- 1975 & 1706 -- 1950 \\
Valid KIDs   & 1459 & 1264 & 1403 & 1183 \\

\hline
Bad\tablefootmark{a}          & 40 & 21 & 75 & 45 \\
Outliers\tablefootmark{b}     & 36 & 20 & 84 & 57 \\
Cross talks   & 211 & 233 & 312 & 337 \\
Pos. uncert.\tablefootmark{c} & 33 & 17 & 55 & 56 \\
Insuf. stat.\tablefootmark{d}     & 92 & 107 & 143 & 141 \\ 
Eccentricity\tablefootmark{e} & 154 & 312 & 138 & 323 \\
\hline
Footprint\tablefootmark{f}    & 41 & 30 & 45 & 34 \\
\hline
\end{tabular}
\tablefoot{
\tablefoottext{a}{KIDs showing no response most of the time.}
\tablefoottext{b}{KIDs placed outside the FoV most of the time.}
\tablefoottext{c}{KIDs position uncertainty greater than 6\arcsec.}
\tablefoottext{d}{KIDs seen less that 10\% of the beammaps.}
\tablefoottext{e}{KIDs eccentricity greater than 0.75.}
\tablefoottext{f}{KIDs is more than 5\arcsec away from constructed position.}
}
\end{table*}

\subsection{Mean geometry and LEKID statistics}
To improve the LEKID array parameters statistics and allow for LEKID performance stability estimation, we stacked all the beam map results of each observing year. To align the individual beam maps, we performed spatial registration using a rigid transformation and used the Iterative Closest Point method \citep{CHEN1992145, 121791}. This allowed us to establish a common reference frame for the beam maps. We could then apply a median and median absolute deviation estimator on all LEKID parameters. An insufficient statistics flag was assigned to LEKIDs that observed less than 10\% of the time. For each single beam map, all the previously defined flags were carried forward using a majority classification. The resulting statistics are presented in Table\,\ref{tab:kidsstat}.\\

The standard deviation of the recovered LEKID individual positions along the individual beam map has a median value of $0.6\pm 0.3$\arcsec, showing excellent repeatability between several observations. This enables the establishment of a flag denoting positional uncertainty for LEKIDs whose position uncertainty exceeds one-fifth of the beam size.
Furthermore, we compared the reconstructed LEKID positions in the field of view (FoV) to their constructed positions in the arrays. Due to the complex optics of CONCERTO, we used a 2D polynomial transformation of degree 4 to account for most of the FoV deformation (see Appendix~\ref{appendix:Fov_Deformation}). This allows us to further flag the misplaced detectors with offset exceeding 5\arcsec\ compared to their designed positions. \\

With 2152 LEKIDs designed for both arrays, of which we could read 1919/1881 (LF/HF) in 2021 and 1998/1960 in 2022, we could robustly define KID parameters for $\sim$85\% of the read tones in 2021 and $\sim$78\% in 2022, excluding the eccentricity flag.
The cross-talk between neighbouring tone frequencies is the first cause of flagging with $\sim$11\% of the read LEKIDs in 2021 and $\sim$16\% of the read LEKIDs in 2022. Additionally, the lack of statistics, i.e. LEKIDs responding only one-tenth of the time, is a significant cause of flagging with about 5\% in 2021 and 7\% in 2022.

\subsection{FoV properties}

\begin{figure*}
\centering
   \includegraphics[width=17cm]{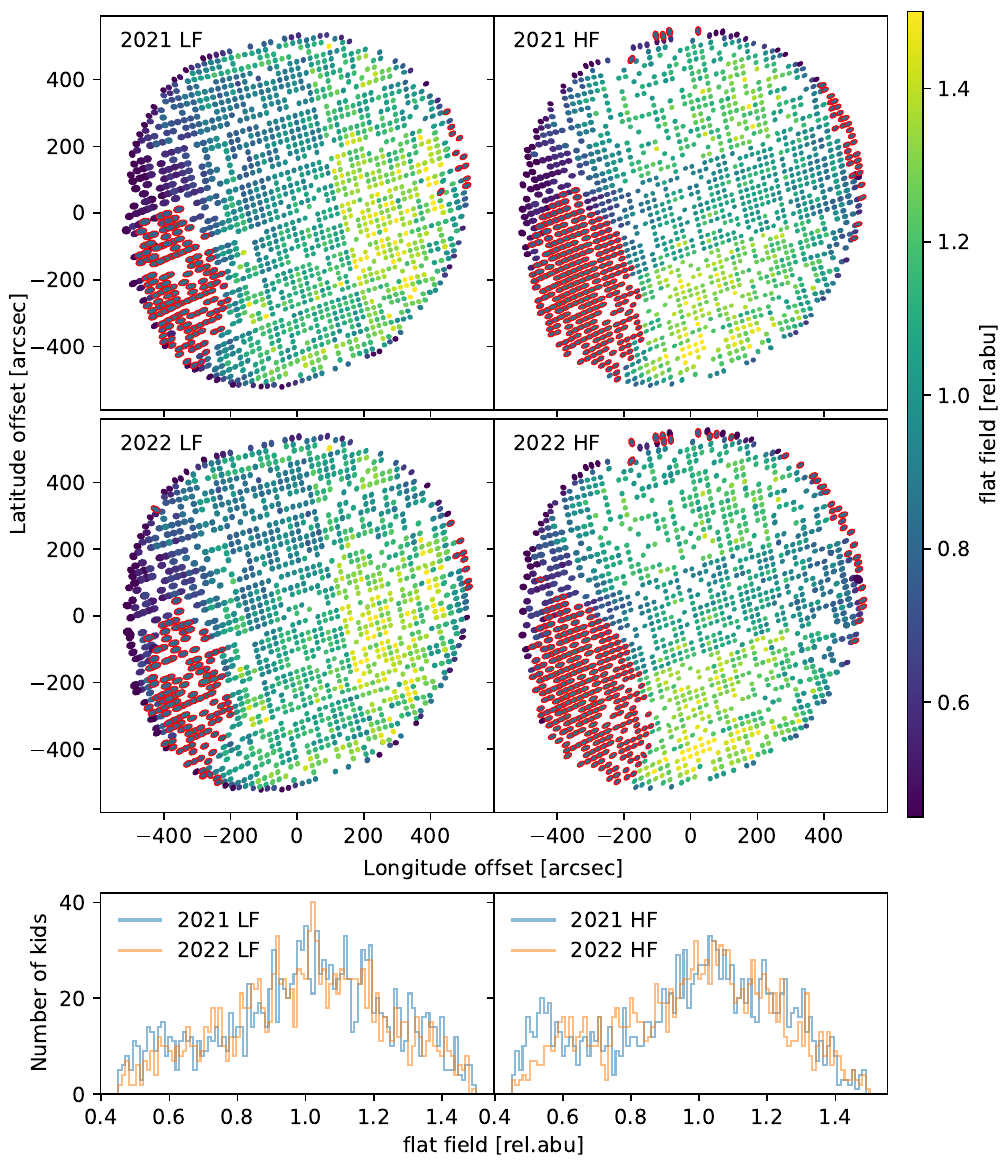}
     \caption{Top: Averaged main-beam flat-field response for the CONCERTO low frequency (left) and high frequency (right) arrays during the 2021 (top) and 2022 (bottom) campaigns. Each valid LEKID is represented by an ellipse with a half-width at half-maximum diameter. The lower left corner of the arrays is experiencing an optical distortion with eccentricity above 0.75 (highlighted in red).
     The flatfield are nearly identical between the two observation periods and their distribution follows the resonance frequencies of the LEKIDs. Bottom: Histograms of the averaged main-beam flat-field response for the CONCERTO low-frequency (left) and high-frequency (right) arrays.}
     \label{fig:flat_field}
\end{figure*}

In Fig.\ref{fig:flat_field}, we represent each KID with an ellipse showing the eccentricity of individual valid kids, defined as $e = \sqrt{1 - \sigma_\mathrm{minor}^2/\sigma_\mathrm{major}^2}$. The eccentricity is relatively homogeneous over the FoV, with a median value of $0.49\pm0.14$, except in the lower left corner, where a region is affected by an increase of eccentricity above 0.75. This represents the second largest flagging cause for 8 and 16 \% of the read LEKIDS for both observing campaigns (see Tab.~\ref{tab:kidsstat}). The cause of this optical aberration is not understood but could be related to a defect in one of the CONCERTO mirrors. These KIDs, presenting a different optical response with respect to the mean properties of the arrays, need to be excluded from any further data analysis. The fraction of usable KIDs compared to the design is thus lowered to 63\% and 60\% for 2021 and 2022.\\

In Fig.\ref{fig:flat_field}, we also present the averaged normalised response to a point source, so-called main-beam flat-field. Overall, the distribution of the main-beam flat field presents a gradient perpendicular to the feed lines with distinct zones corresponding to the distribution of resonance frequencies of the LEKIDs on the feed lines, which are also grouped by electronic sub-bands which are amplified differently to compensate for the increase of the intrinsic KIDs read-out noise with resonance frequencies. The relative response of the LEKID presents a wide distribution between 0.2 and 1.8 for the valid detectors, with 68\% of the LEKIDS in the interval 0.7--1.25. It is very reproducible with a median standard deviation across different observations of $4.6\pm2\%$.

\section{Beam Pattern}
For each observing campaign, we used the beam maps observation to characterize the effective photometric beam pattern of CONCERTO. Excluding all previously flagged LEKIDs, as well as those with eccentricity above 0.75, individual beam maps were co-added onto a map with a refined pixel size of 2\arcsec, allowed by the number of projected LEKIDs and the spatial sampling of those observations. Flat-fielding between pixels was applied based on the averaged main-beam flat field. No correction for opacity was applied, as our focus lies in determining relative quantities. We adjusted a model of 3 elliptical Gaussians, as well as a background level, to each individual beam map, to obtain residual fine pointing offset as well as their peak normalization, marginalizing over the individual potential error beams. The averaged normalized beam map is then obtained by taking a weighted mean of the individual centred and normalized beam maps.

\subsection{Full Beam Pattern}
\begin{figure*}
\centering
   \includegraphics[width=17cm]{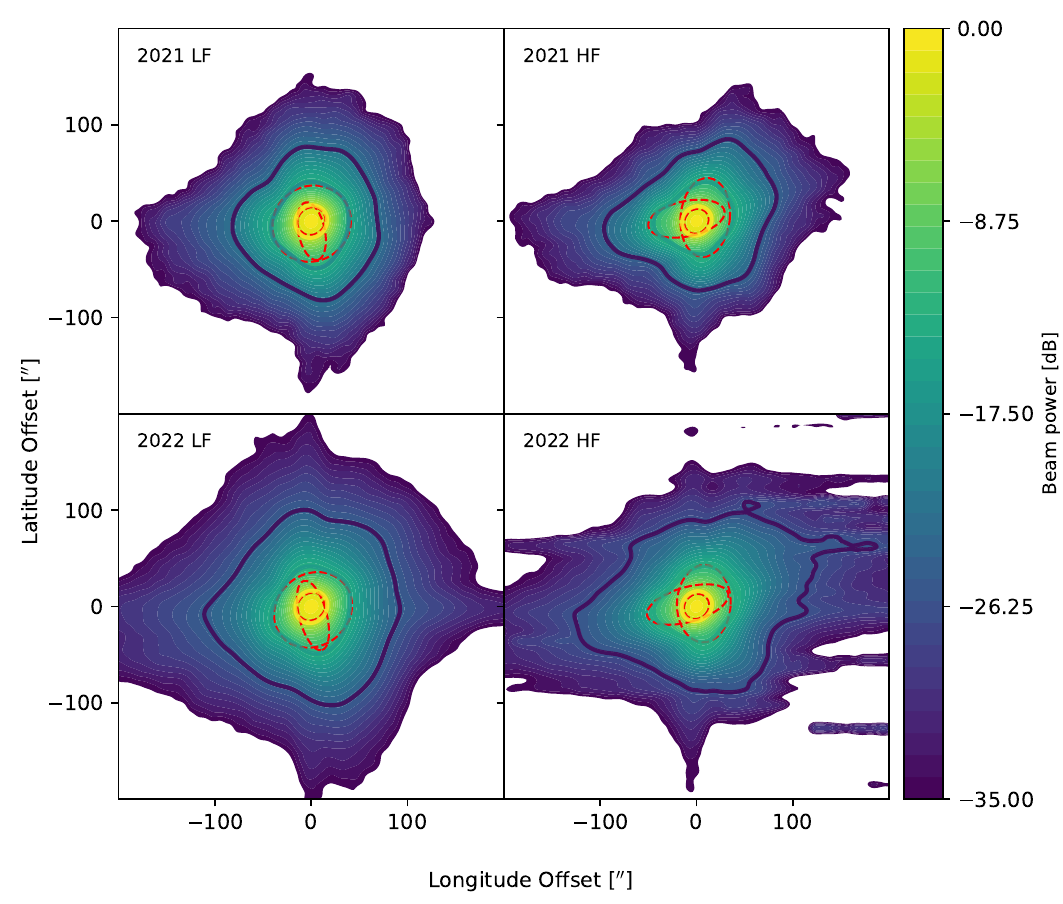}
     \caption{Horizontal average beam patterns of CONCERTO for the low-frequency (left) and high-frequency (right) arrays during the 2021 and 2022 campaigns. These maps were created by combining 13 and 15 beam-map scans on planets and normalizing the results. The thin dashed red lines are the 2D Gaussian decomposition. The thick solid lines (yellow, green and dark blue) depict the half-maximum contours of the main beam and the 2 error beams, when decomposing it into three radial Gaussian functions. The horizontal and vertical extensions are due to the secondary quadrupod legs.}
     \label{fig:beam}
\end{figure*}

The resulting beam patterns are presented in Fig.~\ref{fig:beam} down to $-40$\,dB over an extent of 3\arcmin. The CONCERTO beam maps can be decomposed into several components :
\begin{itemize}
    \item a main beam, which can be described by a 2D Gaussian with a circularized FWHM of 30.2 and 25.9\arcsec for LF and HF, corresponding to the expected values for a diffraction-limited aperture of 11\,m. The main beam is slightly elongated with a mean eccentricity of 0.45 and 0.47 for LF and HF respectively. The mean main beam areas of CONCERTO are 2.0 and 1.6\,$10^{-8}\, \mathrm{sr}$.
    \item two error beams of equivalent amplitudes slightly offset from the main beam. The first error beam, at a level of $\sim -11.5$\,dB relative to the total beam, shows a very elongated component, with an offset of about 10\arcsec and with a 90-degree rotation angle between the two arrays. The second error beam, at roughly the same level, appears at a larger scale with typical FWHM of $\sim 80 \times 60$\arcsec, and encompasses between 28 and 35\% of the total beam solid angle.
    \item the diffraction pattern due to the secondary quadrupod legs of the APEX telescope can be seen between $-35$ and $-40$ dB on a large scale, once the main and error beam patterns are removed.
\end{itemize}
Overall the beam properties are reproducible between the 2021 and 2022 campaigns. All beam parameters, including the modelled beam surface area and an effective single elliptical Gaussian fit, are summarized in Table\,\ref{tab:beam_fit_image}.

\subsection{Radial Beam Pattern \label{radial_beam}}

\begin{figure}
  \resizebox{\hsize}{!}{\includegraphics{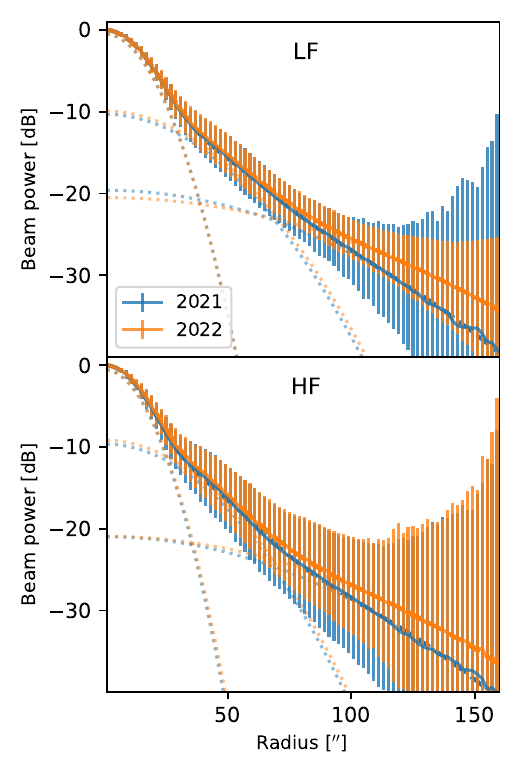}}
  \caption{Radial profiles of the CONCERTO beam for the low-frequency (top) and high-frequency (bottom) arrays during the 2021 and 2022 campaigns. These profiles were derived from the normalized combination of 13 and 15 beam-map scans on planets. The solid line represents the best-fitted model, which combines three Gaussian functions represented by the dotted lines.}
  \label{fig:beam_profile}
\end{figure}

To further quantify the error beams, we constructed the normalized beam radial profiles centred on the main beam position. This allows the study of the error beam up to 150\arcsec, as seen in Fig.~\ref{fig:beam_profile}. The first error beam $\mathcal{A}_2$ is the radial average of the 2D error beams $\mathcal{A}_{2a}$ and $\mathcal{A}_{2b}$ at around -10\,dB. A second error beam is measured at -20\,dB, with a FWHM of 120\arcsec\, in 2021 and 145\arcsec\, in 2022. A similar error beam can also be measured with LABOCA at -26\,dB with a FWHM of $\sim$120 \arcsec \citep{2009A&A...497..945S}. 

\begin{figure}
  \resizebox{\hsize}{!}{\includegraphics{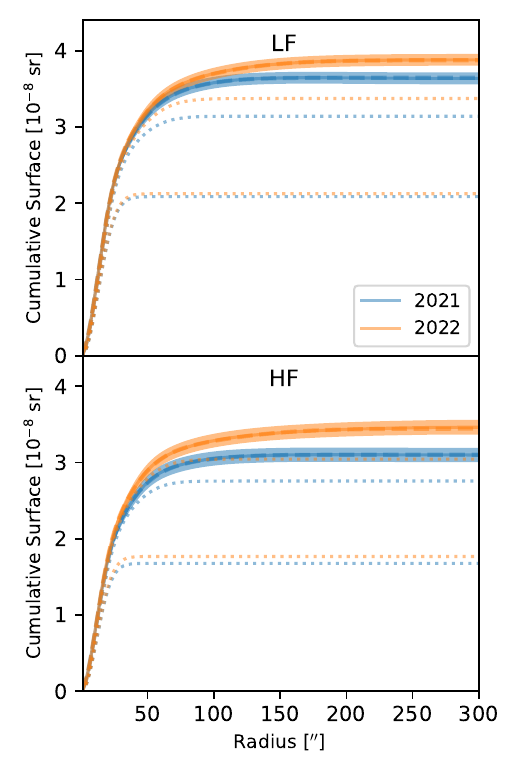}}
  \caption{Cumulative radial profiles of the CONCERTO beam for the low-frequency (top) and high-frequency (bottom) arrays during the 2021 and 2022 campaigns. These profiles were derived from the normalized combination of 13 and 15 beam-map scans on planets. The dashed line represents the best-fitted profile model integrated with the radius. The dotted lines represent cumulatively each Gaussian component, the main beam (lowest) and the first error beam. }
  \label{fig:beam_area}
\end{figure}

The radial profile can be used to derive the total beam area directly from the data, as shown in Fig.~\ref{fig:beam_area}, resulting in a total beam solid angle of $3.6 \pm 0.09$ and $3.1\pm 0.1\, 10^{-8}\mathrm{sr}$ for the LF and HF array in 2021 and $3.9\pm 0.07$ and $3.4\pm 0.1\, 10^{-8}\mathrm{sr}$ in 2022 and can very well be explained by the sum of 3 Gaussians (see Table\,\ref{tab:beam_fit}).
The main beam solid angle, estimated from the radially averaged Gaussian fit, are $2.08 \pm 0.02$ and $1.67\pm0.02 \, 10^{-8}\mathrm{sr}$ in 2021 and $2.12\pm 0.01$ and $1.70\pm 0.02\, 10^{-8}\mathrm{sr}$ in 2022, for the LF and HF array, respectively. The resulting main beam efficiency, defined as the ratio of the main beam solid angle to the total beam angle, is about 0.57 and 0.54 in 2021, and 0.54 and 0.51 in 2022, for LF and HF. Those values are below the main beam efficiency measured with LABOCA of 0.8 \citep{2009A&A...497..945S}. This could be due to the complex optical coupling of CONCERTO with the APEX telescope.

For a diffraction-limited beam (circular aperture), the frequency-dependent beam size is given by 
$
    \theta_{\rm{beam}}(\nu) = {1.22 \times c}/{\nu \times D},
    \label{beamsesize}
$
\noindent where $c$ is the speed of light and $D$ = 11\,m refers to the illuminated aperture of the APEX 12-m antenna. This results in 30.5 and 26.2\,arcsec for LF and HF, respectively, for an Uranus spectrum. The radially averaged measured main beam sizes of $29.6\pm0.1$ and $26.6\pm 0.1$\arcsec are compatible with the diffraction-limited aperture.


These photometric beam properties have been obtained with the MPI moving mirror in a blocked position, far away from the zero optical path difference. Any misalignment between the two MPI arms, will create a double beam in the photometric beam data.
In the spectroscopic mode, when the MPI mirror is moving, the beam will be created by the interfering photons in the overlapping beams. The CONCERTO mirror alignments were optimized to maximize the throughput in the spectroscopic mode at the zero path difference, and not far from it where the continuum beam maps were obtained.

\begin{table}
\caption{Best fit parameters for the 2D beam image for the low frequency (LF) and high frequency (HF) arrays during the 2021 and 2022 campaigns. The Gaussian amplitudes are normalized to the sum of the amplitudes. The main beam parameters are identified with the index 1, while the 2 error beams are presented with the subscript 2a and 2b.
The $\mathrm{FWHM}_\mathrm{eff}$ are the effective full-width half maximum of a single elliptical Gaussian. The beam surface area is derived from the best-fit parameters. Those results are represented as red dashed lines in the average beam patterns of CONCERTO on Fig.\ref{fig:beam}}
\label{tab:beam_fit_image}    

\begin{tabular}{lcc}
\hline
             & \multicolumn{1}{c}{LF}  & \multicolumn{1}{c}{HF} \\

             & \multicolumn{2}{c}{2021} \\
\hline   
$\mathcal{A}_1$ [dB] & $-0.65 \pm 0.01$ & $-0.67 \pm 0.01$ \\
$\mathcal{A}_{2a}$ [dB] & $-11.58 \pm 0.08$ & $-11.22 \pm 0.04$ \\
$\mathcal{A}_{2b}$ [dB] & $-11.61 \pm 0.03$ & $-11.67 \pm 0.04$ \\
$\Delta_{2a}$ [\arcsec]  & $0.32 \times -10.46$ & $ -10.46 \times 2.53$ \\
                        & $(0.04 \times 0.11)$ & $(0.1 \times 0.03)$ \\
$\Delta_{2b}$ [\arcsec] & $0.38 \times -2.55$ & $ 8.98 \times 3.84$ \\
                        & $(0.03 \times 0.04)$ & $(0.05 \times 0.05)$ \\
$\mathrm{FWHM}_1$ [\arcsec] & $27.79 \times 30.67 $ & $27.60 \times 24.34$ \\
                            & $(0.01 \times 0.01)$  & $(0.01 \times 0.01)$\\
$\mathrm{FWHM}_{2a}$ [\arcsec] & $60.95 \times 27.69$ & $80.73 \times 36.60$ \\
                             & $(0.19 \times 0.09)$ & $(0.15 \times 0.09)$ \\ 
$\mathrm{FWHM}_{2b}$ [\arcsec] & $83.28 \times 77.49$ & $82.19 \times 51.17$ \\
                             & $(0.11 \times 0.09)$ & $(0.17 \times 0.11)$ \\ 
${\Omega}$  [$10^{-8} \rm sr$] & $3.51 \pm 0.02$ & $2.96 \pm 0.02$ \\
\vspace{1\baselineskip}\\
$\mathrm{FWHM}_\mathrm{eff}$ [\arcsec] & $31.94 \times 34.63$ & $28.48 \times 32.61$ \\
& ($0.02 \times 0.02$) & ($0.02 \times 0.02$) \\
${\Omega}_\mathrm{eff}$ [$10^{-8} \rm sr$] & $2.80 \pm 0.01$ & $2.34 \pm 0.10$ \\

\hline
& & \\
             & \multicolumn{2}{c}{2022} \\

\hline     
$\mathcal{A}_1$ [dB] & $-0.59 \pm 0.01$ & $-0.69 \pm 0.01$ \\
$\mathcal{A}_{2a}$ [dB] & $-12.89 \pm 0.07$ & $-11.77 \pm 0.06$ \\
$\mathcal{A}_{2b}$ [dB] & $-11.19 \pm 0.03$ & $-10.94 \pm 0.04$\\
$\Delta_{2a}$ [\arcsec] &  $1.75 \times -9.37$ & $-9.16 \times 2.01$ \\
                        &  ($0.05 \times 0.12$) & ($0.11 \times 0.04$) \\
$\Delta_{2b}$ [\arcsec] & $1.83 \times -3.59$ & $7.56 \times 2.90$ \\
                        & ($0.03 \times 0.04$) & ($0.04 \times 0.04$) \\
$\mathrm{FWHM}_1$ [\arcsec] & $31.30 \times 27.50$  & $28.24 \times 24.78$ \\
                            & $(0.01 \times 0.01)$ & $(0.01 \times 0.01)$ \\  
$\mathrm{FWHM}_{2a}$ [\arcsec] & $74.40 \times 28.57$ & $87.42 \times 36.08 $ \\
                               & $(0.28 \times 0.12)$ & $(0.19 \times 0.011$ \\  
$\mathrm{FWHM}_{2b}$ [\arcsec] & $84.84 \times 74.51$ & $80.39 \times 55.75 $ \\
                               & $(0.11 \times 0.09)$ & $(0.16 \times 0.10$ \\  
${\Omega}$  [$10^{-8} \rm sr$] & $3.60 \pm 0.02$ &  $3.17 \pm 0.02$ \\
\vspace{1\baselineskip}\\
$\mathrm{FWHM}_\mathrm{eff}$ [\arcsec] & $32.00 \times 35.36$ & $29.29 \times 33.90$ \\
& ($0.02 \times 0.02$) & ($0.02 \times 0.02$) \\
${\Omega}_\mathrm{eff}$ [$10^{-8} \rm sr$] & $2.85 \pm 0.01$ & $2.48 \pm 0.10$ \\

\hline

\end{tabular}
\end{table}

\begin{table}
 \caption{Best fit parameters for the radial beam profile for the low frequency (LF) and high frequency (HF) arrays during the 2021 and 2022 campaigns. The Gaussian amplitudes are normalized to the sum of the amplitudes. The $\mathrm{FWHM}_\mathrm{eff}$ is the effective full-width half maximum beam profile of a single Gaussian. The beam surface area is derived from the best-fit parameters.}              
\label{tab:beam_fit}      
\centering
\begin{tabular}{l r r}
\hline
             & \multicolumn{1}{c}{LF}  & \multicolumn{1}{c}{HF} \\

             & \multicolumn{2}{c}{2021} \\
\hline             
$\mathcal{A}_1$ [dB] & $-0.48 \pm 0.04$ & $-0.54 \pm 0.05$ \\
$\mathcal{A}_2$ [dB] & $-10.32 \pm 0.35$ & $-9.66 \pm 0.39$ \\
$\mathcal{A}_3$ [dB] & $-19.61 \pm 0.97$ & $-21.00 \pm 1.00$ \\
$\mathrm{FWHM}_1$ [\arcsec] & $29.60 \pm 0.11$ & $26.66 \pm 0.13$ \\
$\mathrm{FWHM}_2$ [\arcsec] & $66.45 \pm 1.08$ & $61.30 \pm 0.95$ \\
$\mathrm{FWHM}_3$ [\arcsec] & $124.26 \pm 2.67$ & $124.54 \pm 2.89$ \\
${\Omega}$ [$10^{-8} \rm sr$] & $3.64 \pm 0.13$ & $3.10 \pm 0.10$ \\
\vspace{1\baselineskip}\\
$\mathrm{FWHM}_\mathrm{eff}$ [\arcsec] & $34.01 \pm 0.77$ &   $30.86 \pm 0.62$ \\
${\Omega}_\mathrm{eff}$ [$10^{-8} \rm sr$] & $3.08 \pm 0.11$ & $2.54 \pm 0.08$ \\

\hline
& & \\
             & \multicolumn{2}{c}{2022} \\

\hline             
$\mathcal{A}_1$ [dB] & $-0.51 \pm 0.03$ & $-0.60 \pm 0.04$ \\
$\mathcal{A}_2$ [dB] & $-9.92 \pm 0.28$ & $-9.18 \pm 0.30$ \\
$\mathcal{A}_3$ [dB] & $-20.51 \pm 0.45$ & $-20.95 \pm 0.73$ \\
$\mathrm{FWHM}_1$ [\arcsec] & $29.58 \pm 0.10$ & $27.10 \pm 0.11$ \\
$\mathrm{FWHM}_2$ [\arcsec] & $67.19 \pm 0.66$ & $61.80 \pm 0.72$ \\
$\mathrm{FWHM}_3$ [\arcsec] & $149.08 \pm 2.20$ & $140.23 \pm 3.42$ \\
${\Omega}$ [$10^{-8} \rm sr$] & $3.87 \pm 0.07$ & $3.44 \pm 0.10$ \\

\vspace{1\baselineskip}\\
$\mathrm{FWHM}_\mathrm{eff}$ [\arcsec] & $34.39 \pm 0.98$ & $31.91 \pm 0.64$ \\
${\Omega}_\mathrm{eff}$ [$10^{-8} \rm sr$] & $3.16 \pm 0.16$ & $2.72 \pm 0.09$ \\
\hline
\end{tabular}
\end{table}

\section{Photometry and Stability}
\label{sec:photometry}

In this section, we utilize the continuum maps of Uranus to assess the consistency of the flux density measurements. Moreover, we validate the robustness of the photometry against variations in atmospheric conditions and temperature using an extensive set of observations.

\subsection{Continuum measurements}

\begin{figure}
   \centering
   \includegraphics[width=8.1cm]{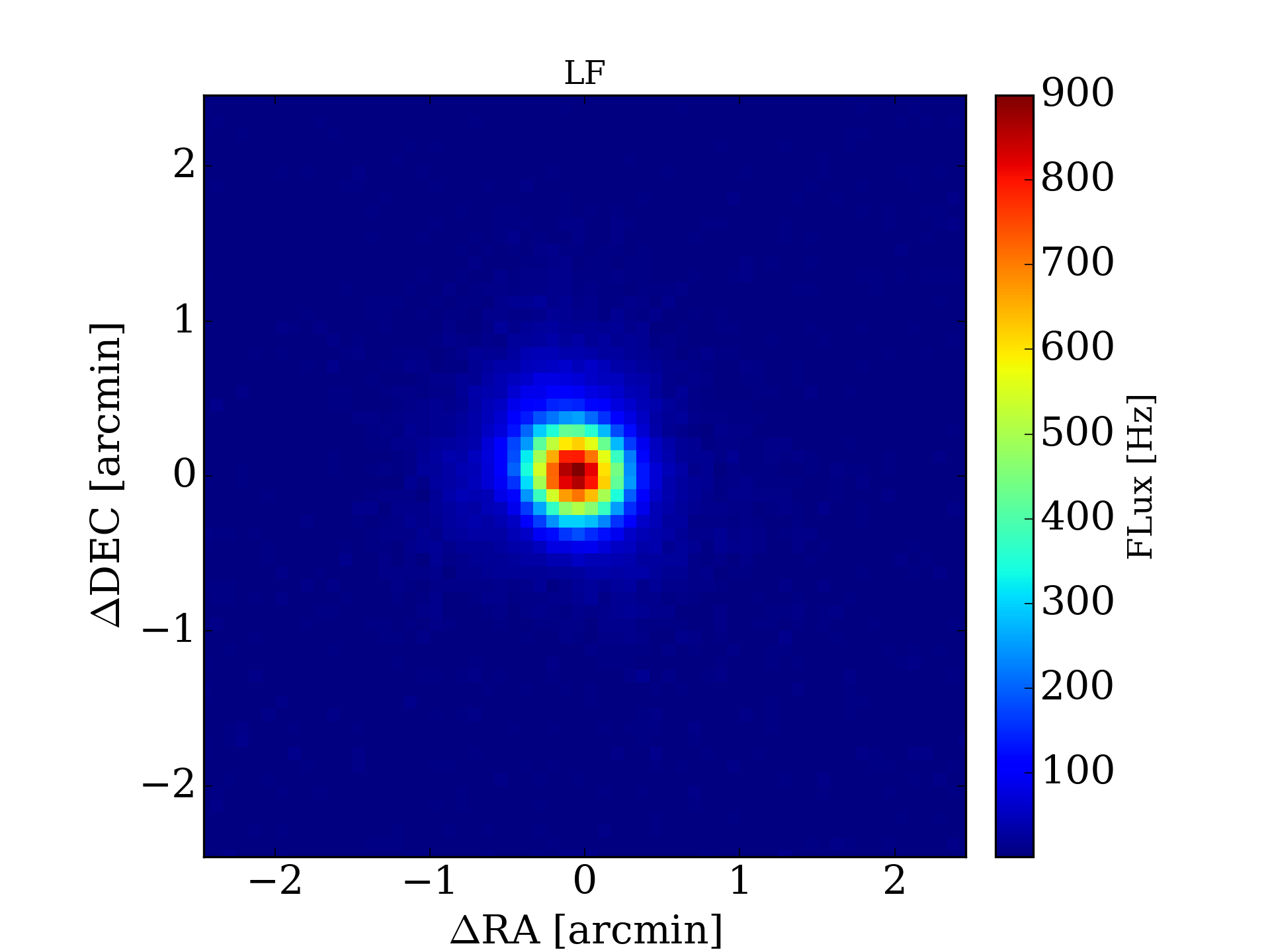}
   \includegraphics[width=8.1cm]{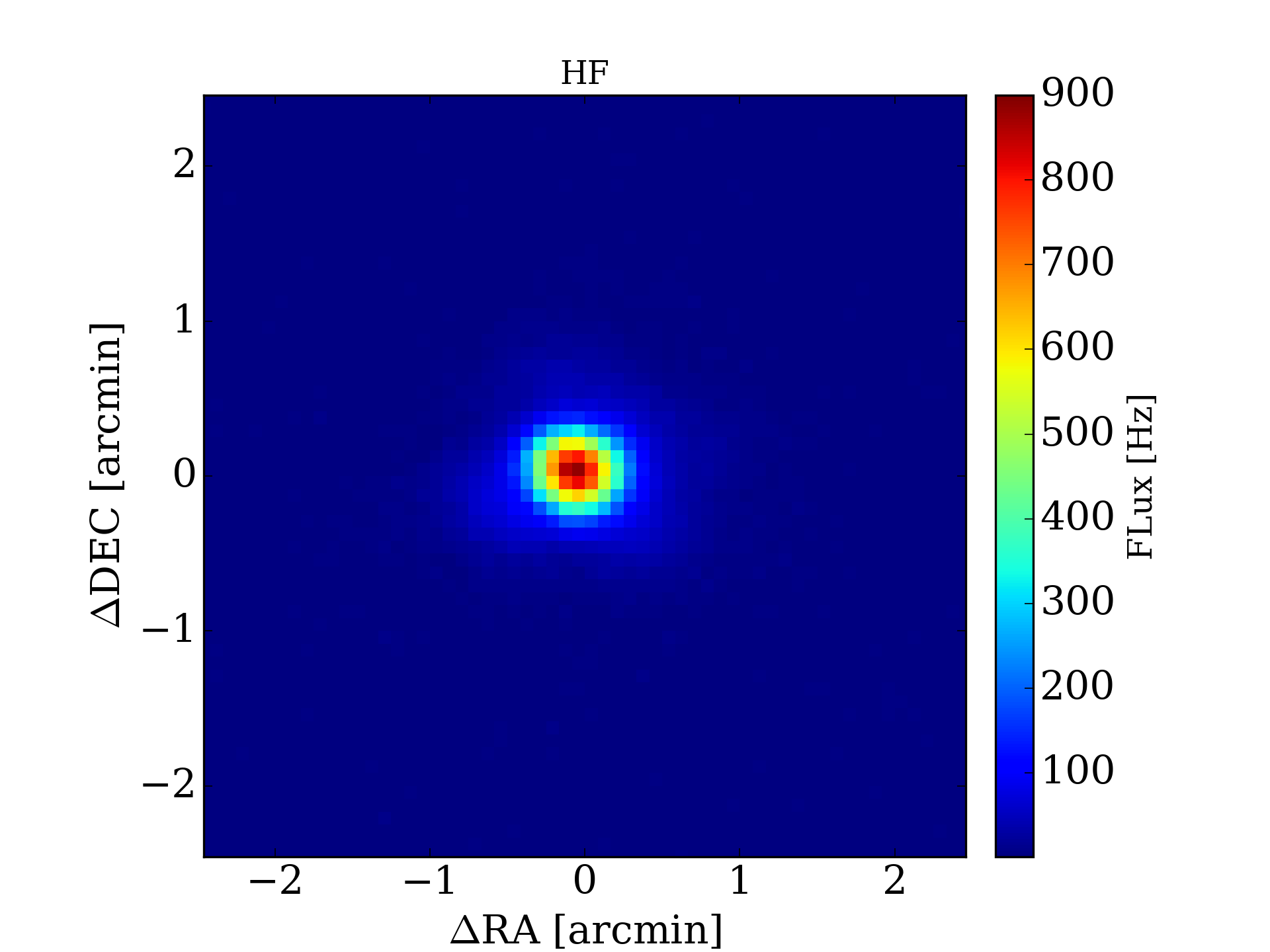}
   \caption{Continuum maps of Uranus from one pointing observation carried out in October 2021. The upper and lower panel refers to the LF and HF, respectively.}
   \label{Uranus_map}
\end{figure}

During the observational campaigns spanning from June 2021 to December 2022, Uranus was utilized as a pointing calibration source. After eliminating low-quality data due to instrumental error or unfavourable weather conditions (i.e. pwv$>$1.5\,mm), we obtained 44 high-quality pointing observations. Using our pipeline, we generated the continuum maps of Uranus. Figure.~\ref{Uranus_map} showcases the LF and HF continuum maps obtained from an observation conducted in October 2021, which displays a clear signal of Uranus in the centre.

Since the apparent angular size of planets varies over a year, we accounted for this effect by utilising the PyEphem software \citep{2011ascl.soft12014R}.

\begin{figure}
   \centering
   \includegraphics[width=8.1cm]{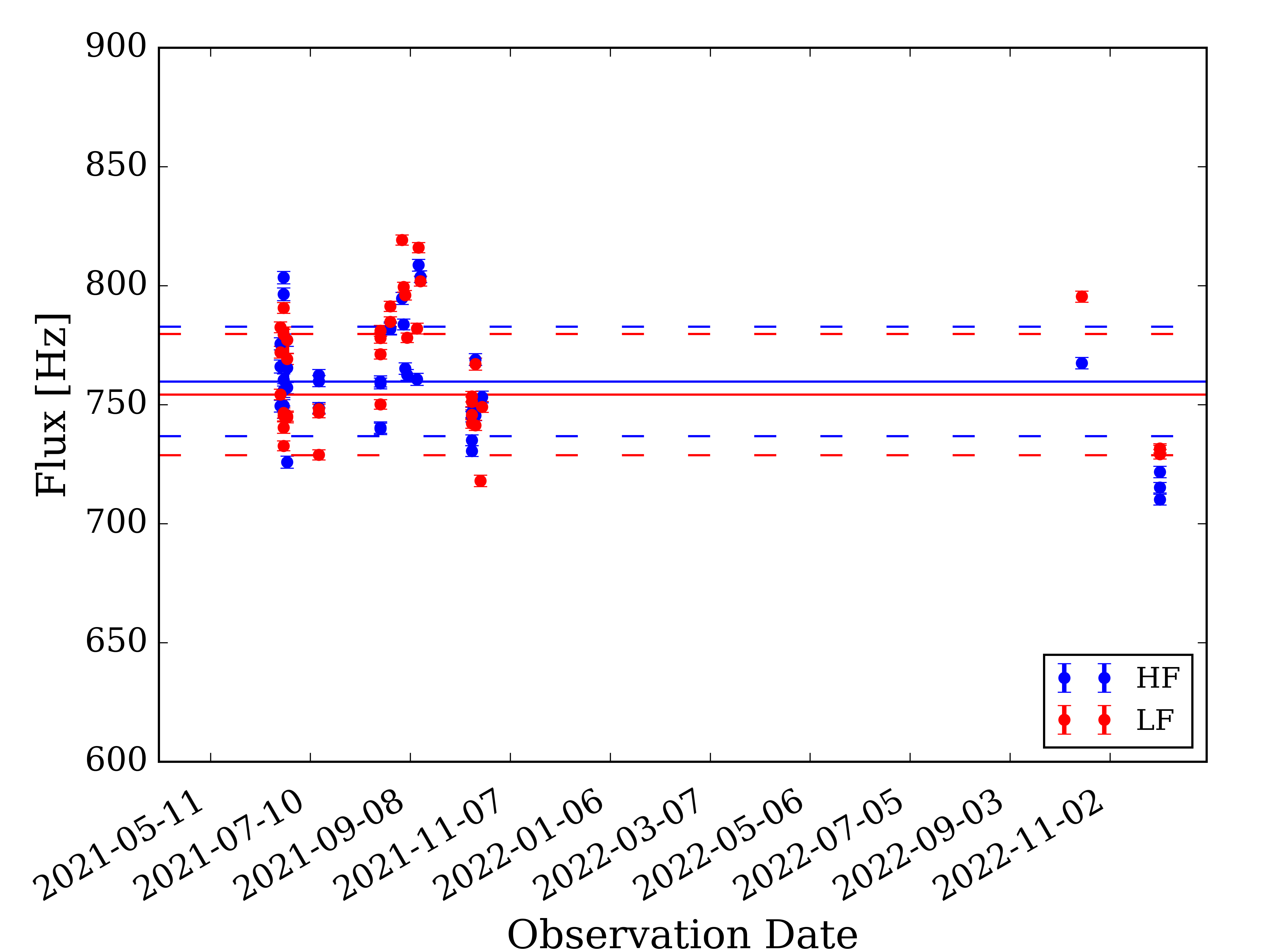}
   \includegraphics[width=8.1cm]{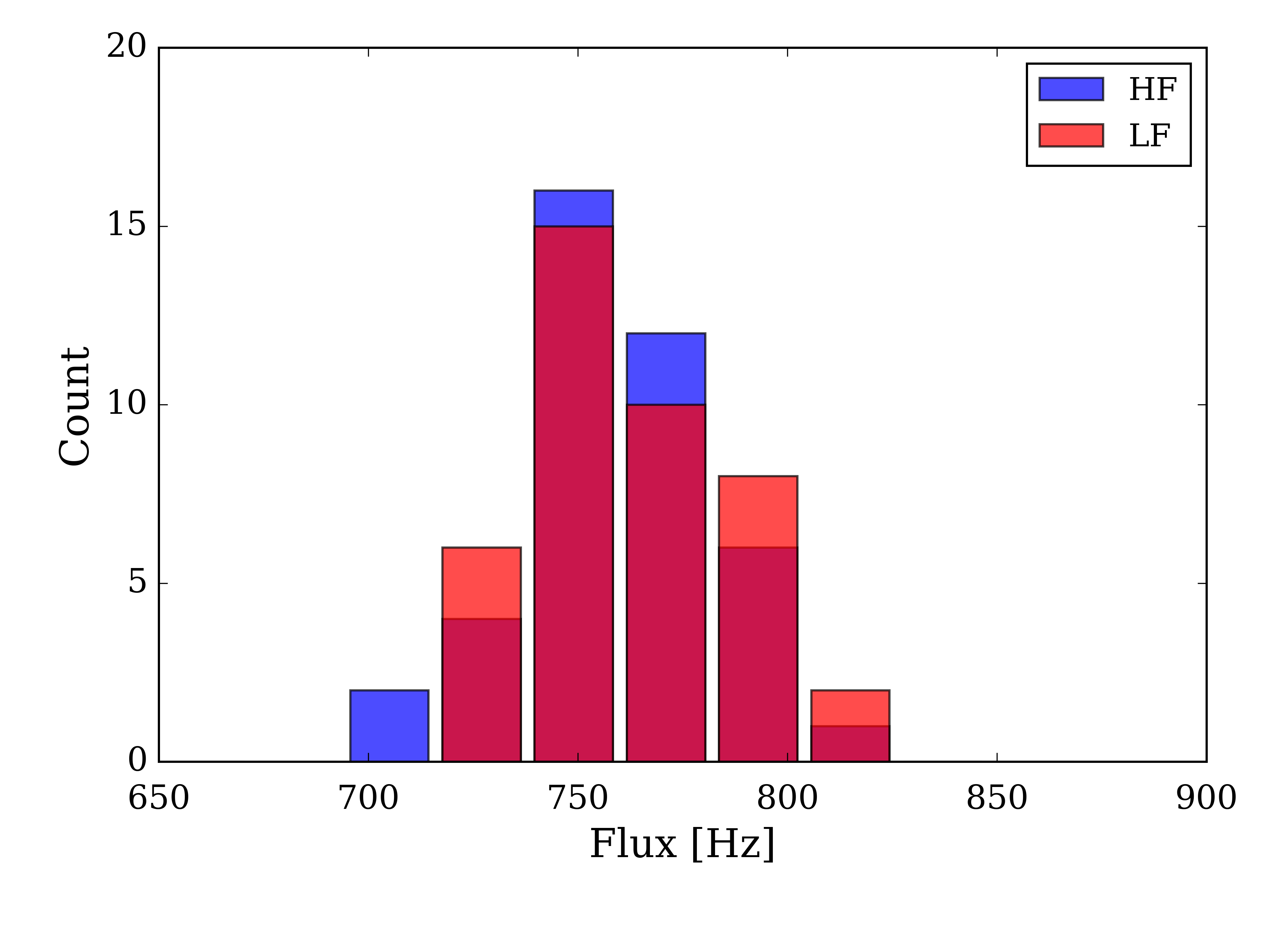}
   \caption{Statistics of CONCERTO-measured flux of Uranus. Upper: the measured flux from Uranus continuum maps. The average and RMS uncertainties are shown as solid lines and dashed lines; Lower: the histogram of the measured flux for LF and HF. The red and blue of both panels refer to LF and HF, respectively.}
   \label{Uranus_flux}
\end{figure}

Flux densities of Uranus are given by the amplitude of a 2D elliptical Gaussian fit to the continuum maps at the effective frequency in the LF and HF bands, respectively. The upper panel of Fig.\,\ref{Uranus_flux} displays the measured flux of Uranus from observations taken in 2021 and 2022. The measurements of flux density are consistent across all observations. The statistical results for the LF and HF bands are 754$\pm$25\,Hz and 760$\pm$23\,Hz, respectively. The lower panel of Fig.\,\ref{Uranus_flux} presents the histogram of the measured flux. The distribution peaks at nearly the same flux, with a 1$\sigma$ dispersion of 3.4$\%$ and 3.0$\%$ for LF and HF, respectively.

\subsection{Flux stability with the precipitable water vapour}

\begin{figure}
   \centering
   \includegraphics[width=8.1cm]{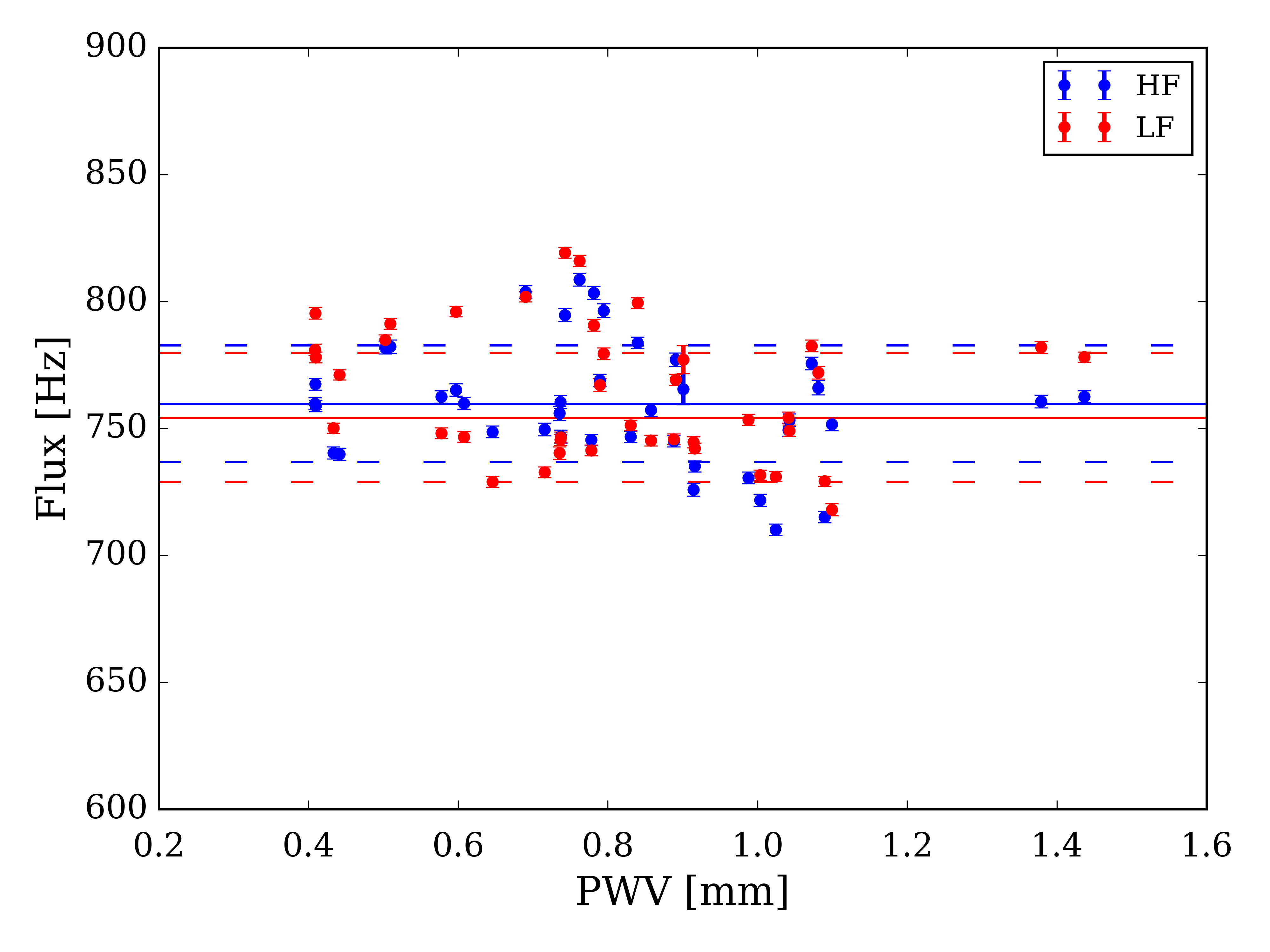}
   \caption{Measured flux of Uranus as a function of the precipitable water vapour (pwv).}
   \label{Uranus_flux_pwv}
\end{figure}

To assess the stability of the continuum flux with different atmospheric conditions, we show the opacity-corrected flux measurement of Uranus versus the pwv in Fig.~\ref{Uranus_flux_pwv}. Our results indicate that the flux remains stable across various atmospheric conditions at both HF and LF wavelengths. This demonstrates the stability of CONCERTO and the effectiveness of our opacity correction method for pwv values ranging between 0.2 and 1.5\,mm.

\subsection{Flux stability against day time}

\begin{figure}[hbt!]
   \centering
   \includegraphics[width=8.1cm]{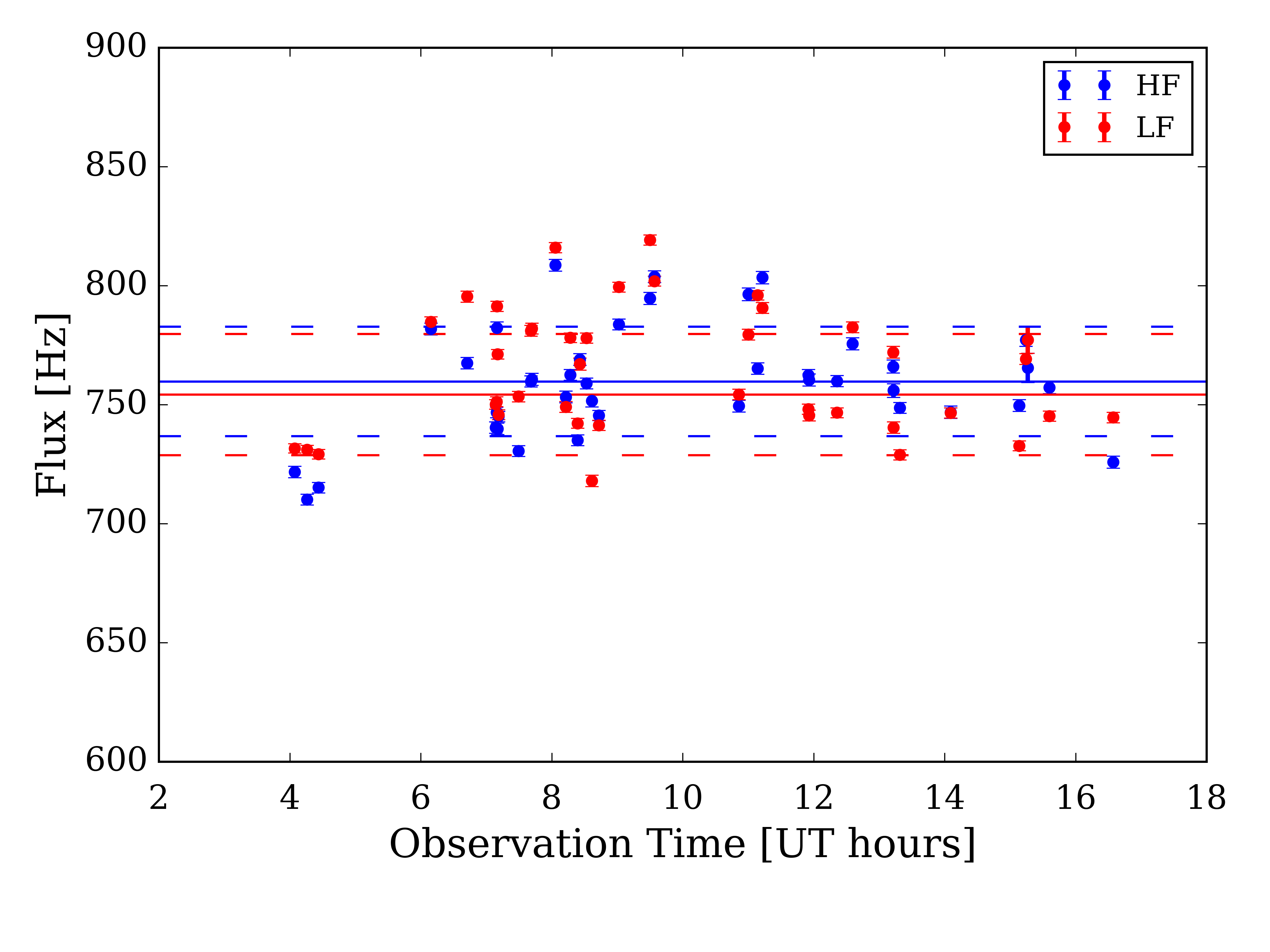}
   \includegraphics[width=8.1cm]{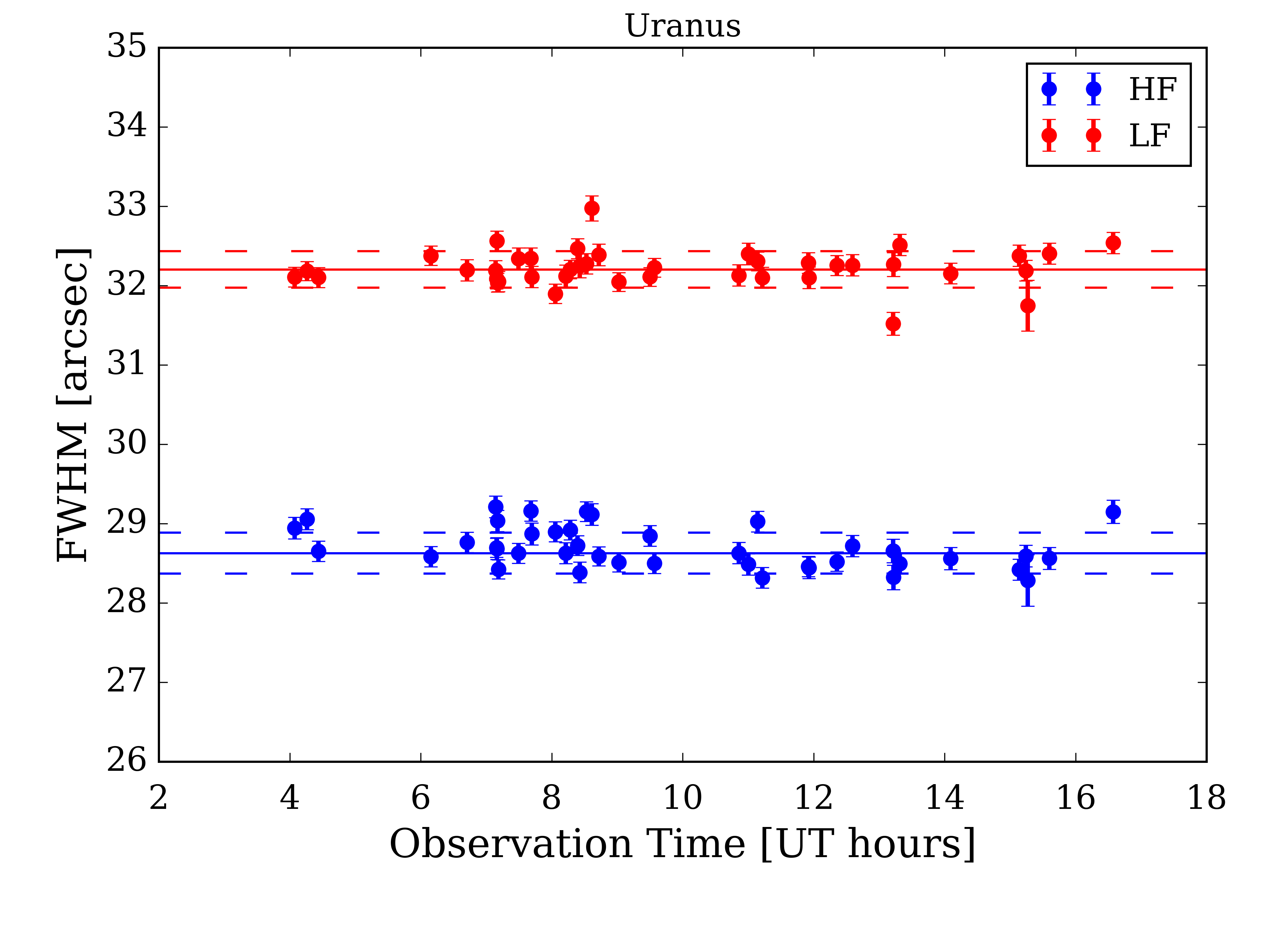}
   \caption{Uranus measured flux density (upper panel) and beam size (lower panel) as a function of the observation time in UT hours. The average and RMS uncertainties are shown as solid and dashed lines, respectively.}
   \label{Uranus_flux_Temperature}
\end{figure}

For the 30-m primary mirror of the IRAM (Institut de Radio Astronomie Millimétrique) telescope, inhomogeneous solar illumination and atmospheric anomalous fluctuations can cause de-focusing of the 30-m mirror. These effects are most pronounced in the afternoon and around sunrise and sunset \citep{2001A&A...374..348O,2020A&A...637A..71P}. To quantify the potential impact of these effects on our observations at APEX, we study the evolution of the measured flux density and beam size of Uranus as a function of observation time in UT hours, as shown in Fig.~\ref{Uranus_flux_Temperature}. We find no clear correlation between the measured flux density or beam size and observation time. The time stability of the flux density and beam size estimates is supported by a relative RMS of 3\% and 1\%, respectively.

\subsection{Flux stability against the effective beam size \label{calib_wrt_beam}}
\begin{figure}
   \centering
   \includegraphics[width=8.1cm]{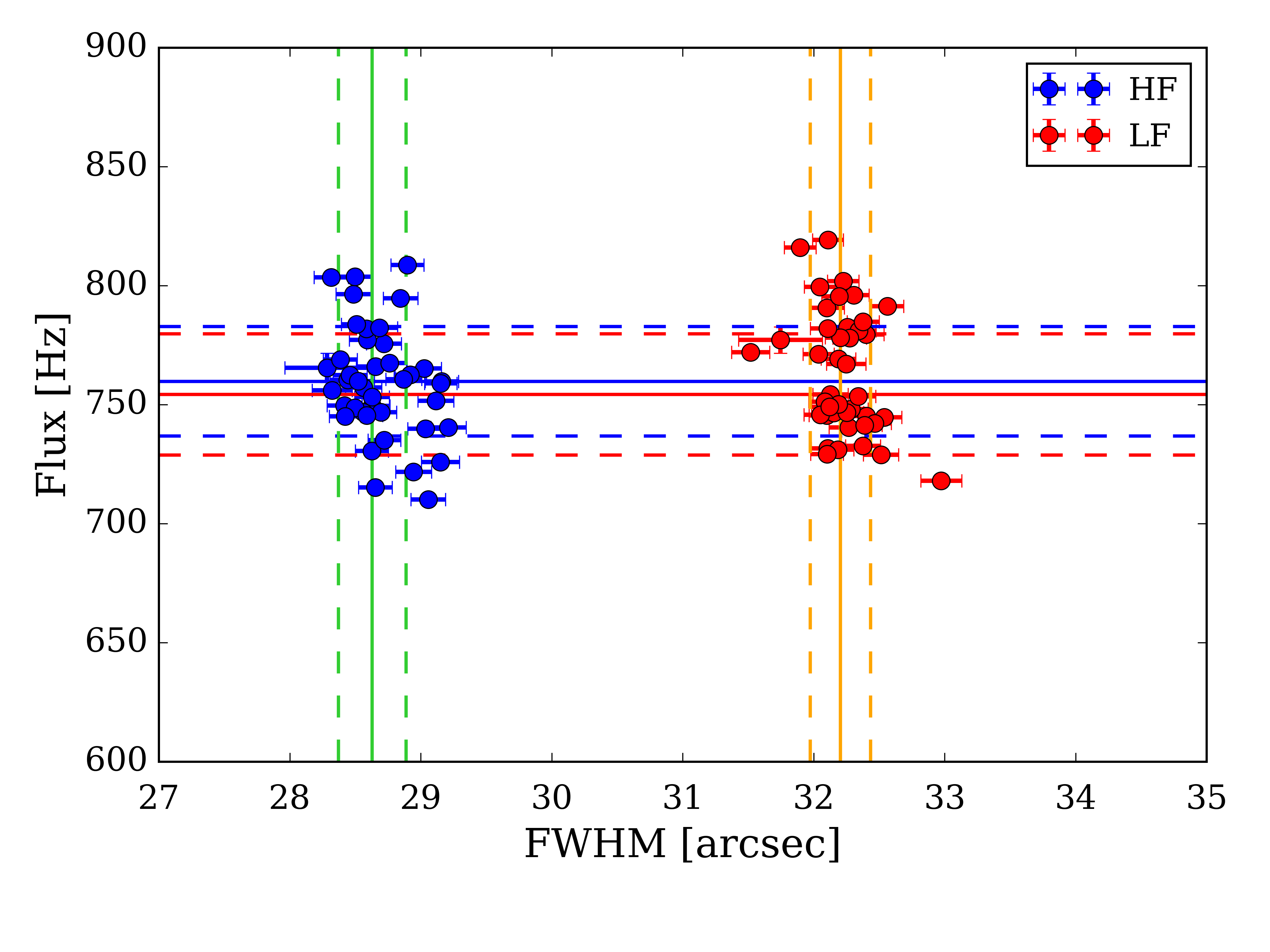}
   \caption{Measured flux of Uranus as a function of the measured effective beam size (which is let free in the fit of Uranus flux here). The average and RMS uncertainties are shown as solid and dashed lines, respectively.}
   \label{Uranus_flux_fwhm}
\end{figure}

In this section, we check for the beam sizes by letting the beam size free in the fit of Uranus flux. We compute the effective beam size from pointing maps by fitting 2D elliptical Gaussians. The geometrical FWHM is computed as $\rm FWHM = 2\sqrt{2\ln\left(2\right)\ \sigma_{x}\sigma_{y}}$,
where $\sigma_{x}$ and $\sigma_{y}$ represent the standard deviations of the 2D elliptical Gaussian. We obtain an effective FWHM of 32.20 $\pm$ 0.23\,arcsec and 28.63 $\pm$0.26\,arcsec for the LF and HF bands, respectively. The effective FWHM values are slightly larger than the $\mathrm{FWHM}_1$ presented in Table~\ref{tab:beam_fit}, as a consequence of using only one Gaussian fitting in this analysis. On the contrary, they are a bit smaller than the $\mathrm{FWHM}_{\rm {eff}}$  measured on the beam map (Table~\ref{tab:beam_fit_image}) because the pointing maps are noisier and cannot properly reveal the extent of the beam.  \\
We present the measured flux density of Uranus as a function of the CONCERTO effective beam size in Fig.~\ref{Uranus_flux_fwhm}. We find no correlation for both LF and HF measurements. 

\section{Absolute photometric calibration}
\label{sec:calibration}

The recorded data needs to be calibrated from Hz to Jy/beam to obtain the astronomical signal measured in radio astronomy units. We derive the absolute photometric calibration factor by comparing the measured flux of Uranus to the planet model. We also verify our photometric calibration using measurements of Mars and quasars. 

\subsection{Calibration on Uranus}
\label{sec:cali_uranus}

The absolute photometric calibration for point sources is applied to Uranus by comparing the recorded signal in Hz to the {\sc ESA2} model \citep{2017A&A...607A.122P}. The predicted Uranus continuum flux measured by CONCERTO can be calculated as the bandpass-weighted integral of the Uranus flux in the CONCERTO frequency band:
\begin{eqnarray}
S_{\rm{Uranus}} = \frac{\int S_{\rm{model}}(\nu)F(\nu)\eta(\nu){\rm{d}}\nu}{\int F(\nu)\eta(\nu){\rm{d}}\nu},
    \label{continnum_flux}
\end{eqnarray}
where $S_{\rm{model}}$ is the flux model of the source (here is the spectral energy distribution of Uranus), $F(\nu)$ is the relative spectral response and $\eta(\nu)$ the aperture efficiency (which is proportional to $\nu^2$ for the absorber-coupled case, see \citet{2013MNRAS.434..992G}. 

\begin{figure}
   \centering
   \includegraphics[width=8.1cm]{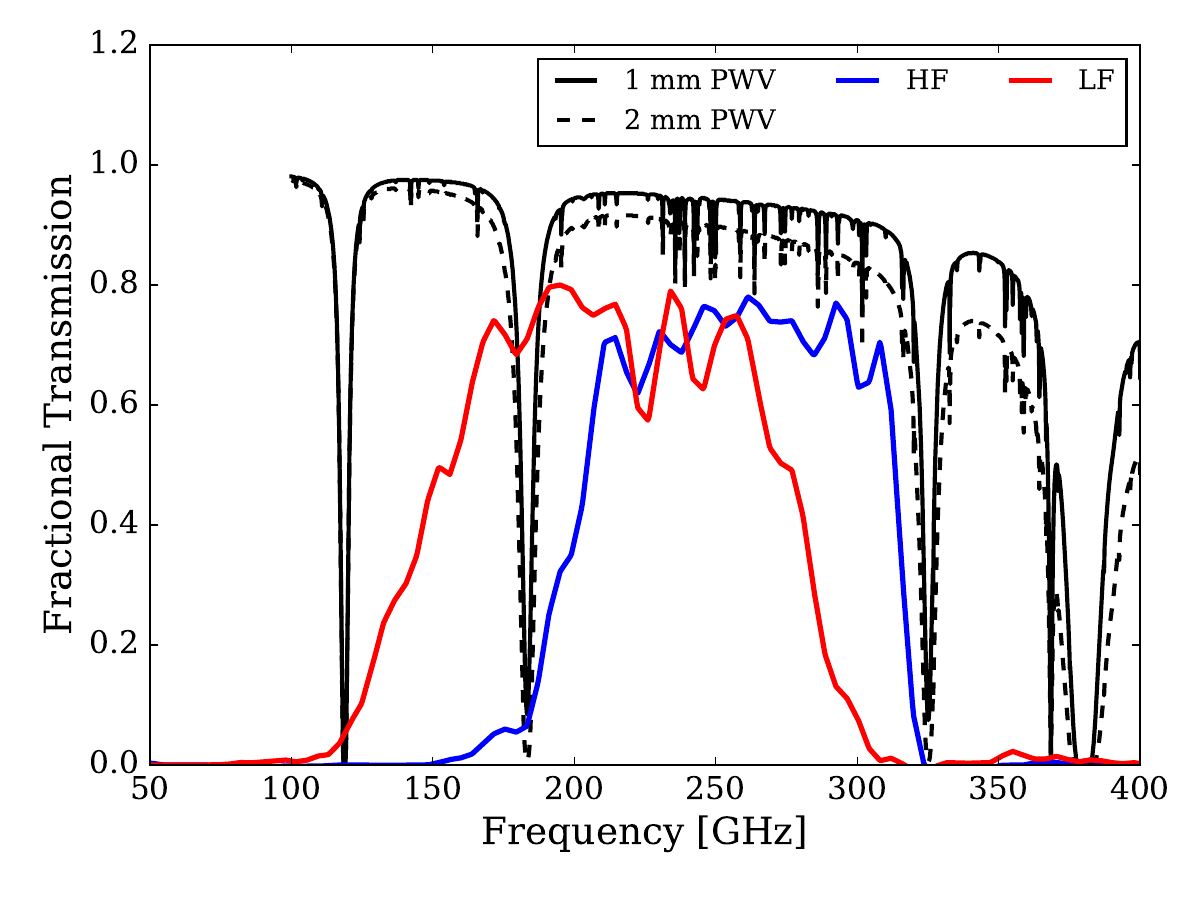}
   \caption{CONCERTO bandpass $F_{meas}=F(\nu)\eta(\nu)$ (see Appendix\,\ref{appendix:bandpass}) for LF (red) and HF (blue), and APEX Chajnantor atmospheric transmission obtained with the ATM model for precipitable water vapour pwv=1\,mm (black solid line) and pwv=2\,mm (black dashed line). Bandpasses are arbitrarily normalised to improve the clarity of the figure.}
   \label{bandpass}
\end{figure}

The measured CONCERTO bandpass $F_{meas}=F(\nu)\eta(\nu)$ for LF and HF is shown in Fig.~\ref{bandpass}. Additionally, for illustrative purposes, we include the APEX Chajnantor atmospheric transmission obtained with ATM model \citep{982447} for pwv of 1 and 2\,mm. The CONCERTO bandpass is computed using the spectroscopic data acquired from the test scans carried out with the shutter in a closed position (i.e. by looking at a flat field). The specifics of the bandpass estimate are outlined in Appendix~\ref{appendix:bandpass}.

The effective frequency, denoted as $\nu_{\rm{eff}}$, is calculated as the frequency-weighted integral of the product of CONCERTO bandpass and Uranus spectral energy distribution\citep{2014A&A...571A...9P,2020A&A...637A..71P}:
\begin{equation}
    \nu_{\rm{eff}}= \frac{\int \nu S_{\rm model}(\nu)F(\nu)\eta(\nu){\rm d}\nu}{\int S_{\rm model}(\nu)F(\nu)\eta(\nu){\rm d}\nu}.
    \label{eq:nu_eff}
\end{equation}
For Uranus, the effective frequencies for LF and HF arrays are $\mathrm{\nu_{\rm{eff}}^{Uranus}}$=225\,GHz and $\mathrm{\nu_{\rm{eff}}^{Uranus}}$=262\,GHz, respectively.

\begin{figure}
   \centering
   \includegraphics[width=8.1cm]{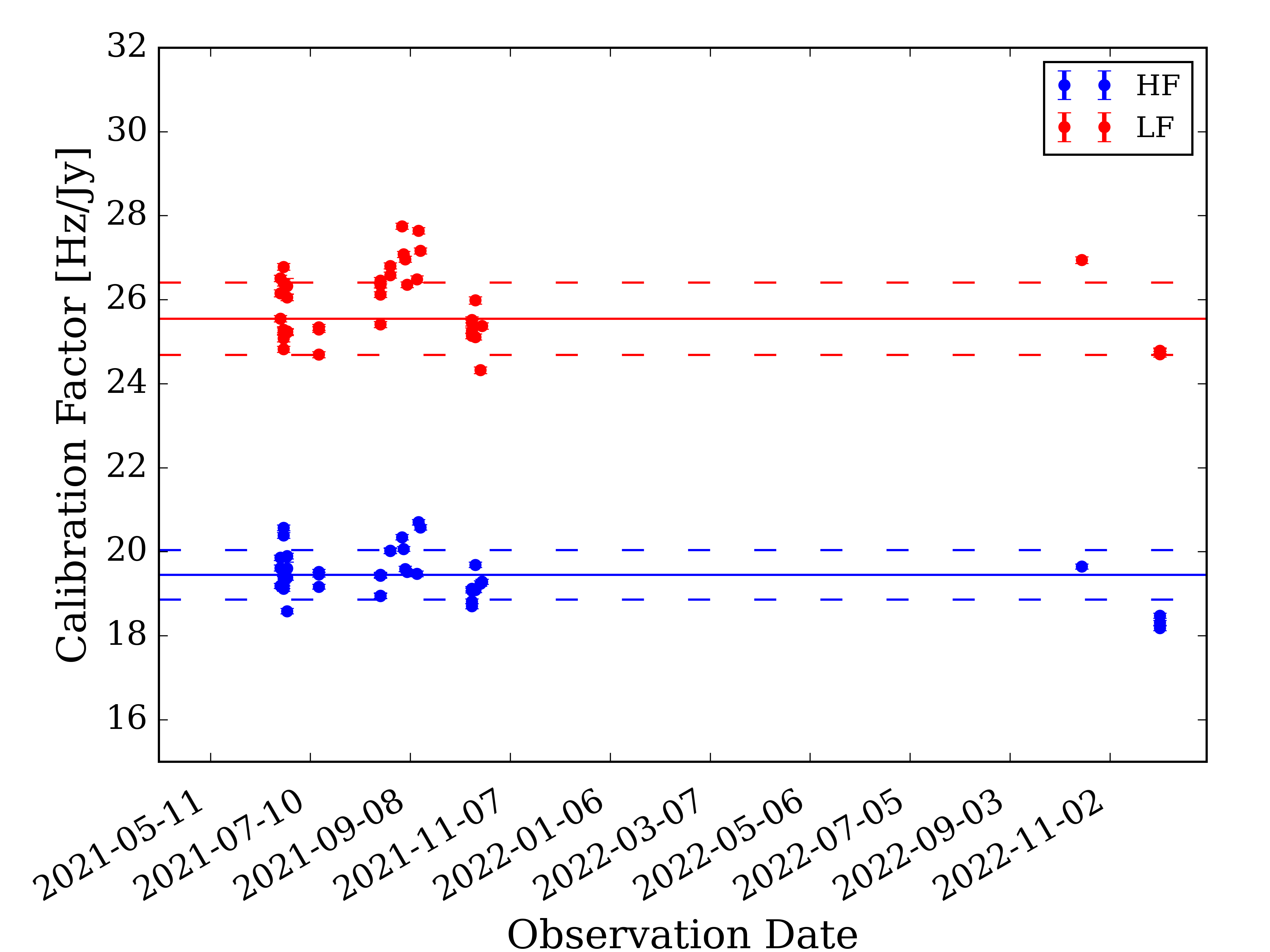}
   \caption{The Jy/beam-to-Hz calibration factor for CONCERTO LF (red) and HF (blue) band. 
   }
   \label{calibration_factor}
\end{figure}

To obtain the Jy-to-Hz calibration factor for CONCERTO, we divide the measured flux density by the flux density value extracted from the Uranus {\sc ESA2} model. The resulting calibration factor is shown in Fig.~\ref{calibration_factor}. The factor is 25.6$\pm$0.9\,Hz/[Jy/beam] and 19.5$\pm$0.6\,Hz/[Jy/beam] for LF and HF, respectively.

\subsection{Implication for Mars}

\begin{figure}
   \centering
   \includegraphics[width=8.1cm]{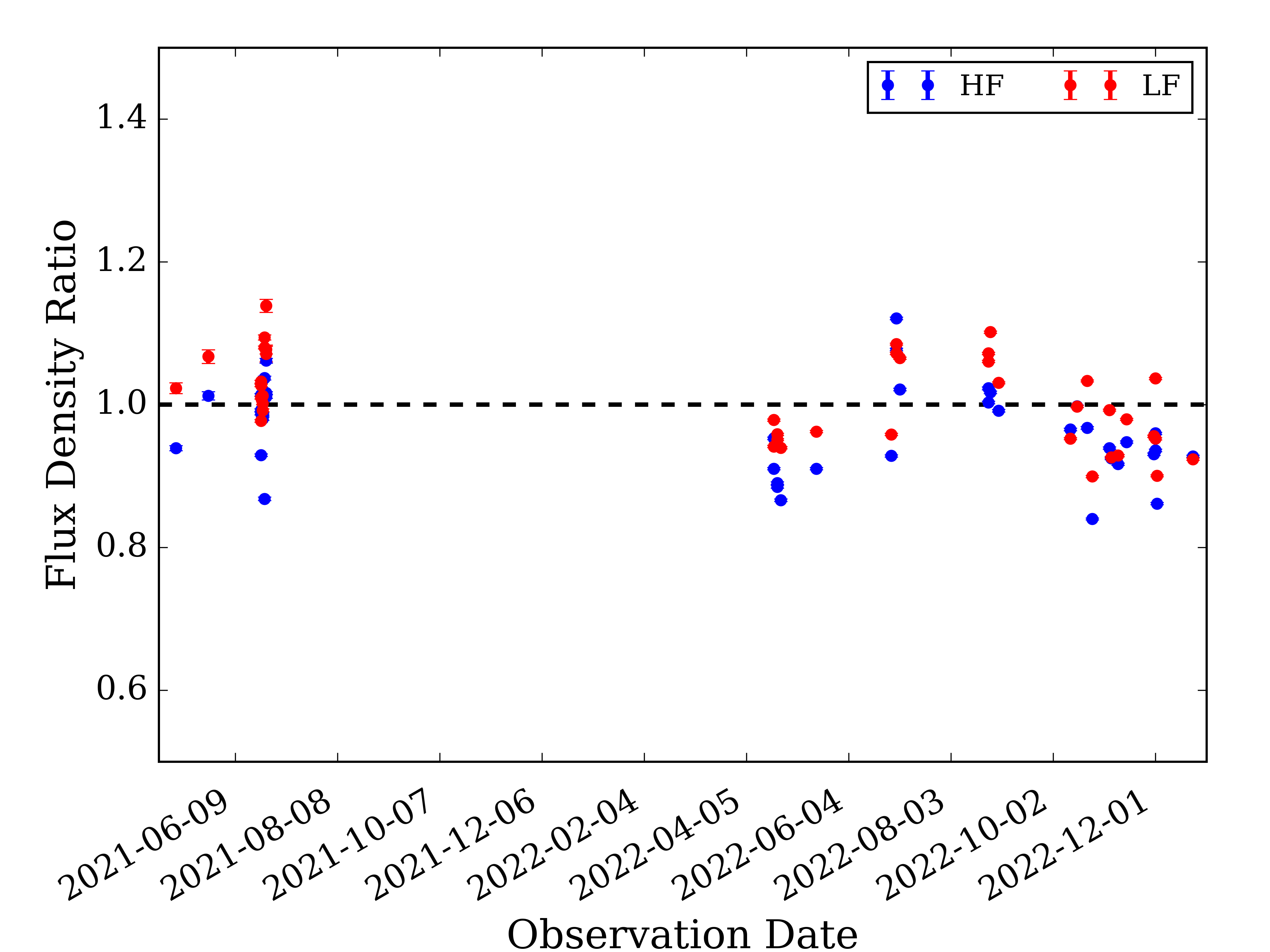}
   \caption{The measured-to-predicted flux density ratio for Mars, calculated by comparing the CONCERTO flux deduced using the calibration factors measured on Uranus and the Mars model.}
   \label{Mars_flux_comapre}
\end{figure}

To verify the accuracy of our absolute photometric calibration, we applied the deduced calibration factors to the Mars continuum signal and compared it with the model from \cite{Lellouch}. Due to the considerable variation in the distance between Mars and Earth over time and Mars's seasonal variations, the flux densities of Mars undergo significant changes, ranging from approximately 100\,Jy (June 2021) to around 2\,000\,Jy (December 2022). The measured-to-predicted flux density ratio is shown in Fig.~\ref{Mars_flux_comapre}. A colour correction between Uranus and Mars has been applied to ensure the consistency of the measurements. The average ratios for the LF and HF bands are 1.01$\pm$0.06 and 0.97$\pm$0.06, respectively. Despite the substantial variations in the flux densities of Mars, the measured-to-predicted flux density ratio remains approximately constant over time, suggesting that our calibration is highly reliable. 

\subsection{Secondary Calibrators \label{quasars}}

In Sect.~\ref{sec:cali_uranus}, we obtained the absolute calibration factors by measuring Uranus, which emits a flux of approximately 30\,Jy. To determine the calibration factor in the sub-10\,Jy range and assess CONCERTO's stability in different brightness ranges, we used quasars, whose flux densities range from 0.5 to 14\,Jy, as secondary calibrators. During the scientific-purpose observation campaigns, quasars are observed as calibrators for the pointing observations. The calibration factors are obtained by comparing the flux measured by the Atacama Large Millimeter/submillimeter Array (ALMA) and CONCERTO. \\

We downloaded the ALMA band 3 (frequency coverage of 84--116\,GHz) and band 7 (frequency coverage of 275--373\,GHz) continuum measurements of the quasars from the ALMA calibrator database. For each quasar, the pair of band 3 and band 7 observed within 26 hours by ALMA and close in time (within 3 days) to the CONCERTO observations are selected. The corresponding quasars and their ALMA measurements in band 3 and band 7 are given in Table~\ref{quasar_table}. To be compared with the Uranus calibration factors, those measurements have to be converted to the Uranus effective frequency of CONCERTO. This is accomplished by first estimating the spectra of quasars $S_{\rm{QSO}}(\nu)$ by fitting the ALMA band 3 and band 7 measurements using a power-law function. 
Next, the predicted flux of the ALMA measurements for the CONCERTO band is calculated by integrating the quasar spectra weighted by the CONCERTO bandpass: $S^{\rm{CONCERTO}}_{\rm{QSO}} = \int S_{\rm{QSO}}(\nu)F(\nu)\eta(\nu){\rm{d}}\nu \big / \int F(\nu)\eta(\nu){\rm{ d}}\nu$. Finally, these fluxes are colour corrected \citep[e.g.][]{2013MNRAS.434..992G,2014A&A...571A...9P} to the CONCERTO effective frequencies derived for the Uranus spectrum (i.e. the fluxes are converted from $\mathrm{\nu_{\rm{eff}}^{QSO}}$ to $\mathrm{\nu_{\rm{eff}}^{Uranus}}$ using a power law).\\

For each quasar observed with CONCERTO, we generated a continuum map and then measured the flux that we also colour-correct from $\mathrm{\nu_{\rm{eff}}^{QSO}}$ to $\mathrm{\nu_{\rm{eff}}^{Uranus}}$.\\

\begin{figure}
   \centering
   \includegraphics[width=8.1cm]{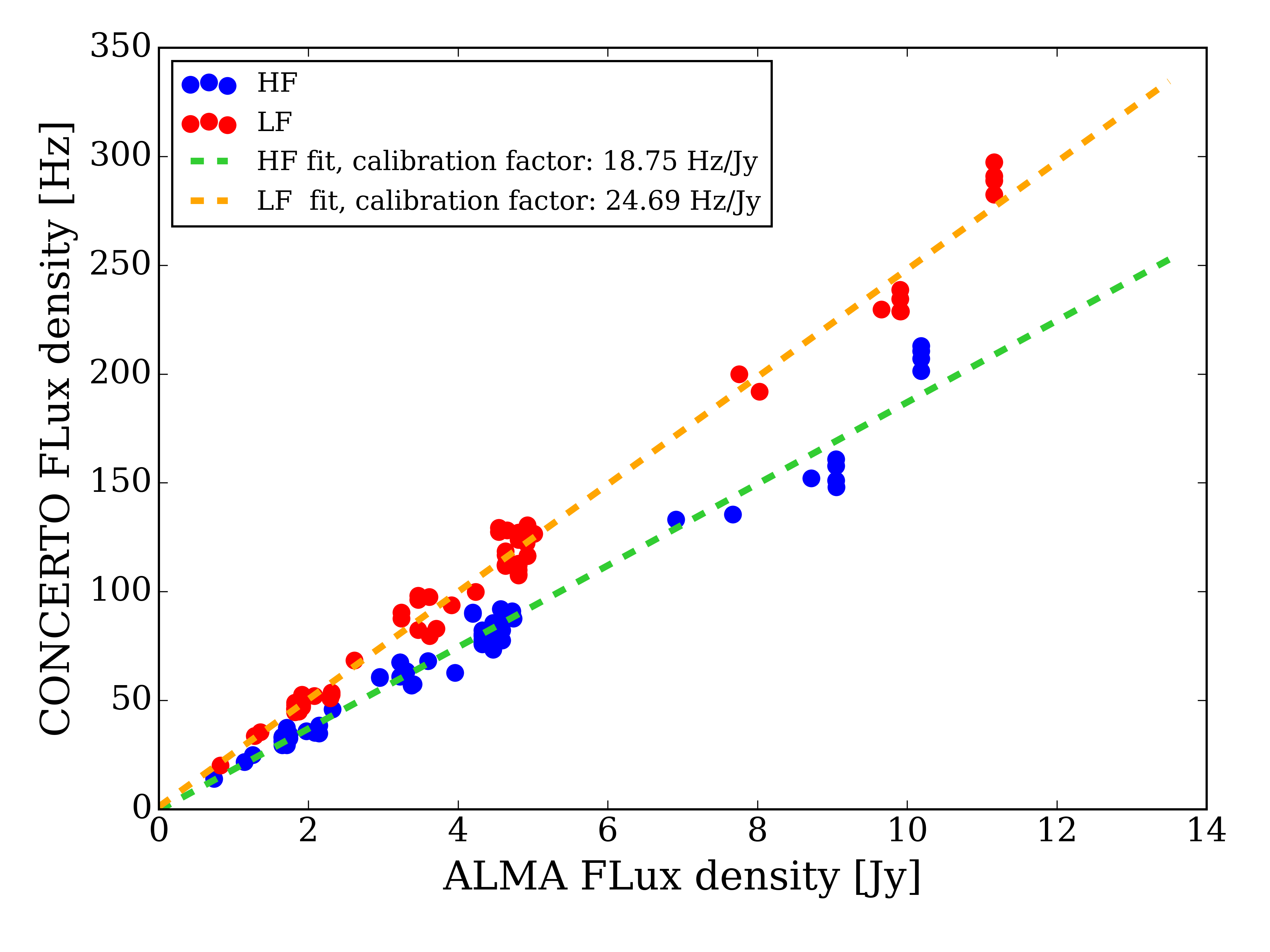}
   \caption{ALMA and CONCERTO measurements for quasars. The ALMA fluxes in LF and HF are deduced from ALMA band 3 and 7 measurements (See Sect.\ref{quasars} for details). CONCERTO and ALMA fluxes are colour corrected to a fixed effective frequency which is derived using Uranus SED ($\mathrm{\nu_{\rm{eff}}^{Uranus}}$.).
   The first-order polynomials fitting (dashed lines) gives Jy-to-Hz calibration factors of  24.7$\pm$0.4\,Hz/[Jy/beam] and 18.8$\pm$0.4\,Hz/[Jy/beam], for LF and HF, respectively.
   }
   \label{calibration_factor_quasar}
\end{figure}

The calibration factors are then determined as the ratio between the CONCERTO measurements and the ALMA flux. Figure~\ref{calibration_factor_quasar} shows the ALMA results and CONCERTO measurements, and the linear fits for LF and HF. The Jy-to-Hz calibration factor derived from quasar pointing observations is 24.7$\pm$0.4\,Hz/[Jy/beam] and 18.8$\pm$0.4\,Hz/[Jy/beam] for LF and HF, respectively. These values agree with the calibration factors obtained from observations of Uranus pointing, which demonstrates the linearity of CONCERTO in measuring sources with flux levels spanning from a few Jy to dozens of Jy and different spectral indices.

\begin{table*}
	\centering
	\caption{ALMA measurements of quasars used for absolute photometric calibration factor calculation. For some quasars (J1256-0547, J1058+0133, J0854+2006, J0538-4405, J1337-1257 and J0522-3627), several ALMA observation pairs observed at different months are used, but we only show one pair as an example here. The third and fourth columns show the flux density measured by ALMA. The fifth column is the slope ($\alpha_{\rm{QSO}}$) of the quasar power-law spectrum ($S_{\rm{QSO}}(\nu) \propto \nu^{\alpha_{\rm{QSO}}}$). The seventh and eighth columns present the interpolation of ALMA measurements to CONCERTO (given at the Uranus effective frequencies). The last two columns give the flux density measured by CONCERTO pointing scans, also corrected to Uranus effective frequencies for HF and LF, respectively.}
	\begin{tabular}{|c|c|c|c|c|c|c|c|c|c|}
		\hline
		   & \multicolumn{2}{c|}{ALMA Band 3} & \multicolumn{2}{c|}{ALMA Band 7} & \multicolumn{3}{c|}{Interpolation} & \multicolumn{2}{c|}{CONCERTO}\\
		\hline
            Quasars & Observation Date & Flux Density & Observation Date & Flux Density & $\alpha_{\rm{QSO}}$ & HF & LF & HF & LF \\
             &  & [Jy] &  & [Jy] & [Jy] & [Jy] & [Jy] & [Hz] & [Hz] \\
            \hline
J2253+1608 & 2021-06-22 & 5.25 & 2021-06-22 & 1.91 & -0.78 & 2.32 & 2.61 & 47.6 & 73.6\\
J2258-2758 & 2021-06-22 & 2.27 & 2021-06-22 & 0.95 & -0.74 & 1.14 & 1.28 & 22.4 & 36.1\\
J0319+4130 & 2021-06-22 & 15.3 & 2021-06-22 & 5.72 & -0.75 & 6.91 & 7.76 & 137.5 & 214.7\\
J1256-0547 & 2021-08-27 & 19.1 & 2021-08-28 & 8.68 & -0.59 & 10.18 & 11.16 & 212.1 & 304.3\\
J1337-1257 & 2021-08-27 & 2.97 & 2021-08-28 & 1.39 & -0.64 & 1.65 & 1.82 & 31.9 & 47.3\\
J1751+0939 & 2021-08-27 & 1.99 & 2021-08-26 & 1.1 & -0.50 & 1.26 & 1.36 & 25.3 & 37.0\\
J1924-2914 & 2021-08-27 & 5.59 & 2021-08-28 & 2.51 & -0.61 & 2.95 & 3.24 & 61.9 & 92.4\\
J1058+0133 & 2021-10-13 & 3.58 & 2021-10-14 & 1.88 & -0.49 & 2.14 & 2.31 & 39.2 & 55.9\\
J0854+2006 & 2021-11-07 & 7.38 & 2021-11-07 & 3.93 & -0.48 & 4.47 & 4.81 & 86.6 & 117.4\\ 
J0522-3627 & 2022-04-19 & 10.1 & 2022-04-19 & 7.12 & -0.30 & 7.67 & 8.02 & 136.8 & 196.2\\
J0433+0521 & 2022-04-21 & 3.67 & 2022-04-21 & 1.77 & -0.61 & 2.09 & 2.29 & 36.0 & 54.0\\
J0210-5101 & 2022-06-30 & 1.42 & 2022-06-30 & 0.62 & -0.70 & 0.74 & 0.82 & 14.4 & 21.6\\
J0538-4405 & 2022-07-02 & 4.99 & 2022-07-01 & 2.84 & -0.47 & 3.22 & 3.47 & 62.1 & 85.6\\
J0423-0120 & 2022-08-31 & 5.95 & 2022-09-01 & 3.11 & -0.54 & 3.60 & 3.91 & 69.5 & 98.2\\
		\hline
	\end{tabular}
	\label{quasar_table}
\end{table*}

\section{Sensitivities}
\label{sec:sensitivity}

\subsection{Instrument noise power spectra}

\begin{table}
    \fontsize{8}{10}\selectfont
	\centering
	\caption{Information on the observations for noise test during commissioning observational campaigns. These scans are obtained with a cap on the cryostat. The corresponding noise power spectrum densities for the photometric mode are shown in Fig.~\ref{noise_psd_MPIOFF}.}
	\begin{tabular}{|c|c|c|}
 \hline
		  Scan name & MPI status & Observation type \\
		\hline
		  20210503\_1852\_S14779 & OFF & pspiral tracking R-Dor\\
		  20210503\_1846\_S14778 & OFF & otf tracking R-Dor\\
		  20210503\_1856\_S14782 & ON 30mm & otf tracking R-Dor\\
		  20210503\_1901\_S14783 & ON 30mm & pspirall tracking R-Dor\\
		  20210503\_1924\_S14792 & ON 30mm & skydip \\
		  20210503\_1907\_S14786 & ON 70mm & otf tracking R-Dor\\
		  20210503\_1911\_S14787 & ON 70mm & pspirall tracking R-Dor\\
		  20210503\_1917\_S14790 & ON 70mm & skydip \\
    \hline
	\end{tabular}
	\label{noise_table}
\end{table}

\begin{figure*}
   \centering
   \includegraphics[width=8.1cm]{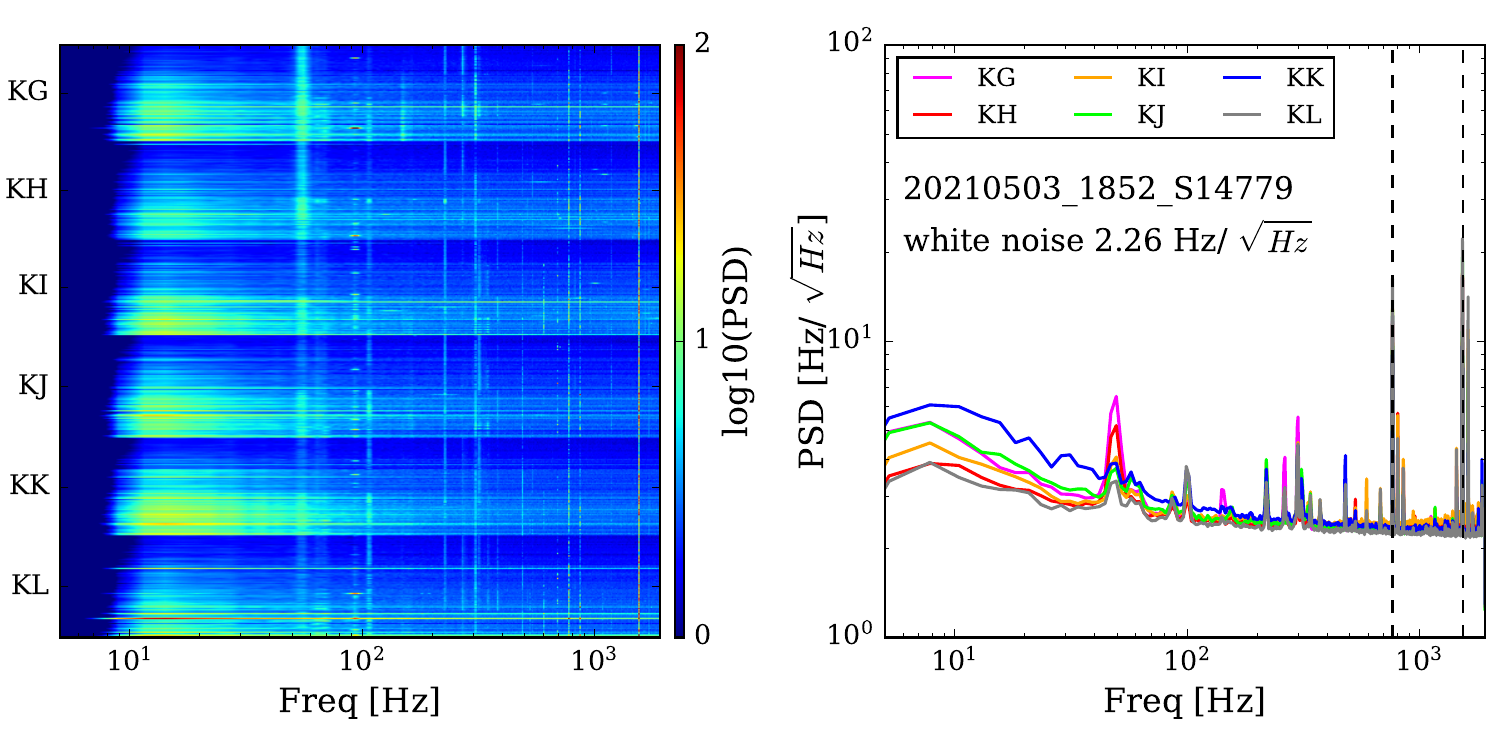}
   \includegraphics[width=8.1cm]{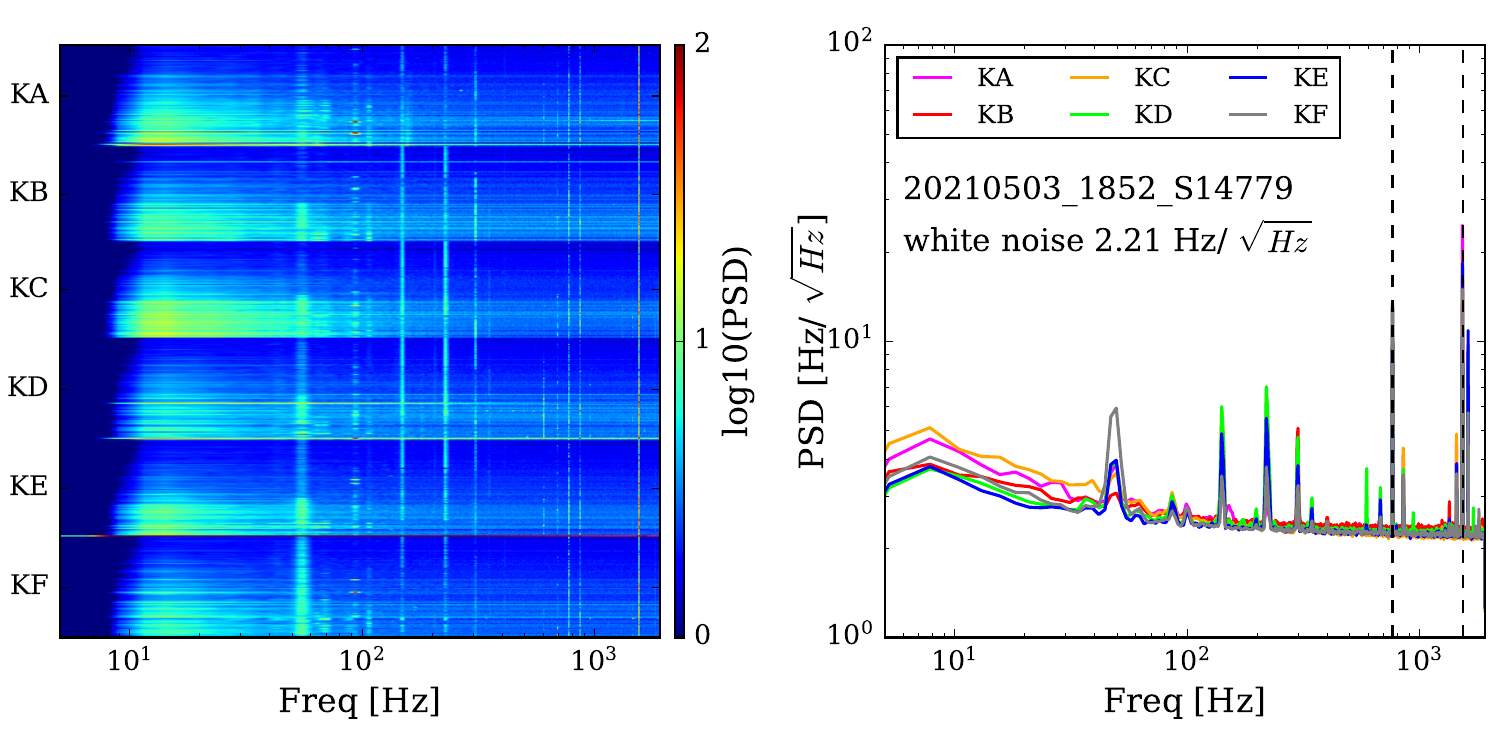}
   \includegraphics[width=8.1cm]{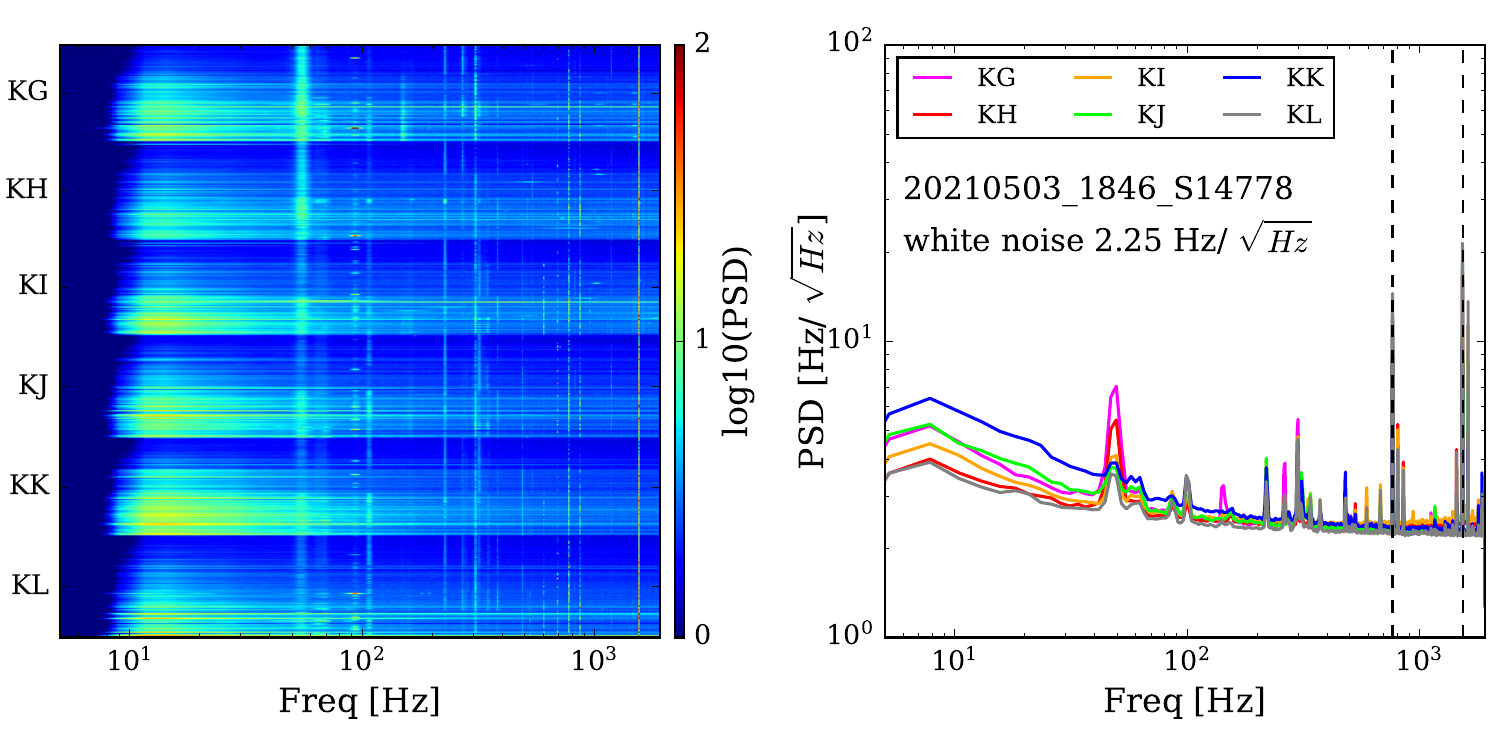}
   \includegraphics[width=8.1cm]{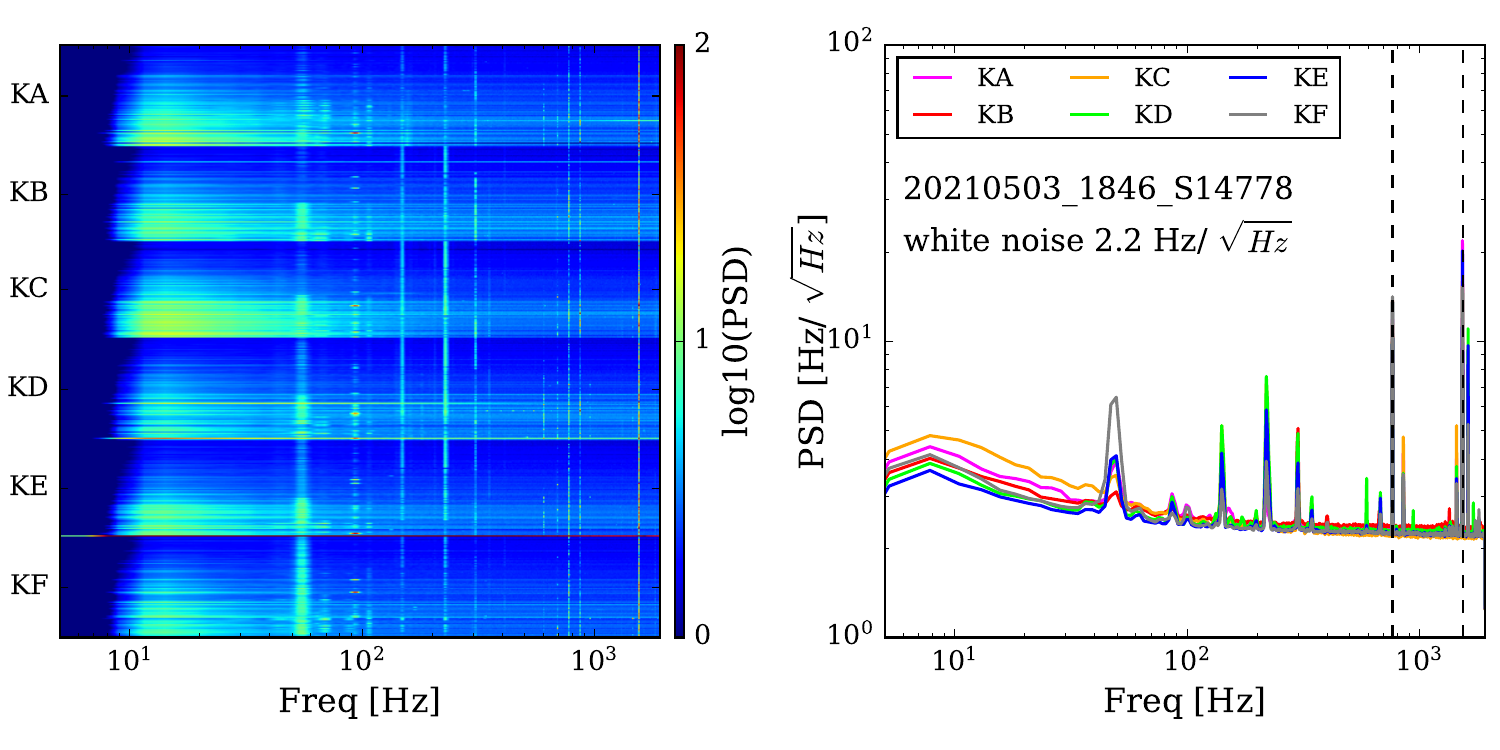}
   \caption{Noise power spectra of the CONCERTO data obtained in photometric mode (MPI OFF) with CAP ON. The odd and even columns show the noise power spectra densities for each LEKID and the median value of the corresponding LEKID array, respectively. A clear peak arises at $\sim$50\,Hz. The black-dashed lines represent the peaks observed at 763\,Hz and 1527\,Hz in the study by \citet{2022JInst..17P0047B}. All scans have similar noise power spectra and white noise of $\sim$ 2.2\,Hz $/\sqrt{\mathrm{Hz}}$. The corresponding observing conditions for the two scans shown here are given in Table.~\ref{noise_table}.}
   \label{noise_psd_MPIOFF}
\end{figure*}

In May 2021, during commissioning observations, we performed some scans under various instrumental conditions to assess the noise performance of CONCERTO by closing the cryostat optical window ("CAP ON"). We conducted test observations in both spectroscopic and photometric modes, including spiral and OTF scans and elevation series of measurements using the telescope that dips from a high to a low elevation (Table~\ref{noise_table}). We calculated the power spectrum densities of the calibrated time-ordered data. We present the noise power spectrum densities in Fig.~\ref{noise_psd_MPIOFF} for MPI OFF. Figures for MPI ON are completely similar. In the figure, the odd and even columns display the noise power spectrum densities for each LEKID, along with the median one calculated over time for the corresponding KIDs array. We obtained similar noise power spectrum densities for scans in different instrumental conditions (depicted in Table~\ref{noise_table}), indicating the stability of the noise across different observing strategies. At high frequency, we measured a white noise level of $\sim$2.2 Hz$/\sqrt{\mathrm{Hz}}$ for both bands.

The noise at low frequency (f$\leq$10\,Hz) is dominated by 1/f noise \citep{2002physics...4033M}, which arises from the electronics. Using a principal component analysis, it is possible to suppress the 1/f noise contamination down to the thermal noise level \citep{2015MNRAS.454.3240B,2020arXiv200701767L,2021MNRAS.508.2897H}. The 1/f noise can be suppressed in the CONCERTO pipeline by removing the common modes.
In addition to the standard 50-Hz noise (mains hum), we also see multiple peaks with frequencies exceeding 200\,Hz. Notably, the two highest peaks at 763\,Hz and 1527\,Hz (indicated by black-dashed lines in Fig.~\ref{noise_psd_MPIOFF}) were also observed in \citet{2022JInst..17P0047B}, where they tested and presented the architecture and performance of the CONCERTO readout and control electronics. These two lines are unexplained but are outside the frequency range of the scientific signal.


\subsection{Instrument noise from elevation series of flat-field measurements \label{inst_noise}}

\begin{figure*}[hbt!]
   \centering
   \includegraphics[width=8.5cm]{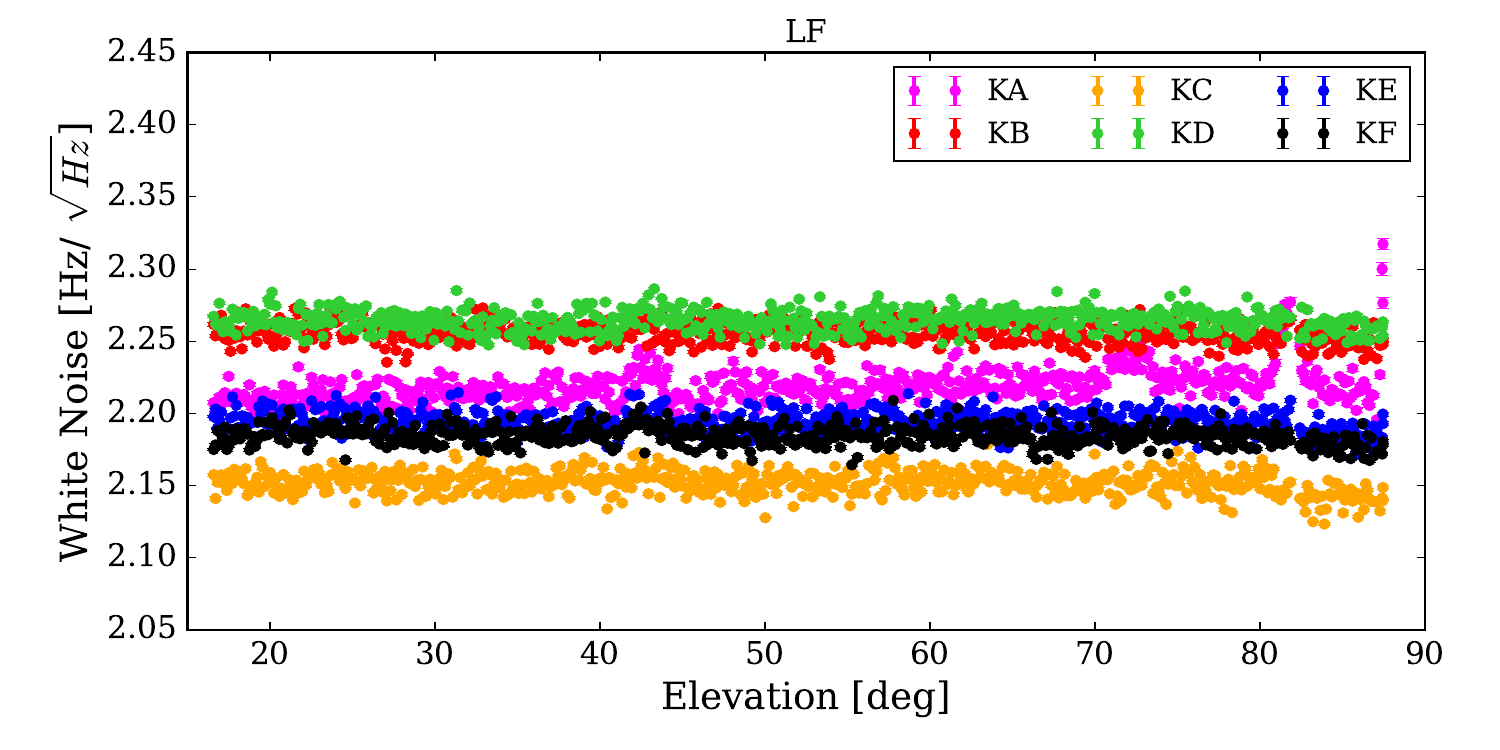}
   \includegraphics[width=8.5cm]{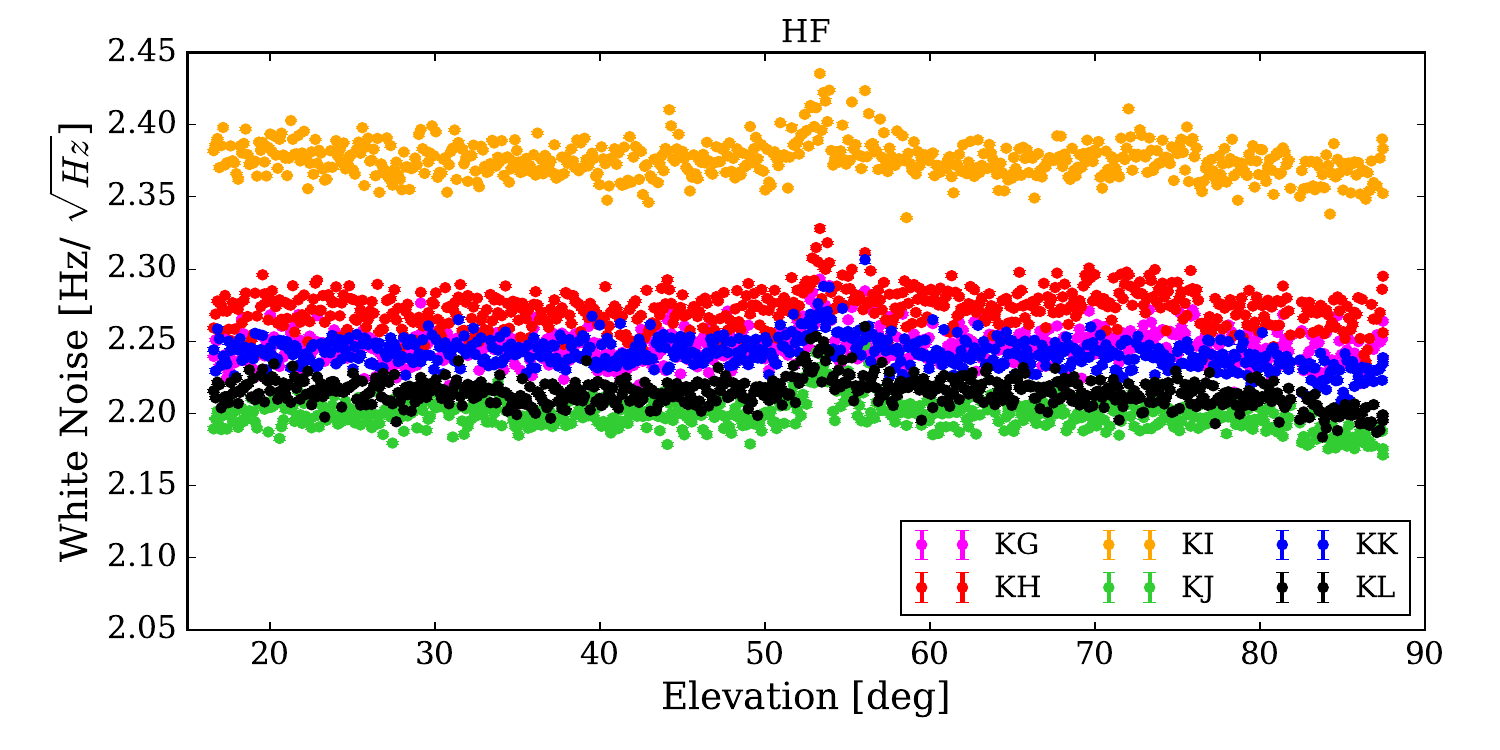}
   \caption{Measured white noise (median value of all LEKIDs inside each electronic box), from noise test observation in elevation series of measurements with the shutter closed, for LEKIDs array in LF (left panel) and HF (right panel). 
   }
   \label{noise_elevation}
\end{figure*}

To test the CONCERTO noise performance at different elevations, we conducted a test observation in spectroscopic mode, scanning across a range of elevations from 88\,deg to 16\,deg at a fixed azimuth with the telescope shutter closed. We measured the noise power spectral densities and presented the white noise level as a function of telescope elevation in Fig.~\ref{noise_elevation}. We observed a white noise level of 2.21$\pm$0.06 Hz/$\sqrt{\mathrm{Hz}}$ for LF and 2.24$\pm$0.04 Hz/$\sqrt{\mathrm{Hz}}$ for HF, which showed very little correlation with the elevation angle. The white noise level at high elevations is slightly lower than at low elevations. For LF and HF, the measured white noise level at 80\,deg elevation is on average 0.004 and 0.006\,Hz/$\sqrt{Hz}$ lower than that at 25\,deg elevation.

\subsection{Noise equivalent temperature (NET) and responsivity \label{sec:NET}}
By combining the white noise levels, the absolute photometric Jy-to-Hz calibration factors, the solid angles, and giving the $C_{\rm{K-to-MJy/sr}}$ conversion factors as given in Appendix\,\ref{uc_cc}, we can determine the NET per KID.
They are equal to:\\
2021: 1.4$\pm$0.1 and 1.4$\pm$0.1\,mK\,s$^{1/2}$\\ 
2022: 1.3$\pm$0.1 and 1.3$\pm$0.1\,mK\,s$^{1/2}$, \\ 
for LF and HF, respectively.
These values closely align with the benchmark value of 1.4\,mK\,s$^{1/2}$ (or 2\,mK$\rm{\sqrt{\mathrm{Hz}}}$) as discussed in \citet{2020A&A...642A..60C}.


Alternatively, we can derive the responsivity by dividing the white noise level by the NET. We obtain:\\
2021: 1.1$\pm$0.1 and 1.1$\pm$0.1\,kHz/K \\
2022: 1.2$\pm$0.1 and 1.3$\pm$0.1 \,kHz/K  \\
for LF and HF, respectively. These numbers give the responsivity to point sources.

Such responsivity can be checked using dedicated measurements. We conducted skydip scans (from 18 to 80 degrees) in MPI ON mode with the external cold black body as a reference (appropriate for diffuse emission studies). The atmospheric signal is modelled with its brightness temperature [K] and corrected for emissivity. The latter is calculated by integrating the atmospheric opacity over the CONCERTO bandpasses.  By comparing the input (atmospheric) with the measured signals, we obtained 1.3$\pm$0.3\,kHz/K and 1.4$\pm$0.3\,kHz/K for the LF and HF bands, respectively. We obtain the same values for two skydips with pwv=1.4 and 1.5\,mm. These numbers give the responsivity on diffuse emission. While they have to be taken with caution as i) in the photon-limited case, the noise is related to the background, which is significantly varying for skydips, and ii) non-linearities can affect skydip measurements), their agreement with the responsivity on point source is remarkable. 

\subsection{Sensitivity from the noise equivalent temperature}
\label{sec:NEFD}
We can use the NET measurements (Sect.\,\ref{sec:NET}) to estimate a first sensitivity (that we will then compare with sky observations). The NEFD (in $\mathrm{mJy/beam\cdot s^{1/2}}$) can be computed following:
\begin{equation}
\label{eq:nefd}
\centering
\mathrm{NEFD_{0}} = \mathrm{NET} \times C_{\rm{K-to-MJy/sr}} \times 10^6 \times \Omega \, , 
\end{equation}
where $\mathrm C_{\rm{K-to-MJy/sr}}$ is the conversion factor from K$_\mathrm{RJ}$ to MJy/sr (see Appendix\,\ref{uc_cc}), and $\Omega$ is the solid angle of the beam as given in Table\,\ref{tab:beam_fit}.
This is the NEFD per KID at zero opacity, i.e. NEFD$_{0}$. \\
From the above numbers and equations, we derive NEFD$_0$ values of 61$\pm$8 mJy/beam$\cdot$s$^{1/2}$ for LF and 81$\pm$9\,mJy/beam$\cdot$s$^{1/2}$ for HF.

In real observation conditions, characterised with a given atmospheric opacity $\tau$ and a given air mass $x$, correcting the flux density for atmospheric attenuation using Eq.\,\ref{opacity_correction} increases the NEFD$_0$ to NEFD= $\rm NEFD_0 \times e^{(\tau_{eff} x)}$. In addition, the NEFD measured on the sky could only be higher than those derived from NET because of additional sky noise.

\subsection{Continuum map on a faint sub-millimetre galaxy in the UDS field}
\label{sec:as2uds}

We have previously quantified the photometric performance of CONCERTO on bright sources with Jy-level flux. To assess its sensitivity on faint sources, we conducted several scans on the UDS field centred on the AS2UDS0001.0 source. We built continuum maps and compared the measured flux with the predicted values. We also demonstrate how the data's root mean square (RMS) uncertainty evolves with integration time.\\

The AS2UDS is a survey of sub-millimetre galaxies (SMGs) selected from the SCUBA2 Cosmology Legacy Survey map of the UKIDsS/UDS field (S2CLS, described in \citet{2017MNRAS.465.1789G}), and further observed at higher angular resolution with ALMA. 
The SCUBA2 survey covers a field of approximately 0.9 deg$^{2}$. We designed a small CONCERTO observation centred on the brightest source, AS2UDS0001.0 (as designated by S2CLS, \citet{2017MNRAS.465.1789G}) at R.A.=2$^{\rm h}$18'30.7" and DEC=5$^{\rm o}$31'31.62". This source has a SCUBA2 flux of 52.7$\pm$0.9\,mJy at 850\,$\mu$m.  We identify this SCUBA2 source with a lensed galaxy at z=3.390 \citep{2013ApJ...762...59W}. Herschel/SPIRE fluxes are 92$\pm$7\,mJy at 250\,$\mu$m, 122$\pm$8\,mJy at 350\,$\mu$m, and 113$\pm$7\,mJy at 500\,$\mu$m. We fit the flux measurements using a modified black body, MBB \citep{2012MNRAS.425.3094C} with $\beta = 1.7$ and obtained T$_{\rm dust}$/(1+z)=7.5\,K. We also fit for the amplitude of the SED template at z=3.4 from the infrared galaxy evolution model of \cite{2015A&A...573A.113B}. Extrapolation to CONCERTO effective frequencies using either the MBB or \cite{2015A&A...573A.113B} model gives different fluxes, ranging from 11 to 17\,mJy for LF and 17 to 26\,mJy for HF. \\

\begin{figure}
   \centering
   \includegraphics[width=7.5cm]{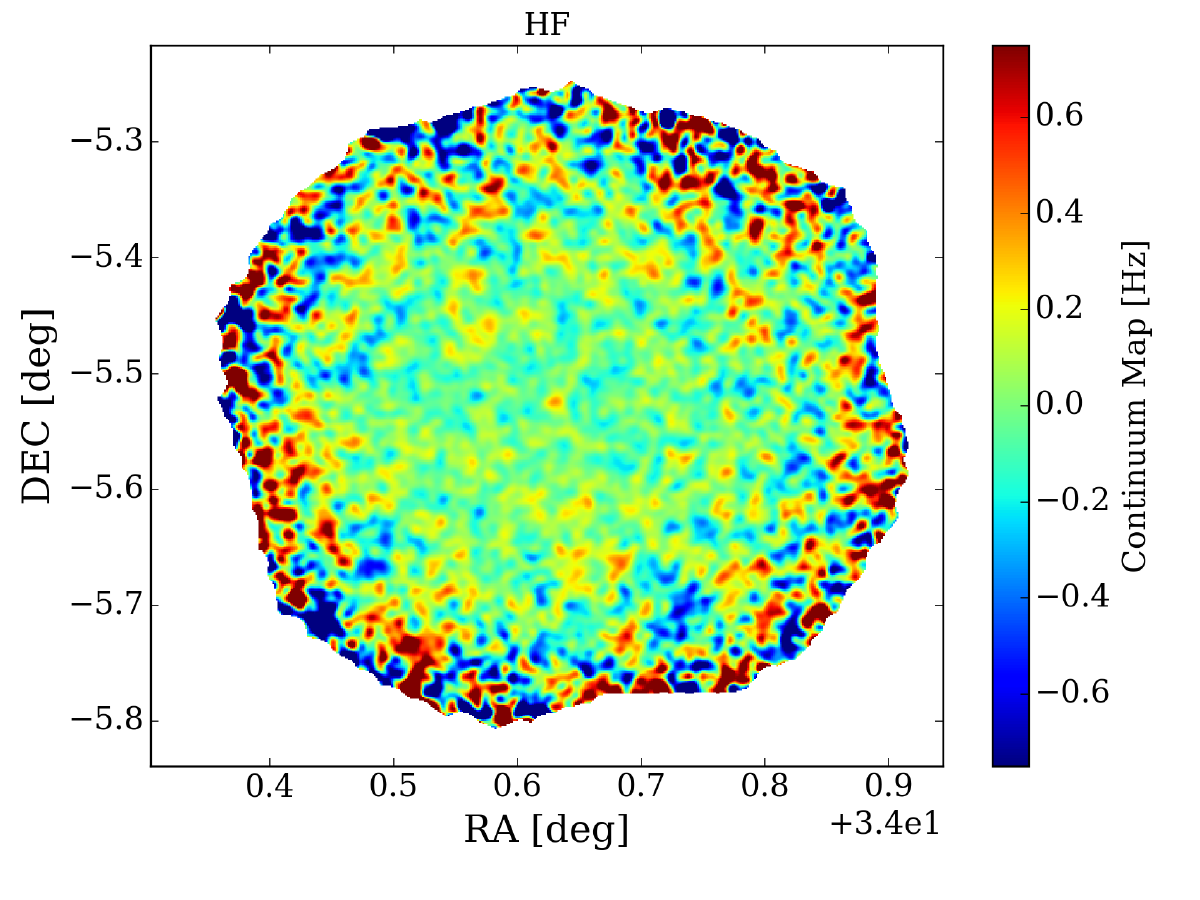}
   \includegraphics[width=7.5cm]{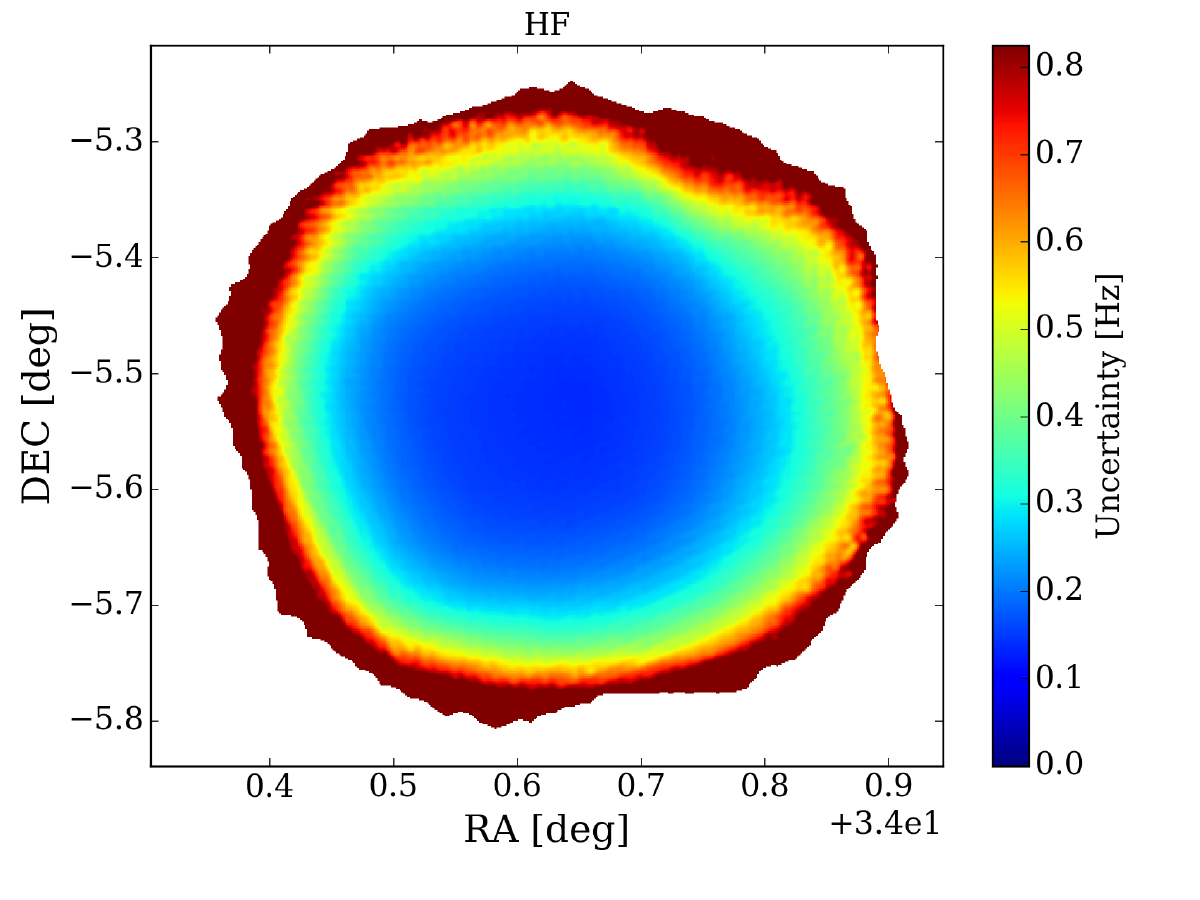}
  \caption{CONCERTO continuum map centred at AS2UDS0001.0 (upper panel) and corresponding uncertainty map (lower panel) for HF.}
   \label{AS2UDS_map}
\end{figure}

The CONCERTO observations of the AS2UDS field were conducted during the early commissioning observation campaign in June 2021. They consisted of 13 observation scans, including 10 rectangular OTF scans and 3 spiral OTF scans, resulting in a total observation time of approximately 4.35\,hours. The observed area was centred on source AS2UDS0001.0 with a radius of approximately 0.25\,degrees. Of the 13 scans, 10 are used to produce the maps (the rest was taken while the cryostat was not at its nominal position). By stacking each continuum map with inverse noise as weight, we produced combined continuum maps of the field, as presented in Fig.~\ref{AS2UDS_map} for HF. The uncertainty map is also shown. The central source, AS2UDS0001.0, is detected with a signal-to-noise ratio of 3.0 and 3.5 for LF and HF bands, respectively. The fluxes measured for LF and HF are 18.9$\pm$6.4\,mJy and 17.6$\pm$5.2\,mJy, respectively (where we estimated the error bars through jackknife resampling). The measured fluxes are consistent with expectations.\\

\begin{figure}[hbt!]
   \centering
   \includegraphics[width=8.1cm]{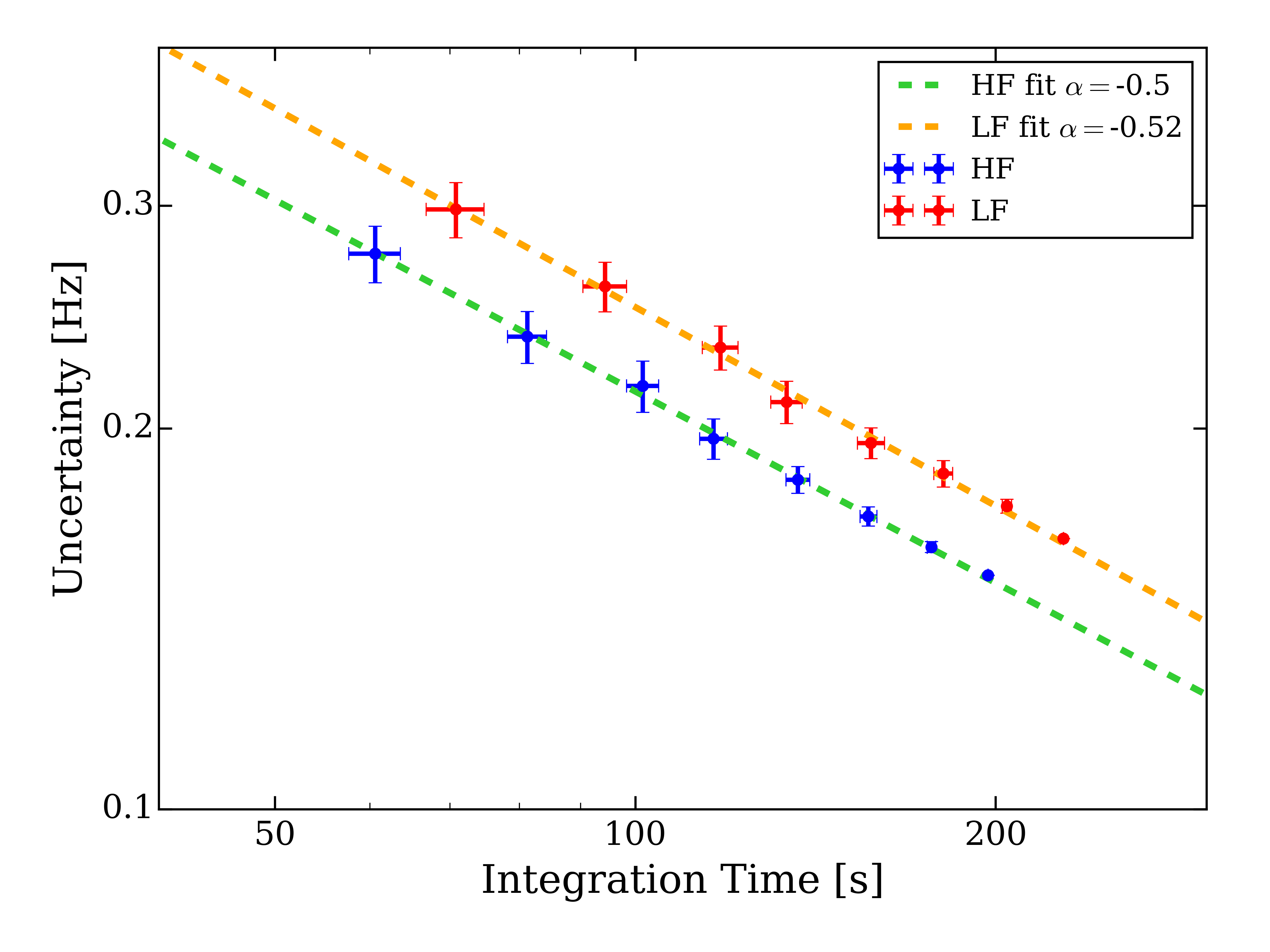}
   \caption{RMS uncertainty of the AS2UDS beam-smoothed continuum map as a function of integration time on pixels of 5-arcsec size (measured on the central 0.1\,deg radius area). As expected, the uncertainty becomes lower with the longer integration time on a pixel. The power-law fitting, shown as dashed lines, gives a slope of -0.52 and -0.50 for LF and HF, respectively.}
   \label{AS2UDS_hits_uncert}
\end{figure}

By measuring the relation between integration time and RMS noise of the AS2UDS map, we can estimate the NEFD. The RMS evolves as follows: 
\begin{eqnarray}
\rm{RMS}(\rm{t_{int}}) = NEFD/\sqrt{\rm{t_{int}}},
    \label{rms_evolve}
\end{eqnarray}
\noindent where $\rm{t_{int}}$ is the effective integration time. We varied the integration time by combining different numbers of AS2UDS scans and then evaluated the flux density uncertainties of each pixel in the combined beam-smoothed continuum map. Figure~\ref{AS2UDS_hits_uncert} displays the uncertainty as a function of integration time, measured for 5-arcsec size pixels in the maps. As anticipated, the uncertainty decreases with a longer integration time. We fitted a power-law function ($y(x) \propto x^{\alpha}$) to the data points, yielding a slope of -0.52$\pm$0.02 and -0.50 $\pm$0.01 for LF and HF, respectively. The measured slopes are consistent with the expected behaviour of $\rm{RMS}(\rm{t_{int}}) \propto 1/\sqrt{\rm{t_{int}}}$, indicating that the CONCERTO observations performed as expected in terms of overall noise. We determined the NEFDs for CONCERTO LF and HF to be 2.83$\pm$0.17\,Hz$\cdot$s$^{1/2}$ and 2.32$\pm$0.11\,Hz$\cdot$s$^{1/2}$, corresponding to 111$\pm$7\,mJy/beam$\cdot$s$^{1/2}$ and 119$\pm$5\,mJy/beam$\cdot$s$^{1/2}$.

The pwv distribution for AS2UDS observations is shown in Fig.\,\ref{AS2UDS_COSMOS_pwv_hist} and leads to pwv=0.75$\pm$0.10\,mm. This translates to $\tau_{\rm{LF}}$=0.08296$\pm$0.00738 and $\tau_{\rm HF}$=0.0688$\pm$0.00648. With a mean elevation $\delta$=62.5$\pm$12.4 degrees, the NEFDs translate to NEFD$_{\rm{0}}$ of 100$\pm$10 and 110$\pm$7\,mJy/beam$\cdot$s$^{1/2}$ for LF and HF, respectively.

\subsection{Sensitivity from the continuum map of the COSMOS field \label{NEFD_COSMOS}}

The primary scientific objective of the CONCERTO project is to map the [CII] and CO emissions in the COSMOS field using intensity mapping. In this section, we present the continuum map of the COSMOS field and evaluate the noise performance of CONCERTO based on hundreds of hours of data.

The COSMOS field, centred at (RA, DEC, J2000)= (150.12, 2.21)\,degrees, covers an area of about 2 square degrees \citep{2007ApJS..172....1S,2022ApJS..258...11W}. This field is particularly advantageous for scientific studies due to its low background levels, absence of bright sources, and visibility from both hemispheres from the ground. As a result, the COSMOS field has been extensively studied across all accessible wavelengths, from X-rays to radio bands \citep[e.g.][]{2004AJ....128.1974S,2007ApJS..172..468Z, 2007ApJS..172..196K,2007ApJS..172...86S,2019A&A...632A..88A}.

\begin{figure}[hbt!]
   \centering
   \includegraphics[width=9.0cm]{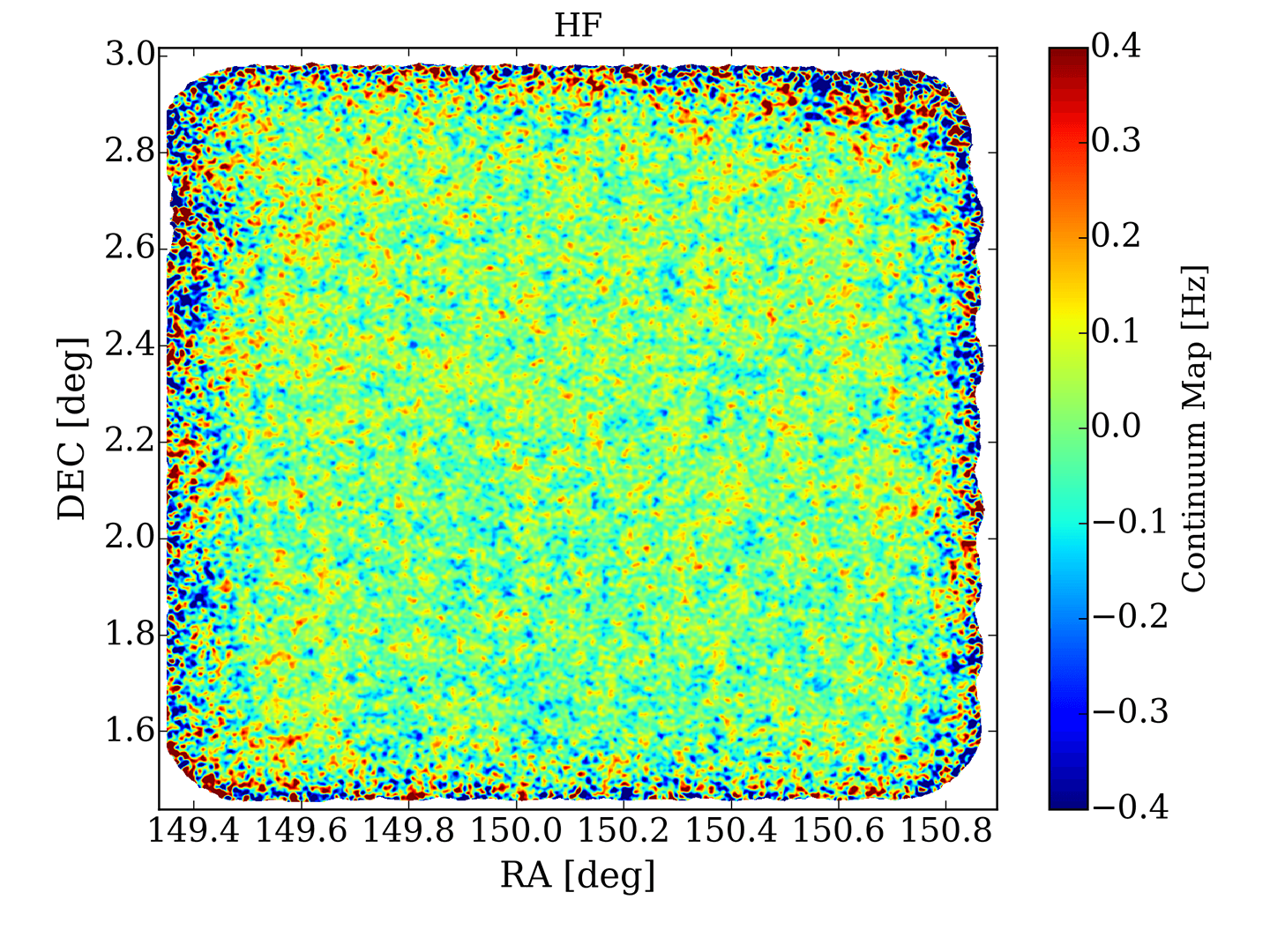}
   \includegraphics[width=9.0cm]{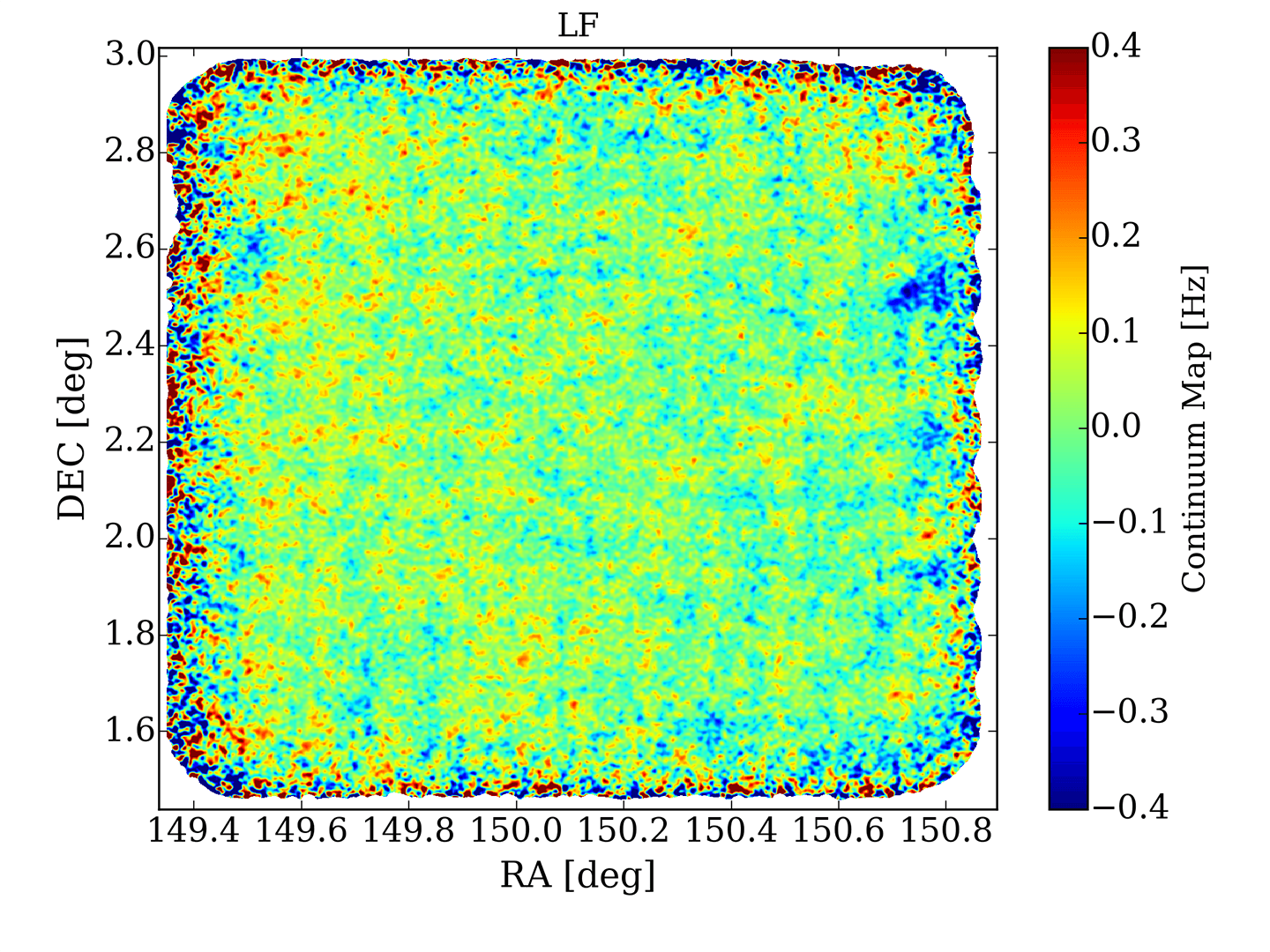}
   \caption{CONCERTO continuum maps of COSMOS field for HF (top) and LF (bottom). The maps have been smoothed by a Gaussian kernel with a FWHM of 30 arcsec.}
   \label{COSMOS_map}
\end{figure}

The CONCERTO project conducted observations of the COSMOS field from July 2021 to December 2022, accumulating a total of 793\,hours of observations, corresponding to 650\,hours on-field. The observations involved rectangular and spiral on-the-fly scans in a raster scanning pattern, totalling 1522 scans with MPI ON. For the analysis presented here, we excluded 72 of the 1522 scans due to poor quality (we removed the scans for which $>$80\% of the KIDs were masked). We processed these spectroscopic data using the continuum pipeline to make continuum maps as shown in Fig.\,\ref{COSMOS_map}.

\begin{figure}[hbt!]
   \centering
   \includegraphics[width=8.1cm]{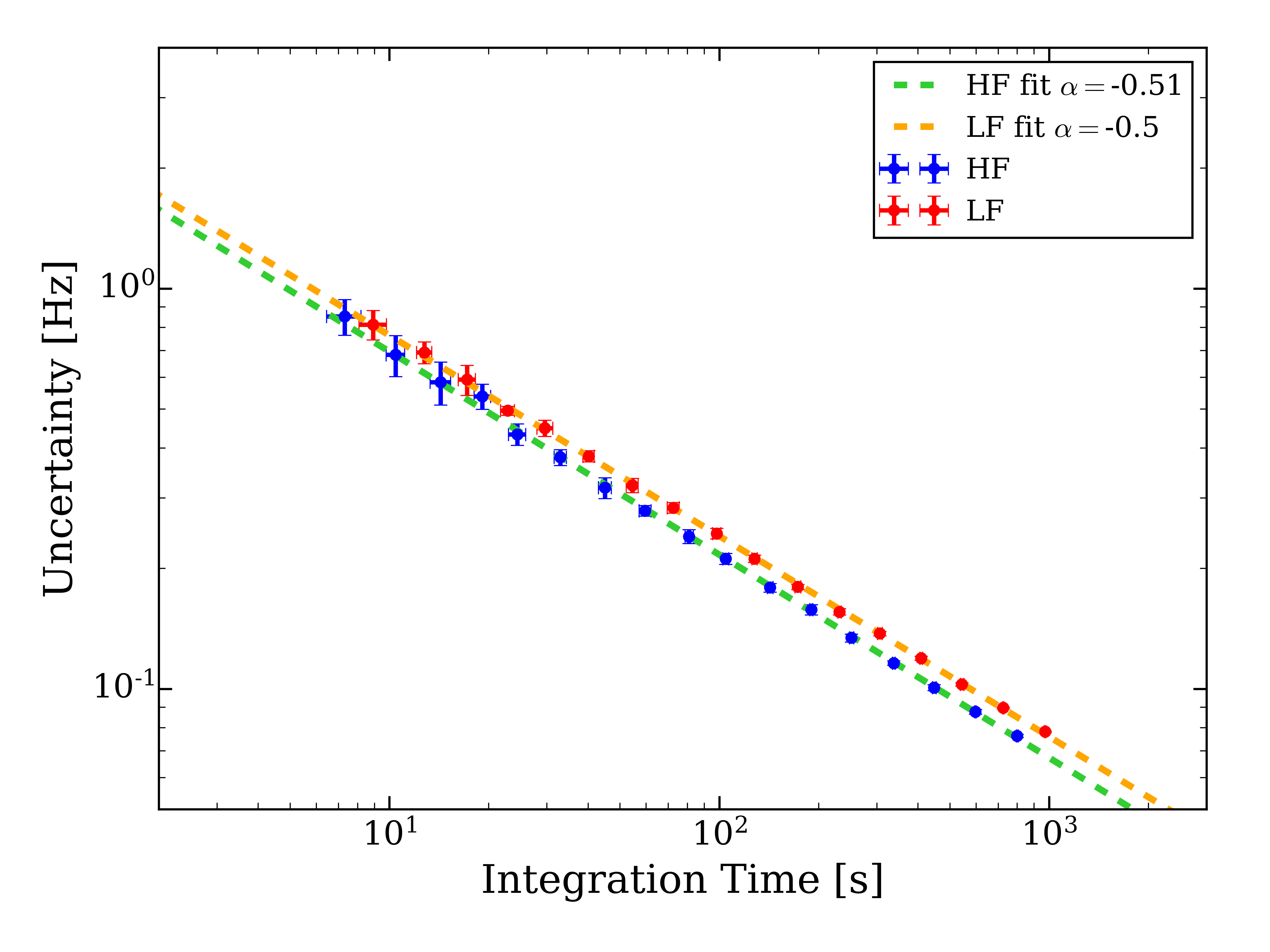}
   \caption{Same as Figure.~\ref{AS2UDS_hits_uncert}, but for COSMOS observations in 2021 and 2022 and measured on the central 0.3\,deg radius area. The power-law fitting, shown as dashed lines, gives a slope of -0.501$\pm$0.002 and -0.506$\pm$0.004 for LF and HF, respectively.}   
   \label{COSMOS_hits_uncert}
\end{figure}

We also measured the RMS uncertainty of the beam-smoothed continuum map as a function of integration time, measured for 5-arcsec size pixels and shown in Fig.~\ref{COSMOS_hits_uncert}. The results from the COSMOS observations indicate that the RMS uncertainty of CONCERTO evolves with integration time as expected, with a slope of $\alpha=-0.501\pm0.002$ and $\alpha=-0.506\pm0.004$ for LF and HF, respectively, as observed in the much shorter AS2UDS observations. The measured NEFDs for CONCERTO using COSMOS data are 2.42$\pm$0.03\,Hz$\cdot$s$^{1/2}$ and 2.24$\pm$0.03\,Hz$\cdot$s$^{1/2}$, corresponding to 95$\pm$1\,mJy/beam$\cdot$s$^{1/2}$ and 115$\pm$2\,mJy/beam$\cdot$s$^{1/2}$ for LF and HF, respectively.\\

\begin{figure}[hbt!]
   \centering
   \includegraphics[width=8.1cm]{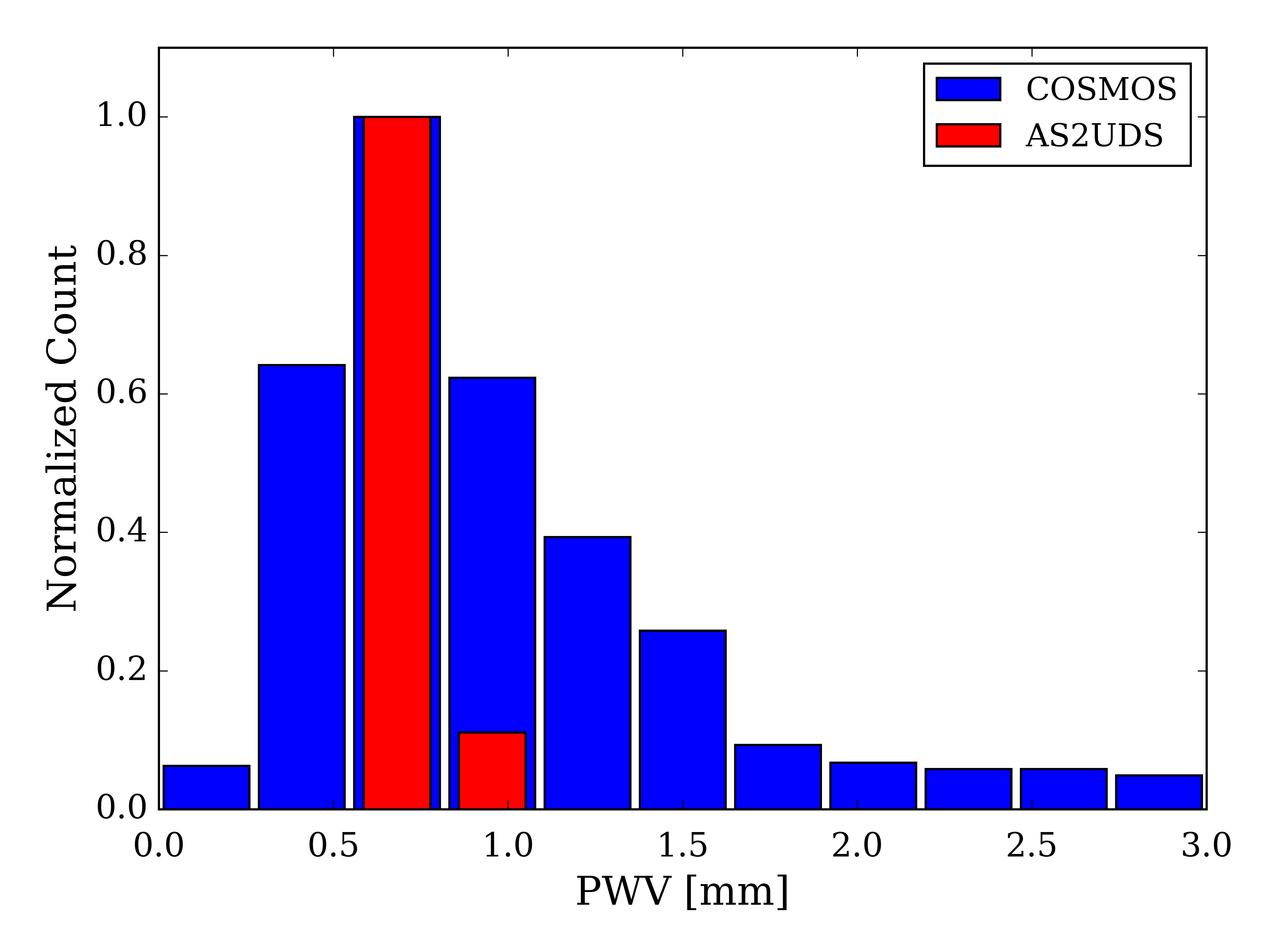}
   \caption{Precipitable water vapour distribution of the utilized AS2UDS and COSMOS observation scans. The histograms have been normalized, with the red and blue columns representing AS2UDS and COSMOS, respectively.}
   \label{AS2UDS_COSMOS_pwv_hist}
\end{figure}

The pwv distribution for COSMOS observations is shown in Fig.\,\ref{AS2UDS_COSMOS_pwv_hist} and leads to  pwv=0.81$\pm$0.63. 
With a mean elevation $\delta$= 55.65$\pm$10.83 degrees, the NEFDs translate to NEFD$_{0}$ of 85$\pm$8 and 105$\pm$10\,mJy/beam$\cdot$s$^{1/2}$ for LF and HF, respectively. They are in very good agreement with those derived from the AS2UDF field (Sect.\ref{sec:as2uds}). Notice that the error bars on these NEFD$_{0}$ take into account the range of pwv and elevation.\\

The on-sky NEFD estimations are 1.4 times higher than those derived from the NET in Sect.\,\ref{sec:NEFD}. This is not surprising as the NEFD of Sect.\,\ref{sec:NEFD} does not comprise any contribution from the sky noise. 

\subsection{Comparison with expectation \label{sect_NEFD_expectation}}

Due to the similarities between the NIKA2 and CONCERTO
detectors, \citet{2020A&A...642A..60C} predicted the CONCERTO NEFD$^{\rm{LF}}$ and NEFD$^{\rm{HF}}$ as if it was a dual-band photometer, based on the measured NIKA2 sensitivity \citep{2020A&A...637A..71P}. By strictly following the method described in \citet{2020A&A...642A..60C}, we calculate the expected sensitivity of CONCERTO as a dual-band photometer (for pwv=2 and an elevation of 60 degrees) at the effective frequencies of Uranus to be: 
NEFD$^{\rm{LF}}$=106[72-145]\,mJy$\cdot$s$^{1/2}$ and NEFD$^{\rm{HF}}$=121[81-162]\,mJy$\cdot$s$^{1/2}$. 
The numbers in the bracket give the ranges of prediction given the ranges of considered NEFD$^{\rm{LF,HF}}_{\rm{NIKA2}}$ as given in \cite{2020A&A...642A..60C}. This translates to NEFD$^{\rm{LF}}_0$=85\,[57-116]\,mJy$\cdot$s$^{1/2}$ and NEFD$^{\rm{HF}}_0$=100\,[67-137]\,mJy$\cdot$s$^{1/2}$. This compares very well to the NEFD$_{0}$ of 85$\pm$8 and 105$\pm$10\,mJy/beam$\cdot$s$^{1/2}$ for LF and HF, measured in COSMOS.
 
\section{Conclusions and Summary}
\label{sec:conclusion}

\begin{table*}
	\centering
	\caption{Summary of CONCERTO main characteristics.}
	\begin{threeparttable}
	\begin{tabular}{|c|c|c|c|}
		\hline
		  & LF Array & HF Array & Section\\
		\hline
		  Frequency range\tnote{a} \,[GHz] & 130-270 &  195-310 & \\
  		  Effective frequency for Uranus [GHz] & 225 & 262 & Sect.\,~\ref{sec:cali_uranus}\\
      	\hline
		  Number of designed detectors & 2\,152 & 2\,152 & Sect.\,\ref{sec:instrument}\\
		  Fraction of valid detectors\tnote{b}\,\, [$\%$] & 78 & 73 & Sect.\,\ref{sec:dataprocessing}\\
		\hline
		  FWHM$_{1}$\tnote{c}\,\, [arcsec] & 29.6$\pm$0.1 &  27.1$\pm$0.1 & Sect.\,\ref{radial_beam}\\
        FWHM$_{2}$\tnote{d}\,\, [arcsec] & 38.1$\pm$5.1 & 35.9$\pm$6.1 & \\
       Solid Angle $\Omega$\tnote{e}\,\, [$10^{-8}$ sr] & 3.87$\pm$0.07 & 3.44$\pm$0.10 & Sect.\,\ref{radial_beam}\\
		\hline 
		  Jy/beam-to-Hz conversion factor\tnote{f}\,\, [Hz/Jy] & 25.6 & 19.5 & Sect.~\ref{sec:calibration}\\
		  Absolute calibration uncertainty\tnote{g}\,\, [$\%$] & 3.4 & 3.0 & Sect.~\ref{sec:calibration}\\
		\hline 
		  White noise level [Hz/$\sqrt{\mathrm{Hz}}$] & 2.21 & 2.24 & Sect.\,\ref{inst_noise}\\
		  $\alpha$ noise vs. integration time & -0.50 & -0.51 & Sect.\,\ref{NEFD_COSMOS}\\
		\hline 
        NET\,\,[mK\,s$^{1/2}$]  & 1.3$\pm$0.1 &  1.3$\pm$0.1  & Sect.\,\ref{sec:NET}\\
		  NEFD\tnote{h}\,\, [mJy/beam$\cdot$s$^{1/2}$]  & 95$\pm$1 & 115$\pm$2 & Sect.\,\ref{NEFD_COSMOS}\\
		\hline 
	\end{tabular}
	\begin{tablenotes}
    \item[a] From \cite{2020A&A...642A..60C}.
    \item[b] Valid detectors are the LEKIDs that met the selection criteria in over 70$\%$ of the 44 Uranus photometric scans.
    \item[c] FWHM of the central Gaussian beam (2022).
    \item[d] FWHM of the Gaussian whose solid angle is equivalent to that of the beam ($\Omega$)
    \item[e] Solid angle derived from the radial profile (2022).
    \item[f] Given for a Uranus spectral energy distribution.
    \item[g] Not taking into account the model uncertainty (which is estimated to be $<2\%$, \cite{Planck2016}).
    \item[h] From COSMOS observations at a mean pwv of 0.81\,mm and elevation of 56\,degrees.
    \end{tablenotes}
	\label{tab:performance_table}
	\end{threeparttable}
\end{table*}

This paper describes the performance assessment in the continuum of CONCERTO at the APEX 12-m telescope using commissioning and scientific observations from April 2021 to December 2022. 
To evaluate its performance, data from bright calibrators (Uranus, Mars, and quasars) and faint sources (AS2UDS and COSMOS) were used, covering the full range of elevations and atmospheric conditions encountered on site. Table.~\ref{tab:performance_table} gives a summary of CONCERTO's main characteristics. The main conclusions are given below.

\begin{enumerate}
      \item We have developed a photometric data processing pipeline that includes data reading and raw-data calibration, bad LEKIDs masking, flat-field normalisation, opacity correction, correlated noise subtraction, and map projection. 
      \item From beam map observations, we derived various parameters of the LEKIDs, their relative position on the focal plane, their relative flat-field gain as well as their individual eccentricity. LEKIDs impacted by the cross-talk effect were identified and flagged. The mean eccentricity of the main beam is 0.45 and 0.47 for LF and HF respectively. The two arrays are suffering from an optical aberration in their lower left corner, increasing the eccentricity to values higher than 0.75. Between 54 and 67\% of the constructed LEKIDs are usable reliably after removing those with cross-talk effect, high eccentricity, or any other problems. The effective beam shows a main beam close to the diffraction-limited values, as well as two large error beams, with a main beam efficiency of about 0.53.
      \item We evaluated the photometric uncertainties using pointing scans of Uranus. For the data processing of these scans, LEKIDs are masked based on their recorded position, beam FWHM, amplitude, and eccentricity information, with only the most stable LEKIDs retained to yield good signal-to-noise continuum measurements. 78.2$\%$ of the LEKIDs in the LF array and 72.5$\%$ in the HF array are recognized as valid detectors, each with a probability $>$70$\%$. We found the uncertainties for the absolute photometric calibration of point sources are 3.4$\%$ and 3.0$\%$, for the LF and HF bands, respectively. These values represent a state-of-the-art performance for a ground-based millimetre-wave instrument. Additionally, we verified the stability of the measured photometric flux against variations in observation date, beam size, pwv, and temperature.
      \item The absolute photometric calibration factor is derived by comparing the measured signal and the ESA2 model for Uranus. The factor is 25.6$\pm$0.9\,Hz/[Jy/beam] and 19.5$\pm$0.6\,Hz/[Jy/beam] for LF and HF. The model uncertainty is estimated to be $<2\%$ \citep{Planck2016}. After applying the photometric calibration, we obtained Mars measured-to-predicted flux density ratios of 1.00$\pm$0.06 and 0.97$\pm$0.06 for the LF and HF bands, respectively. Absolute photometric calibration using secondary calibrators (quasars) produces remarkably consistent results.
      \item We quantified the noise performance of CONCERTO using test observations obtained in photometric and spectroscopic modes. The resulting noise power spectrum densities were similar for all scans. White noise of approximately 2.2\,Hz/$\sqrt{\mathrm{Hz}}$ was measured for CONCERTO, with little correlation with the elevation angle. The measured white noise at 80\,degrees elevation was, on average, 0.0083 and 0.0003\,Hz/$\sqrt{\mathrm{Hz}}$ lower for LF and HF, respectively, than that at 25\,degrees elevation. We verified the ability to detect faint sources by integrating a small region in the UDS field centred on the sub-millimetre galaxy AS2UDS0001.0. We measure flux densities consistent with extrapolation from previous measurements obtained at shorter wavelengths than CONCERTO. The maps of the AS2UDS and COSMOS field were stacked, and it was found that the noise integrated as the inverse of the square root of the integration time.
      \item Utilizing COSMOS data, the NEFDs for CONCERTO LF and HF were measured to be 2.42$\pm$0.03\,Hz$\cdot$s$^{1/2}$ and 2.24$\pm$0.03\,Hz$\cdot$s$^{1/2}$, equating to 95$\pm$1\,mJy/beam$\cdot$s$^{1/2}$ and 115$\pm$2\,mJy/beam$\cdot$s$^{1/2}$ for LF and HF, respectively, for a mean pwv of 0.81\,mm and a mean elevation of 56\,degrees. They are in excellent agreement with expectations from \cite{2020A&A...642A..60C}.
\end{enumerate}

We conclude that CONCERTO had unique capabilities for fast dual-band mapping at tens of arcsecond resolution. A detailed assessment of CONCERTO's spectroscopic performance will be presented very shortly.

\begin{acknowledgements}
      Besides the authors, the technicians and engineers more involved in the experimental setup development have been Maurice Grollier, Olivier Exshaw, Anne Gerardin, Gilles Pont, Guillaume Donnier-Valentin, Philippe Jeantet, Mathilde Heigeas, Christophe Vescovi, and Marc Marton. We acknowledge the crucial contributions of the whole Cryogenics and Electronics groups at Institut Néel and LPSC. We acknowledge the contribution of Hamdi Mani, Chris Groppi, and Philip Mauskopf (from the School of Earth and Space Exploration and Department of Physics, Arizona State University) to the cold electronics. The KID arrays of CONCERTO have been produced at the PTA Grenoble microfabrication facility. We warmly thank the support from the APEX staff for their help in CONCERTO pre-installations and design. The flexible pipes, in particular, have been routed under the competent coordination of Jorge Santana and Marcelo Navarro. We acknowledge support from the European Research Council (ERC) under the European Union’s Horizon 2020 research and innovation programme (project CONCERTO, grant agreement No 788212), from the Excellence Initiative of Aix-Marseille University-A*Midex, a French "Investissements d’Avenir" programme, from the LabEx FOCUS ANR-11-LABX-0013, and from the ECOS-ANID French and Chilian cooperation program.  DQ acknowledges support from the National Agency for Research and Development (ANID)/Scholarship Program/Doctorado Nacional/2021-21212222 and ECOS-ANID N$^{\rm{o}}$ECOS220016. This work has also been supported by the GIS KIDs. We are grateful to our administrative staff in Grenoble and Marseille, in particular Patricia Poirier, Mathilde Berard, Lilia Todorov and Valérie Favre, and the Protisvalor team. We acknowledge the crucial help of the Institut Néel and MCBT Heads (Etienne Bustarret, Klaus Hasselbach, Thierry Fournier, Laurence Magaud) during the COVID-19 restriction period. MA acknowledges support from FONDECYTgrant 1211951 and ANID BASAL project FB210003.
\end{acknowledgements}

\bibliographystyle{mnras}
\bibliography{concerto.bib}
\begin{appendix}
\section{CONCERTO Focal Plane Deformations}
\label{appendix:Fov_Deformation}

\begin{figure*}
\centering
  \includegraphics[width=17cm]{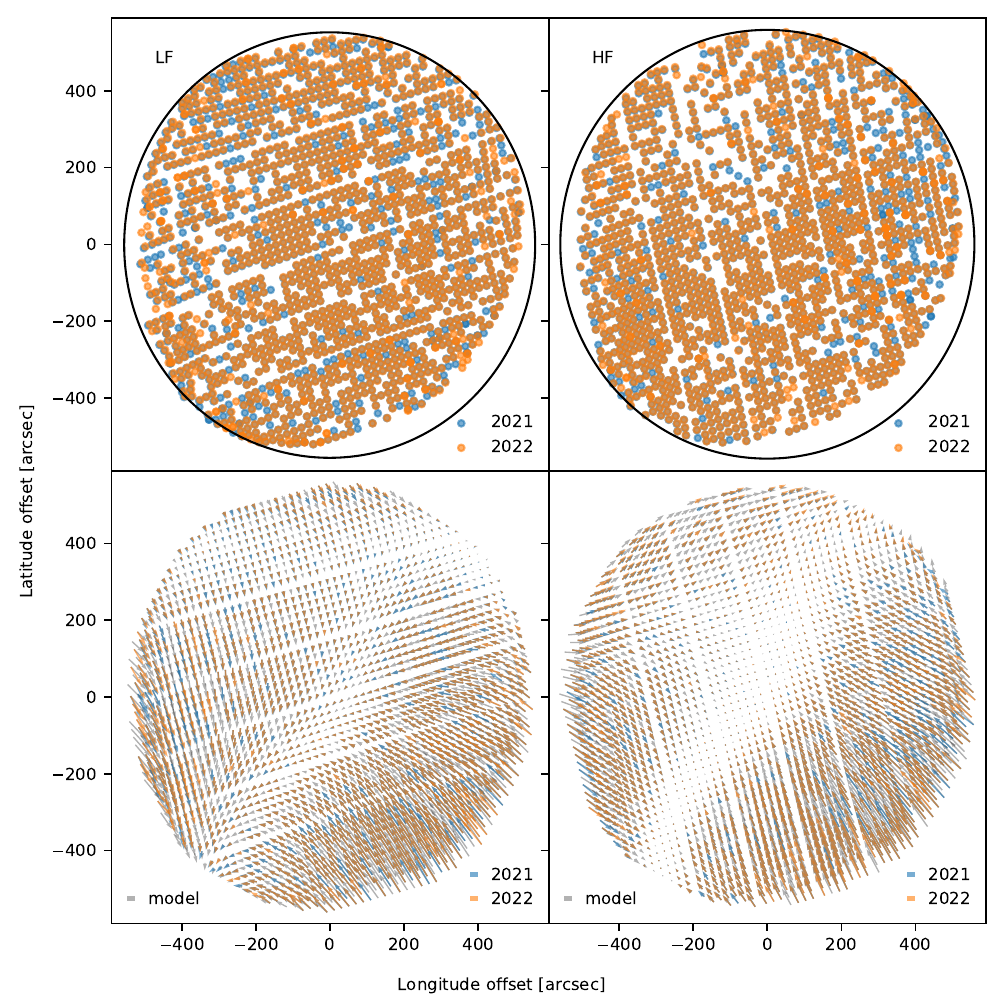}
  \caption{Focal plane deformation for the two CONCERTO arrays. The top panels present the focal plane distribution of the valid kids for the 2021 and 2022 kid sweep parameters, in Cassegrain offsets coordinates, with the LF and HF array on the left and right panels, respectively. Overall the FoV is enclosed by a circle with a diameter of 18.54\arcmin represented as a solid black line. The bottom panel shows the deformation between the constructed offset and the recovered positions in the sky. The underlying black arrows show the fitted two-dimensional polynomial of degree 4 used to match the two positions. A main deformation mode is common to both LF and HF arrays.}
  \label{fig:FocalPlaneDeformation}
\end{figure*}

To explore the focal plane deformations in the CONCERTO instrument, a comprehensive analysis was conducted by aligning the positions of observed pixels on the sky with their corresponding designed positions. Each valid position of the KID pixels was semi-automatically matched to its original geometric location using an iterative approach. Using a 2D polynomial transformation of degree 4, we could match the observed position with the designed position within less than 2\arcsec. The resulting positional offsets exhibited a remarkable consistency between the 2021 and 2022 geometries, demonstrating a high degree of stability over time. The observed focal plane deformations display strong similarities between the low-frequency and high-frequency arrays, indicating a shared source of distortion within the CONCERTO optical chain located outside the cryostat (See Fig.\ref{fig:FocalPlaneDeformation}).   Overall the FoV is enclosed by a circle with a diameter of 18.54\arcmin.

Considering this global deformation, most of the KIDs are placed within $1.6 \pm 0.9$\arcsec\ from their expected design positions for both arrays. We added an additional flag based on this offset with designed positions exceeding 5\arcsec.

Imagery with a large field of view FTS instrument such as CONCERTO is challenging, especially in terms of the optics. From a conceptual point of view, we had to perform a trade-off between the requirements related to the image quality, spectral efficiency and the maximization of the field-of-view producing a diffraction-limited combined beam for each position of the movable roof mirror. In addition to that, technological constraints were imposed to limit the total volume of the Martin-Puplett interferometer (to avoid oversized optical elements and to limit the mechanical space requirements in the APEX Cassegrain cabin).  
All these constraints drove the optical design optimization presented in \citet{2020A&A...642A..60C}, converging finally in 12 off-axis mirrors at room temperature and three polypropylene lenses at cryogenic temperatures and producing a 420\,mm diameter quasi-parallel beam inside the MPI.
Such complex optics led to the need to relax some constraints with low priority for science goals. For example, the distortion of the focal plane with respect to a circle was not considered a priority in setting the optical optimization merit function, as it does not affect the quality of the scientific analysis. We put maximum priority to the image quality and to the spectral efficiency.  

The error beam is higher than expected. This could be due to multiple factors like the variation from the flatness of the first polarizer, misalignment of the optical chain due to the variation of the centre of gravity while the Cassegrain cabin moves in elevation, small misalignments of the mechanical mounting of the mirrors or roughness of some mirrors (or some part of them).

\newpage

\clearpage 
\section{Bandpass estimate}
\label{appendix:bandpass}
To measure the on-site bandpass of CONCERTO, we conducted test scans in the MPI-ON mode with the shutter closed at the APEX site. A first estimate of the bandpass data was generated using the spectroscopic data processing pipeline developed for CONCERTO (Beelen et al., in prep). As illustrated in Fig.~\ref{bandpass}, there are noticeable wiggles in the bandpass, particularly for LF, which could stem from non-linearity induced by the electronics. To evaluate the impact of these fluctuations on our continuum results, we applied a median filtering process with a 25\,GHz wide window to smooth the measured bandpass. Subsequently, we recalculated the relevant calibration parameters. A comparison between the results obtained from the smoothed and unsmoothed bandpasses revealed a relative difference of 0.6\% for LF and 0.5\% for HF for the absolute calibration factor measured with Uranus. This analysis confirms that the effects of bandpass fluctuations are negligible for continuum measurements. Therefore, we have opted to retain the unsmoothed bandpass for this paper.

The measured spectra on the shutter are $\propto {\rm I_{RJ}(\nu)} F(\nu)\eta(\nu)\Omega(\nu)$, where $F(\nu)$ is the relative spectral response and $\eta(\nu)$ the aperture efficiency.
Because $\rm I_{RJ}(\nu)\propto \nu^2$ cancels with $\Omega(\nu) \propto \nu^{-2}$, the measured spectrum is the bandpass, $F_{meas}(\nu)=F(\nu)\eta(\nu)$, applicable for a point source.
 


\section{Unit conversion for CONCERTO \label{uc_cc}}

The conversion of temperature from Kelvin to mega-jansky per steradian (MJy/sr) for CONCERTO can be computed by considering the Rayleigh-Jeans (RJ) law, taking the bandpass into account: 
\begin{equation}
    S = \frac{2k_{B}}{c^2} \left( \frac{\int \nu^{2} F_{\rm ext}(\nu)d\nu}{\int F_{\rm ext}(\nu) d\nu} \right) T_{\rm RJ},
    \label{Rayleigh-Jeans}
\end{equation}
\noindent where $T_{\rm{ RJ}}$ is the RJ temperature,  $k_{\rm{B}}$ is the Boltzmann constant, $\nu$ is the frequency, $S$ is the flux density and $F_{\rm ext}(\nu) = F(\nu)\eta(\nu) \nu^{-2}$ refers to the bandpass for an extended source.
We use $C_{\rm{K-to-MJy/sr}}$ to denote the conversion between K and MJy/sr. Using the measured CONCERTO bandpass, we compute $C_{\rm{K-to-MJy/sr}}^{\rm{HF}}$ = 1\,876.57\,MJy/sr\,K$^{-1}$ and $C_{\rm{K-to-MJy/sr}}^{\rm{LF}}$ = 1\,203.01\,MJy/sr\,K$^{-1}$, for HF and LF, respectively. \\



\end{appendix}
\end{document}